\documentclass[fleqn,usenatbib]{mnras}

\usepackage{newtxtext,newtxmath}

\usepackage[T1]{fontenc}

\DeclareRobustCommand{\VAN}[3]{#2}
\let\VANthebibliography\thebibliography
\def\thebibliography{\DeclareRobustCommand{\VAN}[3]{##3}\VANthebibliography}

\usepackage{graphicx}	
\usepackage{amsmath}	

\usepackage{bm}  
\usepackage{booktabs}  
\usepackage[shortlabels]{enumitem}
\usepackage{balance}
\usepackage[normalem]{ulem} 

\usepackage[table,xcdraw,svgnames]{xcolor}
\hypersetup{
  pdfencoding=auto, bookmarksnumbered,
}

\newcommand{\refsec}[1]{Section~\ref{#1}}
\newcommand{\reffig}[1]{Fig.~\ref{#1}}
\newcommand{\reftab}[1]{Table~\ref{#1}}
\newcommand{\refeqn}[1]{Equation~(\ref{#1})}
\newcommand{\refeqnalt}[1]{Equation~\ref{#1}}

\newcommand{\dif}{\mathrm{d}}
\newcommand{\abs}[1]{\left\lvert #1 \right\rvert}          
\newcommand{\avg}[1]{\left\langle #1 \right\rangle}        

\newcommand{\rvir}{R_\mathrm{vir}}
\newcommand{\vvir}{V_\mathrm{vir}}
\newcommand{\mvir}{M_\mathrm{vir}}
\newcommand{\msun}{M_\odot}
\newcommand{\kpc}{\mathrm{kpc}}

\newcommand{\mi}{\mathrm{i}}
\newcommand{\mt}{\mathrm{t}}
\newcommand{\mf}{\mathrm{f}}

\graphicspath{{fig3/}}

\title[CuspCore II]{The Response of Dark Matter Haloes to Gas Ejection: CuspCore II}

\author[Li et al.]{
Zhaozhou Li,$^{1}$\thanks{E-mail: lizz.astro@gmail.com}
Avishai Dekel,$^{1,2}$
Nir Mandelker,$^{1}$
Jonathan Freundlich,$^{3}$
and
Thibaut L. Fran\c{c}ois$^{3}$
\\
$^{1}$Centre for Astrophysics and Planetary Science, Racah Institute of Physics, The Hebrew University, Jerusalem, 91904, Israel\\
$^{2}$Santa Cruz Institute for Particle Physics, University of California, Santa Cruz, CA 95064, USA\\
$^{3}$Observatoire Astronomique, Université de Strasbourg, CNRS, 11 rue de l'Université, 67000 Strasbourg, France
}

\date{}

\pubyear{2022}

\begin{document}
\label{firstpage}
\pagerange{\pageref{firstpage}--\pageref{lastpage}}
\maketitle

\begin{abstract}
We propose an analytic model, CuspCore II, for the response of dark matter (DM) haloes to central gas ejection, as a mechanism for generating DM-deficient cores in dwarfs and high-$z$ massive galaxies. We test this model and three other methods using idealized N-body simulations. The current model is physically justified and provides more accurate predictions than the earlier version, CuspCore I (Freundlich et al.\ 2020).
The CuspCore model assumes an instantaneous change of potential, followed by a relaxation to a new Jeans equilibrium. The relaxation turns out to be violent relaxation during the first orbital period, followed by phase mixing. By tracing the energy diffusion
$\mathrm{d}E=\mathrm{d}U(r)$ iteratively, the model reproduces the simulated DM profiles with $\sim$10\% accuracy or better.
A method based on adiabatic invariants shows similar precision for moderate mass change but underestimates the DM expansion for strong gas ejection. 
A method based on a simple empirical relation between DM and total mass ratios makes slightly inferior predictions.
The crude assumption used in CuspCore I, of energy conservation for shells that encompass a fixed DM mass, turns out to underestimate the  DM response, which can be partially remedied by introducing an alternative ``energy'' definition. Our model is being generalized to address the differential response of a multi-component system of stars and DM in the formation of DM-deficient galaxies.
\end{abstract}



\section{Introduction}
\label{sec:intro}

\defcitealias{2008gady.book.....B}{BT08}
\defcitealias{2020MNRAS.491.4523F}{F20a}

While the dark matter (DM) dominates the structure formation on large scales,
the baryons can alter the DM distribution in turn on galactic scales through gravity,
e.g., contraction of DM haloes due to the central condensation of baryons \citep{1986ApJ...301...27B},
dynamical heating of DM particles by the dynamical friction of infalling satellite galaxies or gas clumps 
\citep{2001ApJ...560..636E},
and puffing up of DM orbits through the gas mass/potential fluctuations
driven by feedback outflows
\citep[for temporal and spatial fluctuations respectively]{2012MNRAS.421.3464P,2016MNRAS.461.1745E}.
Moreover, the altered mass profiles of the host or satellite haloes
can affect the tidal stripping efficiency of satellites 
(and thus their abundance, e.g., \citealt{2016MNRAS.458.1559Z,2017MNRAS.471.1709G,2017MNRAS.465L..59E}) 
and the dynamical friction heating imposed on hosts \citep{2021MNRAS.508..999D}. 
Such gravitational effects of the baryons on the DM are believed crucial to resolving
the galactic-scale challenges within the current standard 
Lambda Cold Dark Matter ($\Lambda$CDM) cosmological model (see \citealt{2017ARA&A..55..343B,2022NatAs...6..897S} for a review).

One of the challenges is known as the cusp-core problem.
Cosmological simulations without baryons predict universal cuspy DM density profiles
\citep[NFW]{1996ApJ...462..563N}
with a central slope $\alpha=-\dif \log \rho_\mathrm{dm}/\dif \log r \sim 1$.
In contrast, kinematic observations report much flatter DM profiles 
with low central DM densities in many dwarf galaxies,
some of which even favor flat DM cores with $\alpha \sim 0$
\citep[e.g.,][]{1994ApJ...427L...1F,1994Natur.370..629M,1995ApJ...447L..25B,2001ApJ...552L..23D,2008AJ....136.2648D,2011AJ....141..193O,2011AJ....142...24O,2015AJ....149..180O,2016MNRAS.462.3628R,2020ApJ...904...45H}.
The discrepancy is also connected to the ``too-big-to-fail'' problem,
a mismatch in the central densities between observed dwarfs and simulated haloes
\citep{2011MNRAS.415L..40B,2016MNRAS.457L..74D}.
The cored profile in dwarf galaxies 
is commonly assumed to be the result of bursty supernova feedback
\citep[e.g.,][]{1996MNRAS.283L..72N,2012MNRAS.421.3464P,2020MNRAS.491.4523F}.
Because the supernovae energy deposited in the interstellar medium is 
comparable to or larger than the binding energy of the central gas
\citep{1986ApJ...303...39D},
supernovae can effectively eject the gas from the central regions of their host haloes.
Consequently, the DM distribution expands in the shallowed potential well after gas ejection,
leading to a reduction of the central DM density (see detailed discussion later).
This has been confirmed in more sophisticated simulations with baryonic feedback processes included
\citep[e.g.,][]{2010Natur.463..203G,2014MNRAS.437..415D,2016MNRAS.456.3542T,2020MNRAS.497.2393L,2020MNRAS.499.2912F,2022arXiv220612121W}.

The same process for forming DM cores in dwarf galaxies 
might also explain the formation of ultra-diffuse galaxies (UDGs).
UDGs have stellar masses similar to those of dwarf galaxies,
but with significantly lower central surface brightness ($\mu_\mathrm{g,0}>24 \mathrm{mag\,arcsec}^{-2}$)
and larger effective radii ($r_{1/2} > 1.5 \kpc$).
They are ubiquitous and perhaps the dominant population (e.g., \citealt{2021MNRAS.502.4262J}) 
in both the field \citep[e.g.,][]{2016AJ....151...96M,2017MNRAS.468..703R}
and galaxy groups (e.g., \citealt{2015ApJ...798L..45V,2019MNRAS.485.1036M,2020ApJ...899...69L}),
and some of them appear to be highly DM-deficient \citep{2019ApJ...874L...5V,2020NatAs...4..246G,2022MNRAS.512.3230M}.
Hydrodynamic simulations 
\citep{2017MNRAS.466L...1D,2018MNRAS.478..906C,2019MNRAS.490.5182L,2019MNRAS.487.5272J}
suggest that the field UDGs formed by the orbital expansion of stars 
in response to feedback-driven outflow episodes 
(see also \citealt{2021MNRAS.502.5370W} for a different picture involving major mergers at $z>1$).
About half of the group UDGs were field UDGs before accretion, while 
half were normal field dwarfs that turned into UDGs by tidal interactions
after being accreted onto the group.
Even in groups, a cored DM profile puffed up by stellar feedback might be a necessary condition
for boosting the tidal evolution, especially in the formation of 
highly DM-deficient galaxies \citep{2018MNRAS.480L.106O,2019MNRAS.485..382C,2022MNRAS.510.2724O}.

Very surprisingly, a similar DM mass deficit problem has recently been reported for high-$z$ massive galaxies.
Kinematic observations of massive star-forming disc galaxies at $z\sim 2$,
with stellar mass $\sim10^{11}\msun$ and halo virial mass $\sim10^{12.5} \msun$, find a
low central DM fraction ($f_{\mathrm{dm}, R_\mathrm{e}} < 0.3$) with DM cores extending to $\sim 10\kpc$ ($\sim 0.07\rvir$)
in about a third of the sample
(\citealt{2020ApJ...902...98G,2021ApJ...922..143P,2022arXiv220912199N}; 
see also \citealt{2022A&A...658A..76B,2022A&A...659A..40S} for samples with lower mass at $z\sim 1$ for comparison).
Such low central DM fractions and extended cores in massive haloes
are not reproduced in current cosmological simulations
(e.g., FIRE-2, \citealt{2020MNRAS.497.2393L}; TNG, \citealt{2018MNRAS.481.1950L,2021MNRAS.500.4597U}).
The supernova feedback is not energetic enough to expel the central gas
and alter the DM density of massive haloes \citep[e.g.,][]{1986ApJ...303...39D,2014MNRAS.437..415D}.
\citet{2021MNRAS.508..999D} proposed a hybrid scenario where 
compact satellites%
\footnote{%
  The compactness of satellites
  is important because only the compact satellites 
  are capable of penetrating deep into the hosts and heating the central cusps effectively 
  before the satellite mass is entirely stripped.
}
(or giant baryonic clumps, \citealt{2022MNRAS.514..555O})
preheat the DM cusps by dynamical friction, 
making it easier for strong outflows driven by Active Galactic Nuclei (AGN) to generate cores.
Host haloes above a mass threshold of $\sim 10^{12}\msun$ are expected to host both compact satellites and strong AGN 
\citep{2019arXiv190408431D,2021MNRAS.505..172L}.
Each of the two processes seems unable to form extended cores without the other operating in tandem.
\citet{2021MNRAS.508..999D} argue that improvements in numerical resolution and subgrid recipes of feedback models
are required for simulations to reproduce the observed massive DM cores through the proposed hybrid scenario.

Hydrodynamic simulations with feedback are powerful tools for studying galaxy formation.
However, besides suffering from resolution limitations \citep{2018MNRAS.474.3043V} and uncertainties in ad-hoc subgrid recipes of feedback
(see \citealt{2015ARA&A..53...51S} for a review),
simulations do not specify nor isolate the physical mechanisms
through which baryons affect the distribution of DM and stars,
making it difficult to generalize results.
Our goal here is to propose a 
simple analytic model that approximates the response of a non-dissipative spherical system to a rapid mass change within it, 
allowing a parametric study of the effect in different circumstances.

\begin{figure}
\centering
\includegraphics[width=0.95\columnwidth]{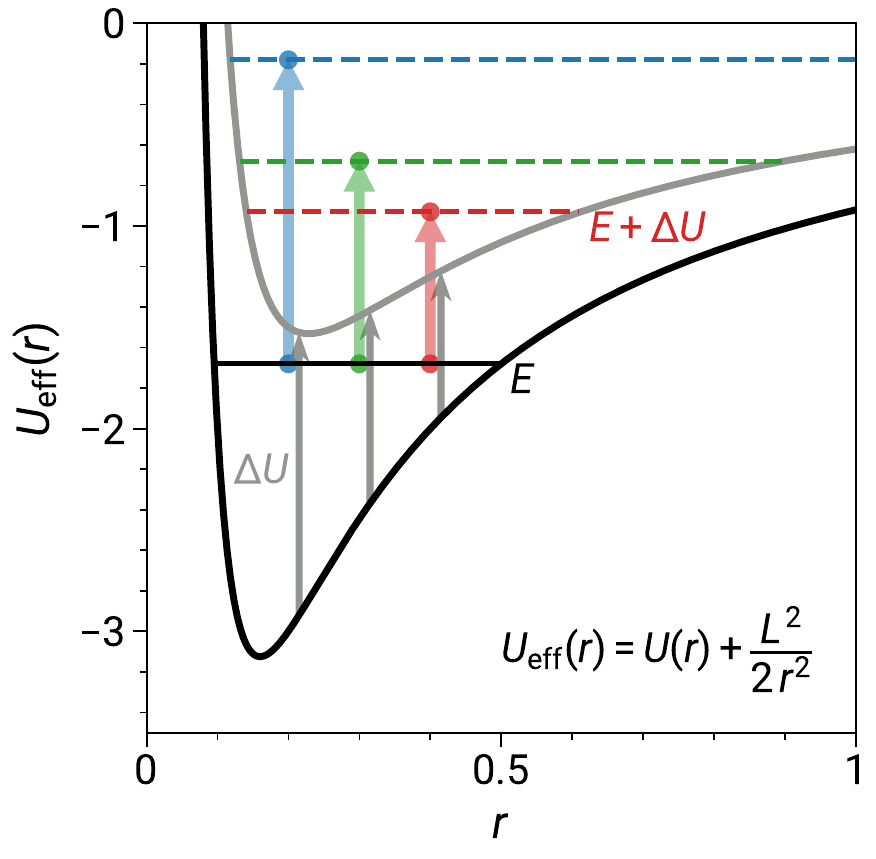}
\vspace{-1em}
\caption{%
A schematic description of the response of particle orbits to the potential change due to gas ejection.
The motion of a particle with energy $E$ and angular momentum $L$
is restricted to a radial interval where $E\geq U_\mathrm{eff}(r)=U(r)+{L^2}/{2r^2}$ (horizontal black line).
When the potential changes from the black curve to the gray curve with $\Delta U(r)$ due to 
a \textit{sudden} removal of a central mass (vertical grey arrows),
particles will move to more extended orbits with higher energy $E+\Delta U(r)$ (horizontal dashed lines).
The energy gain (vertical colored arrows) of a particle depends on the orbital phase, leading to a diffusion of energy.
The orbital expansion is irreversible even if $U(r)$ returns to the initial state 
\citep{2012MNRAS.421.3464P}.
A rigorous analysis of the halo expansion should take into consideration the self-gravity of DM  
using the methods proposed in this paper.
}
\label{fig:sketch}
\end{figure}

The basic idea of puffing up the DM halo%
\footnote{We will only refer to DM hereafter as a shorthand for general collisionless particles
including stars.
}
through baryonic feedback is illustrated in \reffig{fig:sketch}.
The gas loss due to feedback-driven outflows is usually considered as a \emph{sudden} (i.e., impulsive) event.
The resultant sudden change of potential $\Delta U(r)$
will instantaneously move particles into more extended orbits with higher energy
$E' = E  + \Delta U(r) \label{eqn:dEdU}$
in the new potential [\citealt{2008gady.book.....B} (\citetalias{2008gady.book.....B}), eq. 4.283],
leading to halo expansion.
It presents a diffusion process, because particles originally on the same orbit experience different energy gains
depending on their orbital phase.
As first demonstrated by \citet{2012MNRAS.421.3464P},
the DM expansion is irreversible even if 
the system recycles the ejected mass,
because the particles are redistributed to larger radii on average
and thus less affected by $\Delta U(r)$ than the initial state.
A flat DM core may form by either a single strong ejection or repeated outflow/inflow episodes.
A self-consistent model for the DM response with self-gravity included is yet to be performed.

\citet[][hereafter \citetalias{2020MNRAS.491.4523F}]{2020MNRAS.491.4523F}%
\footnote{Public code: \url{https://github.com/Jonathanfreundlich/CuspCore}}
presented a simple approximate analytic model, ``CuspCore'', for the relaxation of a DM halo after an instantaneous mass change.
The CuspCore model generalizes an earlier simplified analysis of an isolated shell
\citep{2016MNRAS.461.2658D} into a continuous series of shells
that encompass a fixed DM mass
(also cf.\ previous shell-based analysis by \citealt{2002MNRAS.333..299G,2002MNRAS.336..159Z}).
The model assumes energy conservation for individual shells  during the relaxation.
This crude assumption was not formally justified.
In fact, as will be shown in \refsec{sec:basic_res},
the energies of particles will continue to evolve because of the redistribution of DM.
This flaw of CuspCore can be partly remedied by 
introducing an alternative ``energy'' definition which is better 
(but not rigorously) conserved for shells (see \refsec{sec:method4}).
We find below that this remedy works well for moderate gas change but fails to reproduce 
the DM density for strong gas ejection (\refsec{sec:result}).

Here we propose a new version of CuspCore (entitled ``CuspCore II'')%
\footnote{Public code: \url{https://github.com/syrte/CuspCore2}}
that treats the relaxation process self-consistently.
It traces the diffusion of orbital energy and updates the phase-space distribution function iteratively (\refsec{sec:method1}).
The current model is physically justified and it accurately reproduces the DM response in idealized N-body simulations.
Moreover, it will allow us to model multi-component systems,
thus enabling the study of the differential response of stars and DM to outflows in the formation of DM-deficient galaxies.

The nature of the problem posed by a time-varying potential depends on how fast the potential evolves.
Though the bursty feedback process is known to be non-adiabatic \citep[][]{2012MNRAS.421.3464P},
adiabatic processes under slowly varying potentials can provide a fiducial reference.
We thus further test the possible validity of two adiabatic methods, 
an exact solution using adiabatic invariants (i.e., actions, \citealt{1980ApJ...242.1232Y})
and an empirical relation between DM and total mass ratios \citep{2020MNRAS.494.4291C}.
This comparison, in turn, also indicates the possible application of our new model in adiabatic problems.

The rest of the paper is organized as follows.
We present the idealized N-body simulations used for testing the models in \refsec{sec:testset}
and analyse the DM density profiles and the relaxation process in the simulations in \refsec{sec:basic_res}.
Then we present CuspCore II and three other models in \refsec{sec:models}
and compare them with simulations in \refsec{sec:result}.
We discuss several general issues
and comment on the details and possible improvements of the models in \refsec{sec:discuss}.
We conclude in \refsec{sec:conclusion}.

\section{Experiment with N-body simulations }
\label{sec:testset}

\begin{figure*}
\centering
\includegraphics[width=1\textwidth]{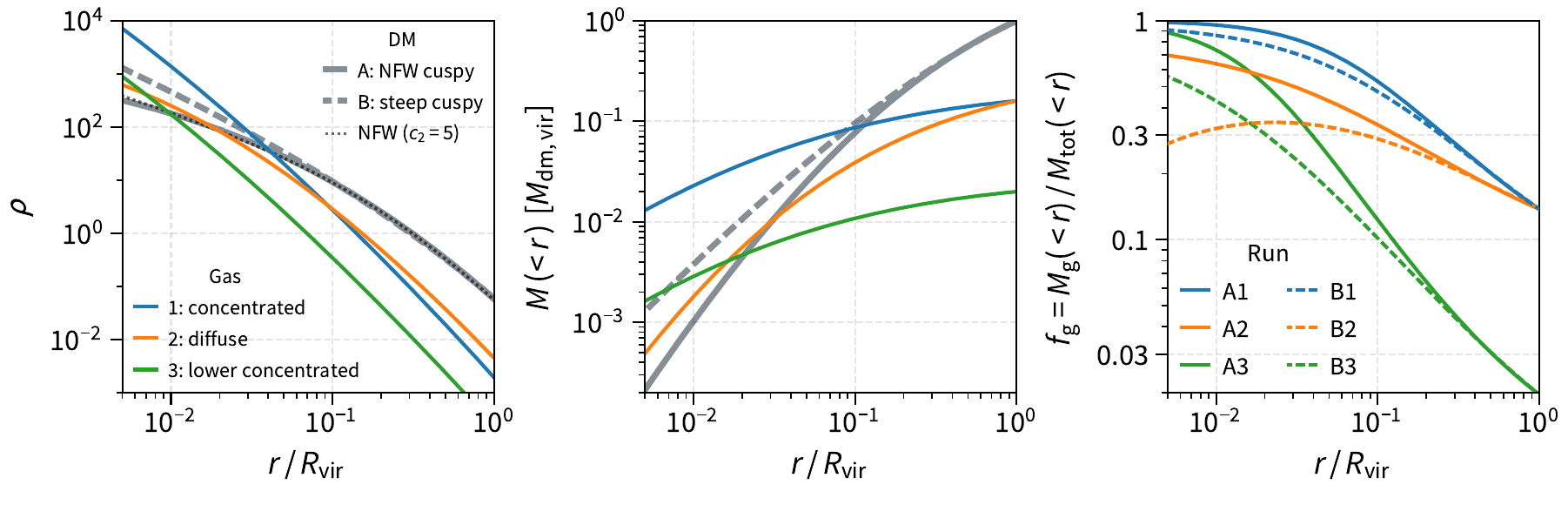}
\vspace{-2.5em}
\caption{%
The initial profiles of density (left), enclosed mass (middle), and enclosed gas fraction (right) profiles of our simulations,
including two DM profiles (A, B), three gas profiles (1, 2, 3), and thus six combinations (A1--B3).
Each density profile is described by a DZ functional form, see \reftab{tab:profile} for their parameters.
The DM profile ``A'' is taken to resemble an NFW profile with $c_2=5$ (shown as dotted line for reference),
while the DM profile ``B'' has a steeper inner slope.
We use the following unit system throughout the paper, $\rvir = 1, M_{\mathrm{dm,vir}} = 1$, and $G = 1$.
\vspace{-0.5em}
}
\label{fig:profile}
\end{figure*}

Following the earlier version of CuspCore \citepalias{2020MNRAS.491.4523F},
we consider an instantaneous change in the potential due to a rapid loss or gain of gas mass, 
followed by relaxation to a new equilibrium.
Throughout this paper,
we use the subscripts, ``dm'' and ``g'', for variables related to the DM and gas components
and ``i'', ``t'', and ``f'', for the initial equilibrium, the transitional state immediately after the change of gas mass,
and the final equilibrium after the relaxation, respectively.

Here we perform a series of N-body simulations with different combinations of initial DM profiles, gas profiles, and gas mass changes.
These systems are taken to be spherical and isotropic
($\beta=0$; see \citetalias{2020MNRAS.491.4523F} for rationale).
For simplicity, the gas profile is assumed to be static after the initial change.

\subsection{Test cases}
\label{sec:cases}

Following \citetalias{2020MNRAS.491.4523F}, the initial density profiles of DM and gas
are described by Dekel-Zhao profiles \citep[DZ]{1996MNRAS.278..488Z,2017MNRAS.468.1005D,2020MNRAS.499.2912F}.%
\footnote{The numerical implementation is available at
\url{https://github.com/JonathanFreundlich/Dekel_profile}}
This family of profiles
has a flexible inner slope and analytic expressions for the profiles of
density, mass, potential, and velocity dispersion.
It has been shown to fit DM
haloes in the NIHAO \citep{2015MNRAS.454...83W} cosmological hydrodynamic simulations better than the
other common two-parameter profiles (e.g., the generalized NFW with variable inner slope and
Einasto profiles, see fig.\ 4 of \citealt{2020MNRAS.499.2912F}). This is being confirmed in the Auriga, Apostle and
EAGLE simulations (Marius Cautun, private communication).

Within a halo of virial mass $\mvir$, radius $\rvir$, 
and mean density $\overline{\rho}_{\mathrm{vir}} = 3 \mvir / 4 \pi \rvir^3$,
a DZ profile is characterized by two shape parameters,
the logarithmic central slope, $\alpha$ ($\geq 0$),%
\footnote{
  For an isotropic system, a non-negative central slope $\alpha$ is required 
  by the non-negativity of the phase-space density \citep{2006ApJ...642..752A}.
  This has been verified explicitly for the DZ profiles by \citeauthor{2021MNRAS.503.2955B}
  (\citeyear{2021MNRAS.503.2955B} and private communication).
} and
the concentration parameter, $c$.
The density as function of radius $r$ is
\begin{gather}
  \rho (r) = \frac{\rho_\mathrm{c}}{(r/r_\mathrm{c})^{\alpha} [1 + (r/r_\mathrm{c})^{1 / 2}]^{2 (3.5 - \alpha)}},
\label{eqn:DZprof}
\end{gather}
where $r_\mathrm{c} = \rvir / c$ and 
$\rho_\mathrm{c} = (1 - \alpha / 3) c^3 c^{\alpha - 3} (1 + c^{1 / 2})^{2 (3 - \alpha)} \overline{\rho}_{\mathrm{vir}}$
are the characteristic radius and density respectively  \citep[eq.~11]{2020MNRAS.499.2912F}.
Note that the shape parameters $\alpha$ and $c$ in the DZ profile
are different from the inner density slope, $s_1$, at the resolution limit, $r_1$
($r_1=0.01\rvir$ in this work; see e.g., \citealt{2016MNRAS.456.3542T,2020MNRAS.497.2393L} for usage of $s_1$),
and the conventional concentration parameter $c_{2} = \rvir / r_{2}$ defined by the radius $r_{2}$
where the slope equals to 2.%
\footnote{
  One may obtain $s_1$ and $c_2$ through
  $s_1 = {(\alpha + 3.5 \sqrt{x_1})}/{(1 + \sqrt{x_1})}$ and $c_2 = c [ {1.5}/{(2 - \alpha)} ]^2$,
  where $x_1=r_1/r_\mathrm{c}$ (\citealt{2020MNRAS.499.2912F}, sec.\ 2.1.3).
}

Here we consider two DZ components in the halo, DM and gas.
For each component, the amplitude $\rho_\mathrm{c}$
is thus scaled with its mass enclosed within the virial radius,
$M_\mathrm{dm,vir}$ or $M_\mathrm{g,vir}$, instead of the total mass $\mvir$.
We use the following internal unit system for the simulations and throughout the paper: 
$\rvir = 1, M_{\mathrm{dm,vir}} = 1$, and $G = 1$.
Accordingly, the units for velocity, time, density, specific energy, and specific angular momentum 
are $\vvir=\sqrt{G M_{\mathrm{dm,vir}}/\rvir}$,
$t_\mathrm{vir}=\rvir/\vvir$, $M_{\mathrm{dm,vir}}/\rvir^3$, $\vvir^2$, and $\rvir\vvir$, respectively.

\begin{table}
\label{tab:profile}
\caption{
  Initial conditions of the simulations: 
  DM and gas profiles (\refeqnalt{eqn:DZprof}) 
  and fractional gas changes.
}
\begin{center}
\setlength{\tabcolsep}{0.7ex}
\begin{tabular*}{1\columnwidth}{c@{\extracolsep{\fill}}cccccl}
\toprule
Label & $M(<\!\rvir)$     & $c$             & $\alpha$     & $c_2$    & $s_1$   & Note                \\ 
\midrule
\multicolumn{5}{l}{Initial DM profile}\\
A     & 1               & 7.1             & 0.22           & 5.0      & 0.91    & NFW cuspy           \\
B     & 1               & $\mathbf{1.33}$ & $\mathbf{1.3}$ & 6.1      & 1.5     & Steep cuspy         \\ 
\midrule
\multicolumn{5}{l}{Initial gas profile}\\
1     & 0.16            & 50              & 1.7            & 1250     & 2.4     & Concentrated        \\
2     & 0.16            & 50              & $\mathbf{0}$   & 28       & 1.4     & Diffuse             \\
3     & $\mathbf{0.02}$ & 50              & 1.7            & 1250     & 2.4     & Lower concentrated  \\
\midrule
\midrule
\multicolumn{6}{l}{Fractional gas change $\eta \equiv \Delta M_\mathrm{g} / M_\mathrm{g, i}$} \\
&&      {$-1$  }      &&&&      Completely removed      \\
&&      {$-0.5$}      &&&&      Half removed            \\
&&      {$0$   }      &&&&      Unchanged               \\
&&      {$1$   }      &&&&      Doubled                 \\
\bottomrule
\end{tabular*}
\end{center}
The differences among each profile group are highlighted in bold.
The combinations of DM and gas profiles are labeled as Run A1--B3.
The total number of simulations is $2 \times 3 \times 4 = 24$.
\end{table}

To cover different conditions, we adopt the following combinations of initial DM and gas profiles.
Their DZ parameters ($c, \alpha$) are summarized in \reftab{tab:profile},
where the conventional parameters ($c_2$, $s_1$) are also listed for reference.
\setlist{nosep}
\begin{itemize}[leftmargin=*]

\item DM profile:
  \begin{enumerate}[A.,leftmargin=1.5em]
    \item {NFW cuspy}: resembling a NFW halo of $c_2=5$ 
      in the radial range $[0.01, 1] \rvir$ (see \reffig{fig:profile}),
      {which is guided by the typical hosts of observed high-$z$ massive cores \citep{2020ApJ...902...98G}.}
    \item {Steep cuspy}: representing a compact halo with a steeper inner slope
      due to baryonic contraction {(though the specific slope is chosen arbitrarily)}.
      {For ease of comparison,}
      it has the same half mass radius and thus a similar outer profile as Case A. 
  \end{enumerate}

\item Gas profile:
  \begin{enumerate}[1.,leftmargin=1.5em]
    \item {Concentrated}: a very cuspy and concentrated profile {with a gas mass close to the cosmic baryon fraction,
      which represents an extreme case where the inner halo is highly gas-dominated}.
    \item {Diffuse}: a gas profile with lower central density.
    \item {Lower concentrated}: same as Case 1 but with a lower gas mass.
  \end{enumerate}

\end{itemize}
The above DM and gas profiles are illustrated in \reffig{fig:profile}.
We label their combinations as Run A1--B3.

For each combination, we perform four separate simulations 
with different values of the fractional gas change,
$\eta \equiv M_\mathrm{g, f} / M_\mathrm{g, i} - 1 = -1$, $-0.5$, 0 and 1,
which represent the gas mass being completely removed, half removed, unchanged and doubled, respectively.
For simplicity,  we only consider a constant fractional gas change at all radii,
but the methods presented in this work can be generalized to other forms of gas mass change as well.

Therefore, we have $2 \times 3 \times 4 = 24$ simulations in total,
including several extreme cases with a complete removal of a total gas mass 
which can be as high as 16\% of the DM mass
(close to the cosmic baryon fraction).
The fraction is even higher for the inner halo, because the gas distribution is much more concentrated than the DM 
(\reffig{fig:profile}, right panel).%
\footnote{
  An exception is Run B2 where the gas fraction
  is not monotonic with radius due to the combination of a very cuspy DM halo profile and a flat gas profile.
  It is unclear if such combination exists in real galaxies.
  Anyway, this does not affect our methodological analysis.
}

\subsection{N-body simulations}
\label{sec:nbody}

The N-body simulations are performed using the public codes \texttt{NEMO} \citep{1995ASPC...77..398T}%
\footnote{\url{https://teuben.github.io/nemo}}
and \texttt{Agama} \citep{2018arXiv180208255V,2019MNRAS.482.1525V}%
\footnote{\url{https://github.com/GalacticDynamics-Oxford/Agama}}.
\texttt{NEMO} is a comprehensive stellar dynamics toolbox with a fast N-body code \texttt{gyrfalcON}
\citep{2000ApJ...536L..39D,2002JCoPh.179...27D} included, while \texttt{Agama} is a powerful and
flexible C++/Python package for dynamical modeling.

In order to generate finite haloes, we truncate the DZ profiles squared-exponentially at $r\gtrsim 4\rvir$ for both DM and gas
as input.
The radius $4\rvir$ is somewhat arbitrary 
and is taken to be large enough to ensure that the profile within $\rvir$ is only slightly affected by the truncation, but small enough to reduce unnecessary numerical calculations.

We generate the initial conditions using \texttt{Agama}. 
\texttt{Agama} computes the spherical isotropic distribution function using the
\citet[see also \refeqnalt{eqn:eddington}]{1916MNRAS..76..572E} inversion for
the input DM density profile in the total potential (DM+gas),
and then it samples DM particles in the phase space of position and velocity accordingly.
The generated DM particles are in equilibrium by construction.
The gas component is presented as an external analytic potential profile.

The DM particles are then evolved using the N-body code \texttt{NEMO}/\texttt{gyrfalcON}
under their self-gravity and an analytic external gas potential. 
The latter is implemented through the \texttt{Agama} plugin for \texttt{gyrfalcON}.
The simulation starts after the gas removal (or addition), 
when the gas potential is different from its initial configuration and thus the system is out of equilibrium.

\begin{figure*}
\centering
\includegraphics[width=1\textwidth]{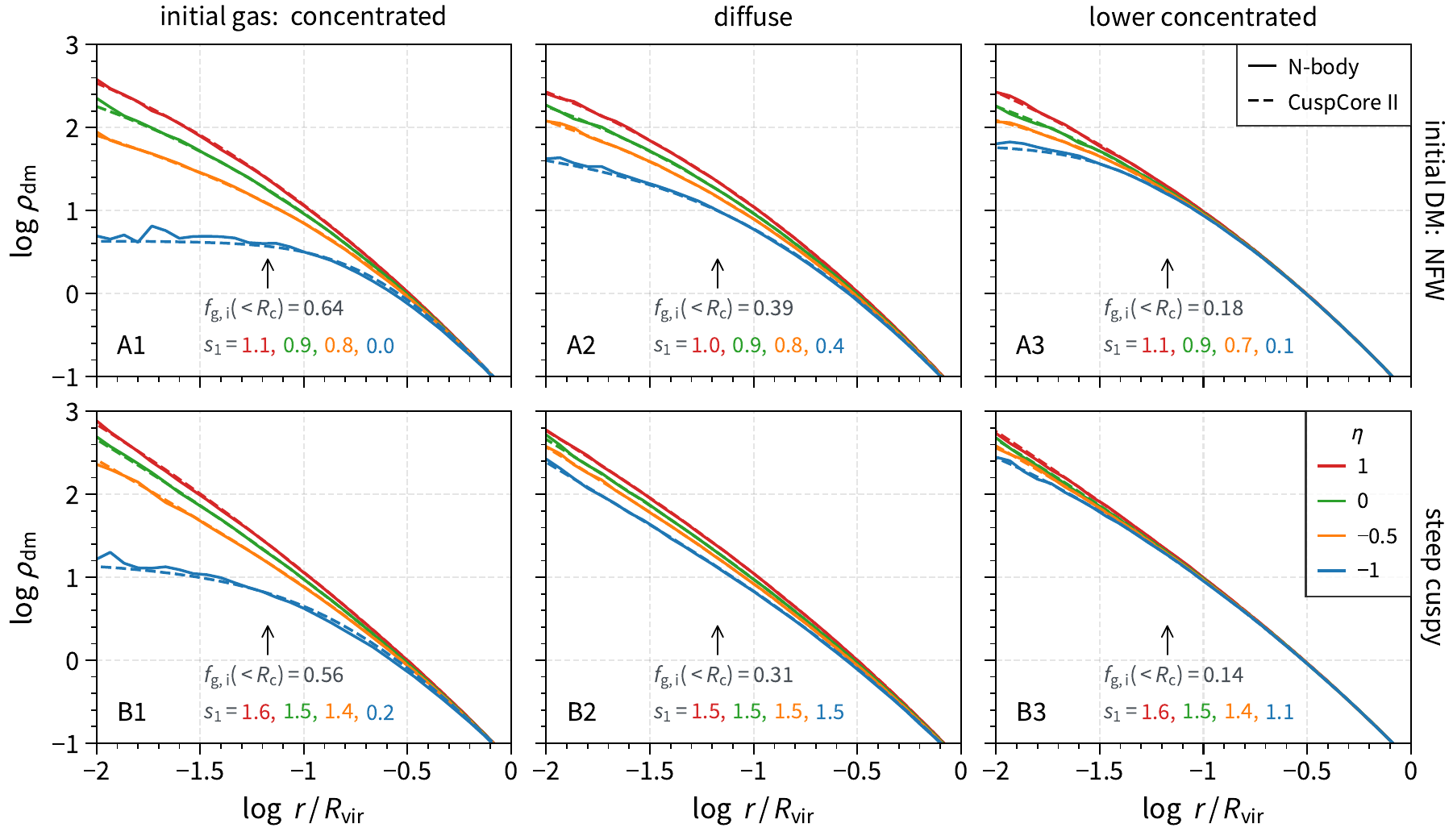}
\vspace{-1.5em}
\caption{%
The relaxed DM profiles after a change of gas mass.
We show the DM profiles of the final simulation snapshots (solid lines), 
in comparison with the predictions by CuspCore II (dashed lines; \refsec{sec:method1}).
The simulations have different initial DM profiles (rows), gas profiles (columns),
and fractional gas mass changes $\eta \equiv \Delta M_\mathrm{g} / M_\mathrm{g, i} - 1$ (colors).
Specifically, $\eta=1, 0, -0.5$, and $-1$ represent
gas mass being doubled, unchanged, half removed, and completely removed, respectively.
In the bottom of each panel,
we show $s_1$ (the negative slope at $r=0.01 \rvir$) of the predicted profiles, distinguished by colors.
We also show the initial gas fraction $f_\mathrm{g,i}$ within the typical core size of high-$z$ massive galaxies, $R_\mathrm{c}=0.067\rvir$
{(indicated by vertical arrows)} for reference.
\vspace{1em}
}
\label{fig:prof_nbody}
\end{figure*}

The mass of a single DM particle is $m_\mathrm{p} = 10^{- 6} M_{\mathrm{dm,vir}}$,
so that $\rvir$ contains $N = 10^6$ particles initially, while 
the total number of particles (including those beyond $\rvir$) is about twice as large.
We adopt a softening length $\varepsilon = 0.0015 \rvir$ with the \citet{2001MNRAS.324..273D} $P_1$ kernel%
\footnote{
  This corresponds to a Plummer-equivalent gravitational softening kernel of
  $\varepsilon_{\mathrm{plum}} = 0.001 \rvir$,
  see \citet{2001MNRAS.324..273D} for details. 
  It is why we add an additional 1.5 in the time step criteria of \citet{2003MNRAS.338...14P}
  which adopts the Plummer kernel originally.
}
and a variable time step 
$\tau = 0.2 \sqrt{\varepsilon / 1.5 | \bm{a} |} = 0.16 \sqrt{\varepsilon / | \bm{a} |}$ 
\citep{2003MNRAS.338...14P}, where $\bm{a}$ is the acceleration of a particle.
We find that the above setting ensures convergence at $r \gtrsim 0.015 \rvir$
(cf.\ \citealt{2021MNRAS.505...18E}).
The above initial conditions and numerical configuration are justified by the cases without gas change ($\eta=0$),
where the final density profiles align well with the initial analytic profiles (\reffig{fig:prof_nbody}).

For each simulation, we stop the run at $t_{\mathrm{stop}} = 12$.
As shown later in \refsec{sec:profile},
this stopping time is sufficient for the complete relaxation of the inner halo ($\lesssim 0.2 \rvir$)
and for preliminary relaxation of the halo outskirts ($\sim\rvir$).

{At each simulation snapshot, we measure the DM mass profile
at 40 radii equally spaced in logarithm within $[10^{-3}, 10^1] \rvir$ 
relative to the halo center (at the coordinate origin by construction) and interpolate $M(<\ln r)$ by a cubic spline.
The density profile is then calculated from the derivative of the spline, $4\pi r^3\rho(r)=\mathrm{d}M/\mathrm{d}\ln r$.
}

\section{Results from N-body simulations}
\label{sec:basic_res}

Before comparing the simulations with the models,
we first present some results directly obtained from N-body simulations,
which can help us understand the relaxation process after gas removal/addition
and provide useful insights for developing the theoretical models.

\subsection{Density profile}
\label{sec:profile}

\begin{figure*}
\centering
\includegraphics[width=1\columnwidth]{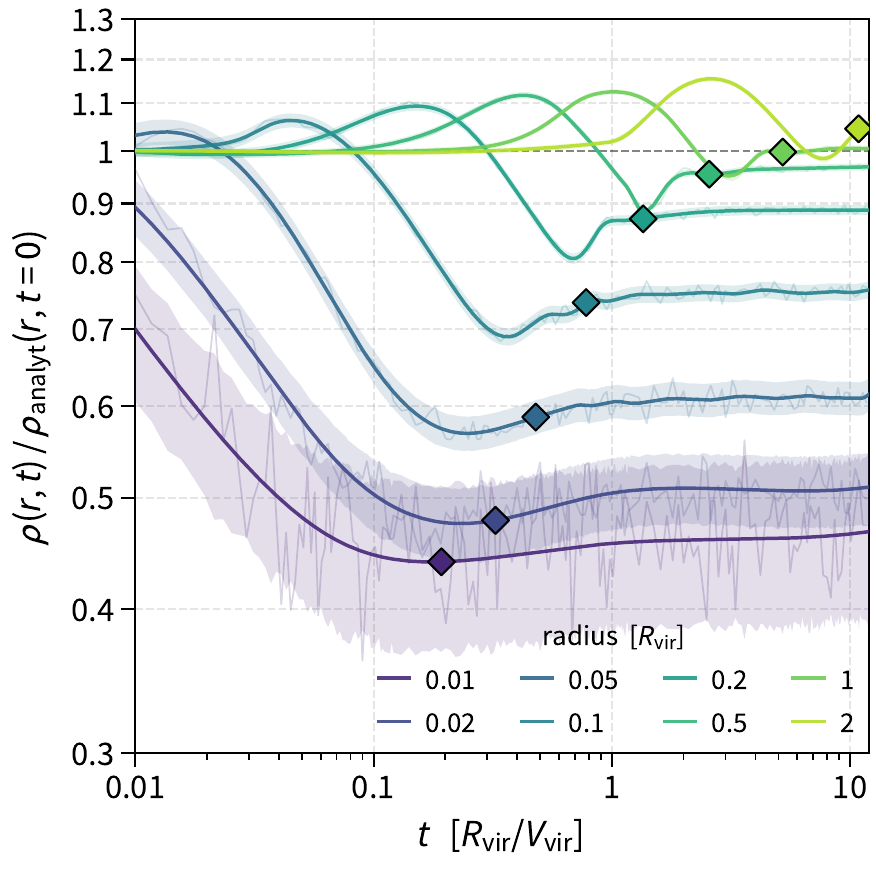}\hfill
\includegraphics[width=1\columnwidth]{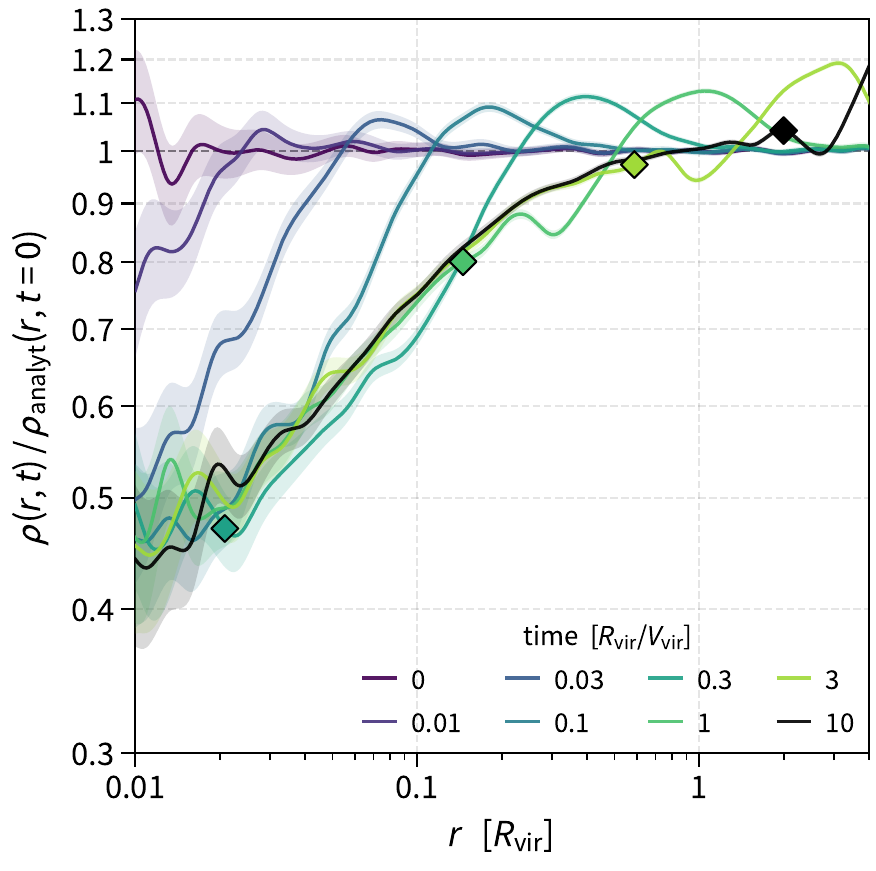}
\vspace{-0.5em}
\caption{%
Time evolution of the DM density profile relative to the initial {\textit{analytic}} profile, $\rho(r,t)/\rho_\mathrm{analyt}(r,t=0)$,
for the simulation A1 with $\eta=-0.5$.
\emph{Left panel}: the densities at given radii as a function of time.
\emph{Right panel}: the density profiles at different times.
For visual clarity, the curves in the left panel are smoothed using penalized splines.
The shades specify rough estimates for the uncertainty of density due to Poisson fluctuation.
The square on each curve indicates the average radial period $\avg{T_r}$ of the particles at given radius 
in the final snapshot for reference.
The DM density at a given radius oscillates during the first average orbital period and becomes roughly stabilized afterwards.
}
\label{fig:rhot_t}
\end{figure*}

\reffig{fig:prof_nbody} shows the relaxed DM profiles in the final simulation snapshots.
As expected, the inner DM profiles become lower (or higher)
than their initial values in response to the gas removal (or addition).
The response is stronger at radii where the gas mass change relative to the total mass
is higher (cf.\ \reffig{fig:profile} for the initial gas fraction profile).

It might be interesting to examine the gas removal from the inner halo 
that is required to reproduce the observed flat cores of $\sim10\kpc$ in high-$z$ massive galaxies \citep{2020ApJ...902...98G}.
For an NFW halo at $z=2$ with $\mvir=10^{12.5}\msun$ and typical $c_2=5$ (hence $\rvir\sim150\kpc$),
a core of $10\, \kpc$ corresponds to $R_\mathrm{c}=0.067\rvir$.
We thus provide the initial gas fraction within this radius in \reffig{fig:prof_nbody} for reference.
A gas mass equal to $14\%$--$64\%$ of the total mass within $R_\mathrm{c}$
is removed from simulations when $\eta=-1$ and half the numbers when $\eta=-0.5$.
As shown later in \refsec{sec:result}, though somewhat arbitrary,
we may use the mass fraction removed within $R_\mathrm{c}$
as a good indicator for the strength of gas ejection for the inner halo.

Among all the simulations, 
the complete removal of concentrated gas in a relatively shallow DM halo (Run A1 with $\eta=-1$) manifests the maximum impact.
In this case, after removing a gas mass as high as 64\% of the total mass within $0.067\rvir$,
a flat core forms and extends to $\sim 0.1\rvir$ where $\rho_\mathrm{dm}$ becomes 0.1 dex lower than the central density.
This core size is comparable to and larger than the typical core size of high-$z$ galaxies \citep{2020ApJ...902...98G}.
In contrast, a contracted DM halo (e.g., Run B1) is more resistant to the same gas mass change.
As mentioned in \refsec{sec:intro},
to create cores in a such halo it might be necessary to
preheat the halo by dynamical friction of compact satellites,
which can make the DM more responsive to gas ejection \citep{2021MNRAS.508..999D}.
We do not address preheated DM in this paper.

In \reffig{fig:prof_nbody}, we also quote the inner density slopes $s_1$ at $r=0.01\rvir$.
The slope after relaxation exhibits a large diversity among the simulations.
We note that the change in DM density is larger in Run A2 than in A3, but the change in $s_1$ is in the opposite order.
It is because the change of the inner slope is mainly determined by the local $\Delta M_\mathrm{g}(<0.01\rvir)$
rather than the overall $\Delta M_\mathrm{g}$.
Therefore, the inner slope does not necessarily change at the same level
as the DM deficit in the core region.

To trace the process of the relaxation,
we show in \reffig{fig:rhot_t} the time evolution of the DM profile $\rho(r,t)$ for 
Run A1 with half the gas removed ($\eta=-0.5$, a moderate mass change).
The DM density is presented relative to the initial analytic profile, $\rho(r,t)/\rho_\mathrm{analyt}(r,t=0)$.
For reference, we mark the average radial period $\avg{T_r}$ of the particles at given radii in the final snapshot,
where the radial period of a particle is twice the time moving between its pericenter and apocenter, 
$T_r=2\int_{r_\mathrm{per}}^{r_\mathrm{apo}} \dif r/\abs{v_r}$.
As shown in the left panel,
at a given radius, the density first goes up, down, and up again, then roughly stabilizes
after the first orbital period $T_r$.
The same trend is also clear in the right panel.
For gas addition cases, we obtain similar results but in reverse direction correspondingly.
The quick stabilization of density profile suggests that
the stopping time $t_\mathrm{stop}=12$ is sufficient to obtain 
a well relaxed profile for the inner halo $\lesssim0.2\rvir$ 
and an approximate stabilization at $\rvir$.

We interpret the density oscillations and settling along the following lines (see also \citealt{2018MNRAS.473..498P}).
After the sudden gas removal,
the system starts to expand in the shallowed potential.
The orbital period of particles in an inner shell is usually smaller than that in an outer shell.
Consequently, when an inner shell expands and approaches its apocenter,
an outer shell is still expanding.
This makes the mass between the two shells condense, leading to an overdense caustic-like feature propagating in the phase space.
Similarly, when the inner shell turns around at apocenter and falls back,
an underdensity is created outside the shell.
However, we see that such an oscillation in density at a given radius is prominent only in the first orbital period
and it decays fast due to the gradual phase mixing.
This suggests that the violent relaxation that is associated with the time-varying potential is effective
only during the first orbital period \citep{1967MNRAS.136..101L}, while the phase mixing dominates at later times.
This reminisces of the {fast} relaxation after a gravitational collapse or major merger 
(\citetalias{2008gady.book.....B}, sec.\ 4.10.3; \citealt{2004MNRAS.349.1117B}).

\subsection{Orbital evolution of sample particles}
\label{sec:example_orbit}

\begin{figure*}
\centering
\includegraphics[width=0.99\textwidth]{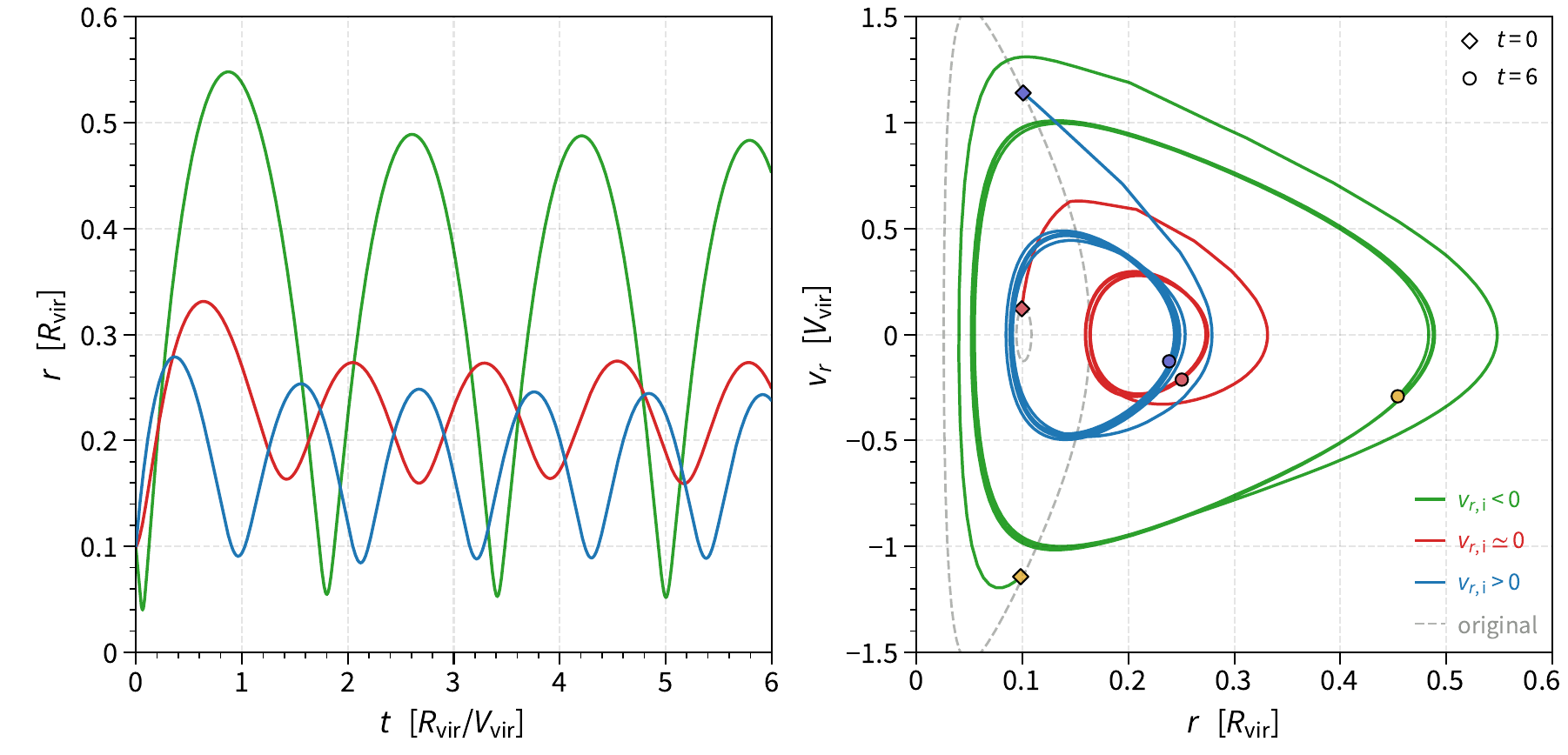}\\ \vspace{1mm}
\includegraphics[width=0.99\textwidth]{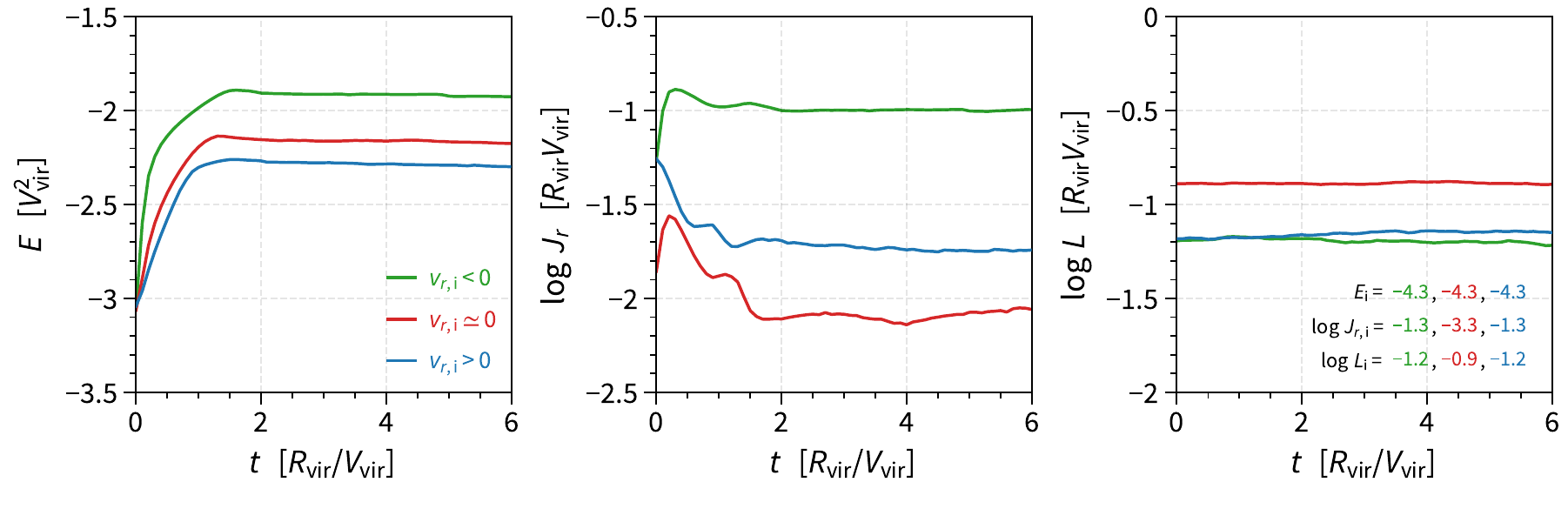}
\vspace{-1em}
\caption{%
Orbits of three example particles in the simulation A1 with complete gas removal ($\eta=-1$).
Upper left and lower panels: 
the time evolution of radial position $r$, orbital energy $E$, radial action $J_r$, and angular momentum $L$.
Upper right panel: the trajectories in the $v_r$-$r$ space, where $v_r$ is the radial velocity.
Note that the area enclosed by an orbit in the $v_r$-$r$ space is $2\pi J_r$.
The particles are selected to have the same initial radius and total velocity but different $v_{r,\mathrm{i}}$.
Two particles (blue and green) shared the same orbit in the initial potential \emph{before} the gas removal,
and the third (red) was in a nearly circular orbit with $v_{r,\mathrm{i}}\simeq 0$.
The values of $E$, $J_r$, and $L$ in the initial potential are quoted in the bottom of the lower-right panel for reference.
The orbits and associated orbital integrals, $E$ and $J_r$, evolve significantly
only during the first radial period, which suggests that the violent relaxation is effective
only in the early phase of the relaxation.
}
\vspace{0.5em}
\label{fig:eg_orbit}
\end{figure*}

\reffig{fig:eg_orbit} shows the orbits of three example particles in 
the simulation A1 with complete gas removal ($\eta=-1$, a strong mass change).
The particles are selected to have the same initial radius and speed (but different moving directions)
and thus the same energy in the initial potential.
Two particles share the same original orbit with orbital circularity $\epsilon=L/L_\mathrm{cir}=0.5$,
and the third is in a nearly circular orbit with $\epsilon=0.995$,
where $L_\mathrm{cir}$ is the angular momentum of the circular orbit with the given energy.
Their orbits expand due to both the initial removal of gas and redistribution of DM,
with both the pericenter and apocenter increased \citep[cf.][]{2021ApJ...921..126B}.

These orbits expand and ``overshoot'' in the first period and roughly stabilize afterwards
with apocenter radii slightly smaller than their first apocenter radii
(cf.\ \citealt{1996MNRAS.283L..72N}; also \citetalias{2008gady.book.....B}, 
{sec.\ 4.10.3, for similar ``overshoot'' in collapse}).
The quick stabilization suggests again that the violent relaxation is only effective in the first orbital period,
followed by the phase-mixing stage, in consistency with \reffig{fig:rhot_t}.
This is also confirmed by the little evolution of the orbital integrals after the first period 
(to be seen in the lower panels of \reffig{fig:eg_orbit}).

\begin{figure*}
\centering
\includegraphics[width=0.99\textwidth]{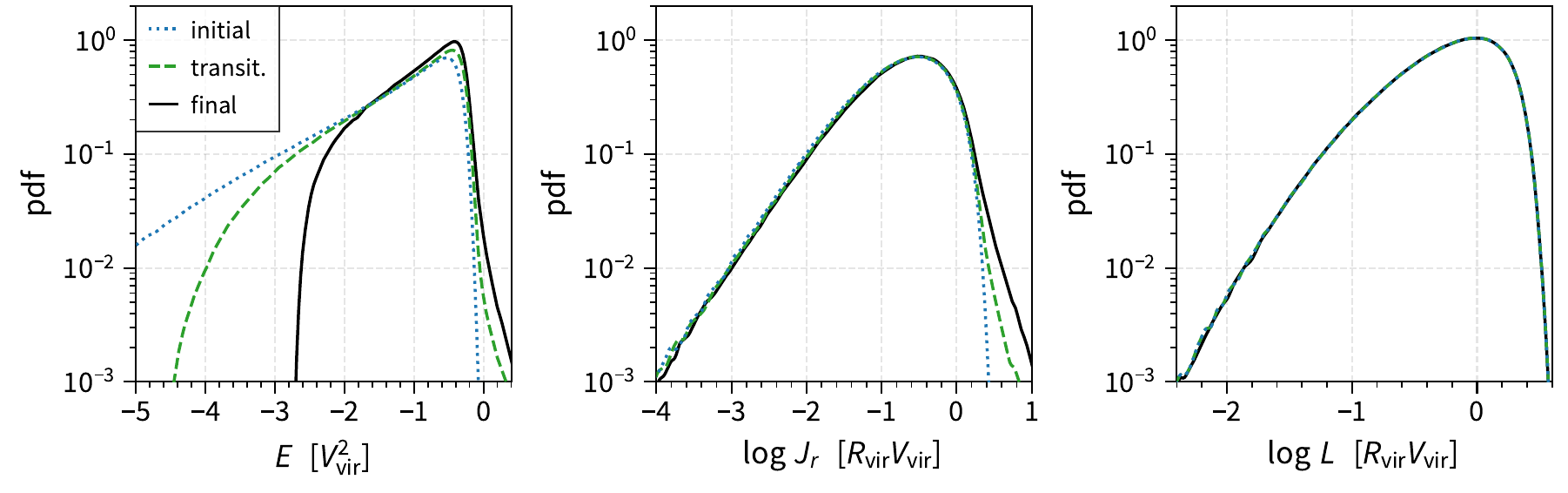}
\vspace{-1em}
\caption{%
Relaxation in a simulation.
The panels show the ensemble distribution of orbit integrals, 
including the specific orbital energy $E$, radial action $J_r$, and angular momentum $L$,
at the initial (dotted), transitional (dashed), and final states (solid)
of the simulation A1 with complete gas removal ($\eta=-1$).
}
\label{fig:orbint_pdf}
\end{figure*}

\begin{figure*}
\centering
\includegraphics[width=0.88\textwidth,trim={0 0 0 1em},clip]{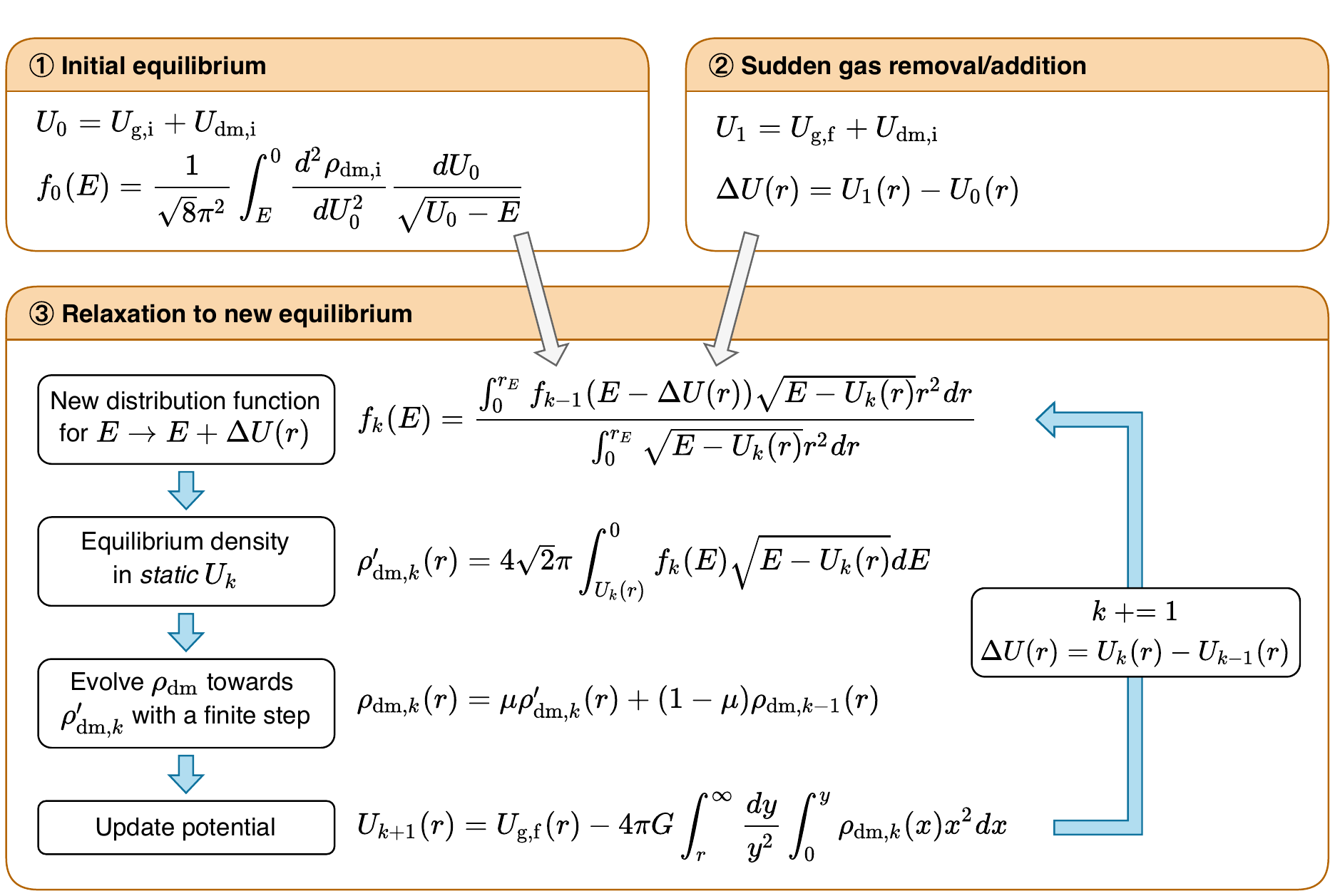}
\vspace{-0em}
\caption{%
Flowchart of CuspCore II, which solves the final DM density profile by iteratively tracing the energy diffusion 
$E'=E+\Delta U(r)$.
}
\vspace{0.5em}
\label{fig:flowchart}
\end{figure*}

\subsection{Evolution of orbital integrals}
\label{sec:orb_integ}

For a spherical system in equilibrium, the phase-space \emph{distribution function},
$f(\bm{r},\bm{v})\equiv \dif^6 M / \dif^3 \bm{r} \dif^3 \bm{v}$, 
can be expressed as a function of integrals of motion (\citetalias{2008gady.book.....B}, sec.\ 4.2),
e.g., $f(E,L)$ or $f(J_r, L)$,
where $E$, $L$, and $J_r$ are the specific orbital energy, angular momentum, and radial action respectively.
The radial action of an orbit (\citetalias{2008gady.book.....B}, eq.\ 3.224)
is the integral of the radial velocity $v_r$ along this orbit from the pericenter 
to the apocenter, $J_r = \frac{1}{\pi} \int_{r_\mathrm{per}}^{r_\mathrm{apo}} |v_r| d r$,
which is proportional to the area enclosed by the orbit in the $v_r$-$r$ space.
If we can quantify the ensemble evolution of the orbital integrals,
we are able to predict the final density profile by integrating the distribution function over velocities
(see \refeqnalt{eqn:rhof}).

The lower panels of \reffig{fig:eg_orbit} show the evolution of the orbit integrals of 
the same three example particles as \refsec{sec:example_orbit}.
\reffig{fig:orbint_pdf} shows the ensemble distribution of orbital integrals 
at the initial, transitional, and final states for all DM particles in the same simulation.
We also show the change of integrals of all individual particles in Appendix \ref{sec:integral_change}, \reffig{fig:orbint_diff}.
In all cases, $L$ is trivially conserved due to the spherical symmetry,
while $E$ and $J_r$ exhibit significant diffusion.

The energies of the three particles have increased significantly 
due to the evolving potential but not at the same level (\reffig{fig:eg_orbit}).
For a particle, $\Delta E$ depends on both the initial position and velocity.
The change of potential is higher in the inner halo.
Therefore, a particle initially located close to the center or 
a particle moving inwards at the time of gas removal 
will experience a larger cumulative change of potential along its orbit history,
$\Delta E= \int \tfrac{\partial}{\partial t} U (r (t), t) \dif t$,
thus exhibiting a greater orbit expansion.
By tracing $\Delta E$ statistically,
it is possible to predict the final distribution function and thus the density,
which motivates the CuspCore II model (\refsec{sec:method1}).

In spherical systems, $J_r$ and $L$ are known as adiabatic invariants
which are conserved if the potential changes sufficiently slow (\citetalias{2008gady.book.....B}, sec.\ 3.6).
Clearly, the sudden mass change and subsequent relaxation in our case are 
not adiabatic \citep[e.g.,][]{2012MNRAS.421.3464P,2019MNRAS.485.1008B},
so $J_r$ of a particle can either increase or decrease significantly (easily by a factor of 2 or more as shown in \reffig{fig:eg_orbit} and \ref{fig:orbint_diff}).
Interestingly, in contrast to the large variation of $\Delta J_r$ for individual particles,
the ensemble $p(J_r)$ varies much less, except for the tail being broadened greatly.
This is because the increases and decreases of $J_r$ for individual particles largely balance each other, 
making the mean $\Delta J_r$ small (\reffig{fig:orbint_diff}).
The median value of $J_r$ has increased by 4\% due to the initial gas removal,
and by another 6\% during the subsequent relaxation, thus by 10\% in total.
Note that this simulation (A1 with $\eta=-1$) represents the case with the strongest gas ejection in our sample.
It suggests that the adiabatic approximation might be valid to a certain extent
at least for cases with weaker gas ejection,
which motivates the possible usage of the adiabatic methods presented in Sections \ref{sec:method2} and \ref{sec:method3}.

\section{Modeling the DM response}
\label{sec:models}

In this section, we present four different methods for modeling the relaxation process
of DM after a sudden gas mass change.
We first describe our new model based on energy diffusion (\refsec{sec:method1}),
then we present two adiabatic methods  (\refsec{sec:method2} and \ref{sec:method3})
and finally the earlier version of CuspCore (\refsec{sec:method4}) for comparison.

\subsection{Method I: Energy diffusion assuming \texorpdfstring{$\Delta E = \Delta U(r)$}{ΔE=ΔU(r)}}
\label{sec:method1}

We propose a new method (entitled "CuspCore II")
modeling the DM relaxation based on iteratively tracing the energy diffusion.
We describe the iterative procedure in the following (see \reffig{fig:flowchart}).
The validity of the underlying assumptions will be discussed in Section \ref{sec:dEdU_assumps},
and a note on the numerical implementation is provided in Appendix \ref{sec:implement}.

For a \emph{spherical} and \emph{isotropic} system in equilibrium, 
the phase-space distribution function, 
$f (\bm{r}, \bm{v}) \equiv \dif^6 M / \dif^3 \bm{r} \dif^3 \bm{v}$
can be expressed as a function of energy, $f(E[\bm{r}, \bm{v}])$.
Given a DM density profile, $\rho_\mathrm{dm,i}(r)$,
subject to its self-gravity and an external gas potential, $U_0(r)=U_\mathrm{dm,i}(r)+U_{\mathrm{g,i}}(r)$,
the distribution function of DM particles can be obtained though the
Eddington (\citeyear{1916MNRAS..76..572E}; also \citetalias{2008gady.book.....B}, eq. 4.46) inversion,
\begin{gather}
  f_0 (E) = \frac{1}{\sqrt{8} \pi^2}  \int_E^0 \frac{\dif^2 \rho_{\mathrm{dm, i}}}{\dif U_0^2}  \frac{\dif U_0}{\sqrt{U_0 - E}}.
\label{eqn:eddington}
\end{gather}

Then this equilibrium state is broken by an instantaneous potential change due to gas removal/addition, 
$\Delta U(r)$.
The energy of a particle in a varying potential is changing
with time as $\dif E = \frac{\partial}{\partial t} U (r(t), t) \dif t$ or equivalently
$\Delta E = \Delta U (r)$ for a short time interval (\citetalias{2008gady.book.....B}, eq. 4.283).  
It describes an energy diffusion process,
where the particles with the same initial energy but located at different radii now have different energies (see \refsec{sec:orb_integ}).
By tracing this diffusion,
we can derive the consequent DM distribution function and density profile.
Because the DM contribution to the potential is  varying itself during the relaxation,
we have to resort to an iterative procedure.

For each iteration step, we consider a very short time interval $\Delta t$, 
in which the potential changes from $U_{k-1}$ to $U_{k}=U_{k-1}+\Delta U$. 
In our problem, $\Delta U$ is initially computed from the instantaneous gas mass change, 
$\Delta U=U_1-U_0 = U_{\mathrm{g,f}}-U_{\mathrm{g,i}}$,
and then updated by the difference between adjacent steps during the subsequent DM relaxation.
Following $\Delta E = \Delta U (r)$, the energy distribution of DM particles, $N(E)\equiv \dif M/\dif E$, becomes
\begin{gather}
  N_{k} (E) = 16 \sqrt{2} \pi^2 \int_0^{r_E} f_{k-1} (E -
  \Delta U (r)) \sqrt{E - U_k (r)} r^2 \dif r, 
\label{eqn:Nvar}
\end{gather}
where $r_E$ is the radius satisfying $U_k
(r_E) = E$ (see Appendix \ref{sec:der_Nvar} for derivation).

Assuming that the potential $U_{k}$ is static
and the system remains isotropic during relaxation (see discussion in \refsec{sec:dEdU_assumps}),
the phase-mixed distribution function (\citetalias{2008gady.book.....B}, eq. 4.58) under $U_{k}$ would be
\begin{gather}
  f_k (E) = {N_k (E)}/{g_k (E)},
\label{eqn:f=N/g}
\end{gather}
where $g_k (E)$ is the volume of phase space per unit energy (\citetalias{2008gady.book.....B}, eq. 4.56)
associated with the potential,
\begin{gather}
  g_k (E) = 16 \sqrt{2} \pi^2 \int_0^{r_E} \sqrt{E - U_k (r)} r^2 \dif r.
\label{eqn:g} 
\end{gather}

The equilibrium density that we would have for a tracer population with energy distribution $N_k (E)$
under a static potential $U_k$ is
\begin{gather}
  \rho_{\mathrm{dm},k}' (r) = \int f_k \dif ^3 \bm{v} = 4 \sqrt{2} \pi \int_{U_k (r)}^0 f_k (E) \sqrt{E - U_k (r)} \dif E
\label{eqn:rhof}
\end{gather}
(\citetalias{2008gady.book.....B}, eq. 4.43).
However, $\rho_{\mathrm{dm},k}' (r)$ is not the correct density because the potential is not really static. 
As the DM density evolves towards $\rho_{\mathrm{dm},k}'$, the underlying potential due to its self-gravity is changing as well.
Therefore, we have to update the density profile by finite steps before the potential changes significantly.
We thus update the density with a small step parameter $\mu\in(0,1)$,
\begin{gather}
  \rho_{\mathrm{dm},k} (r) = \mu \rho_{\mathrm{dm},k}' (r) + (1 - \mu) \rho_{\mathrm{dm},k-1} (r).
\label{eqn:newrho}
\end{gather}
We find that the final results are nearly identical for any $\mu \lesssim 0.25$
and therefore adopt $\mu = 0.125$ in this work
(see \refsec{sec:extension} for more discussion).

Based on the Poisson equation, the new potential $U_{k+1}$ is
\begin{gather}
  U_{k+1} (r) = U_{\mathrm{g,f}} (r) - 4 \pi G \int^{\infty}_r \frac{\dif y}{y^2} \int_0^y \rho_{\mathrm{dm},k} (x) x^2 \dif x.
\label{eqn:newU}
\end{gather}
The difference between $U_{k+1}$ and $U_{k}$ will in turn be used to trace the energy diffusion via Equation (\ref{eqn:Nvar}).
Repeating the procedure described in Equations (\ref{eqn:Nvar} -- \ref{eqn:newU}) until convergence,
we can obtain the final relaxed DM profile, $\rho_\mathrm{dm,f}$.

The above procedure is based on tracing the detailed diffusion of $E$ (\refeqnalt{eqn:Nvar}).
Alternatively, one can try to evolve $N(E)$ by the mean energy change $\avg{\Delta E}$ as function of $E$.
We find $\avg{\Delta E} \simeq \Delta U(r_\mathrm{cir})$ to be 
a good approximation (where $r_\mathrm{cir}$ is the circular orbit radius,
see Appendix \ref{sec:integral_change}),
which enables another solution with similar precision (provided in Appendix \ref{sec:method1_var} for interested readers).

\begin{figure*}
\centering
\includegraphics[width=1\textwidth]{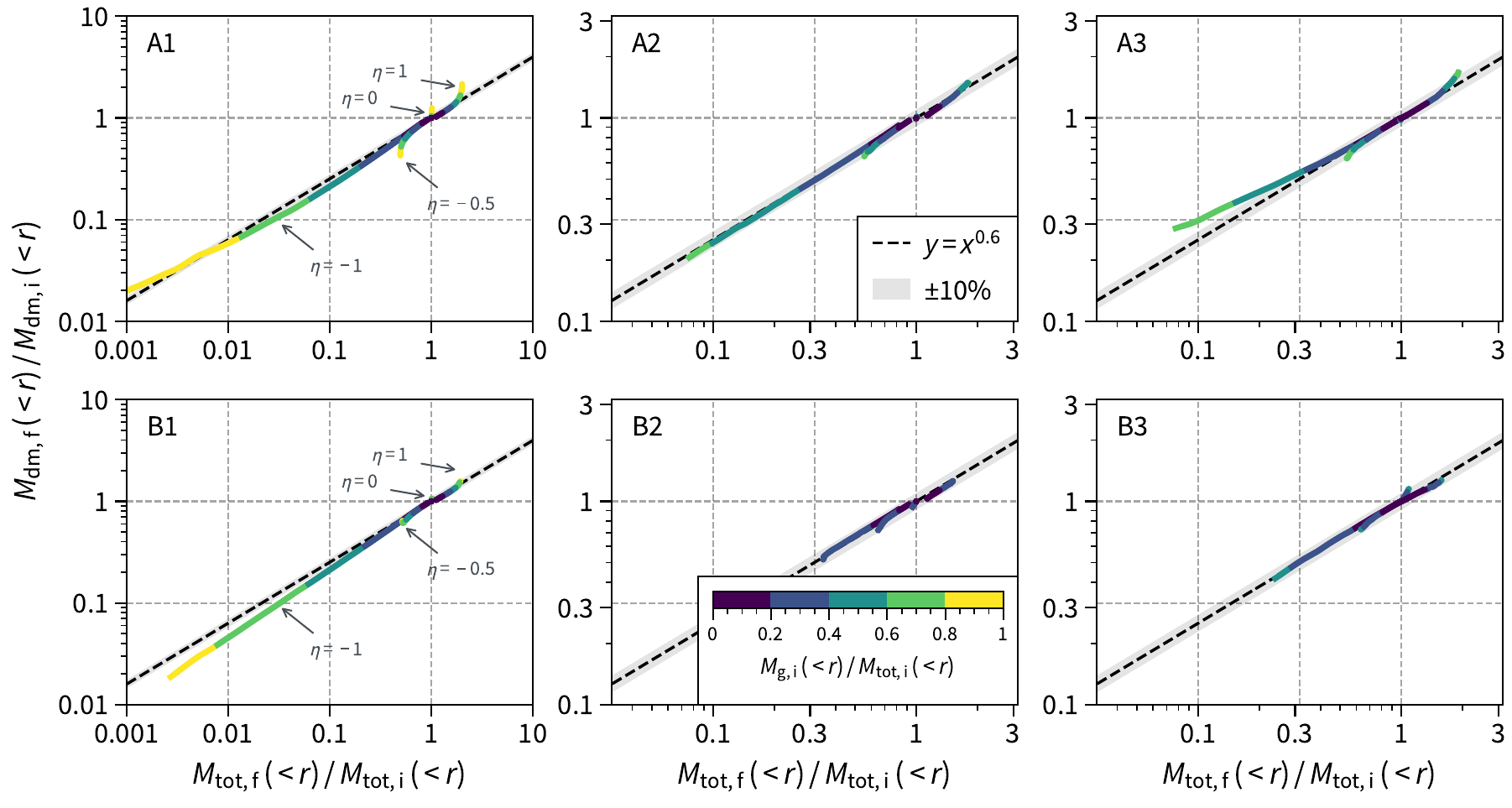}
\vspace{-2em}
\caption{%
Empirical basis for Method III,
the relation between DM and total mass ratios in the simulations.
The ratios are calculated for the masses in the 
initial equilibrium before gas change (``i'') and the final equilibrium after relaxation (``f'').
Each panel contains 4 curves, corresponding to the fractional gas change $\eta=-1, -0.5, 0$, and 1 respectively,
as annotated in the left panels. 
Each curve shows the relation between the mass ratios
for a series of $r \in [0.01, 3] \rvir$, colored by the initial gas fraction 
$f_\mathrm{g,i}(r)=M_\mathrm{g,i}(<r) / M_\mathrm{tot,i}(<r)$.
A power-law relation $y=x^{0.6}$ is shown as dashed line with a 10\% error band for reference.
}
\label{fig:mass_ratio}
\end{figure*}

\subsection{Method II: Adiabatic invariants}
\label{sec:method2}

As shown in \refsec{sec:orb_integ},
there is a large variation in the radial action $J_r$ of individual particles
during the relaxation to the new equilibrium.
However, the ensemble distribution $p(J_r)$ is much less affected 
even in the case with strong gas ejection.
Together with the conservation of the angular momentum $L$,
the small change in $p(J_r)$ seems to motivate the usage of an adiabatic method
which assumes that the distribution of actions, $p(\bm{J})\equiv p(J_r , J_\theta=|\bm{L}|-|L_z| , J_\phi=L_z)$, is invariant.
Here $J_r$, $J_\theta$, and $J_\phi$ are the radial, latitudinal, and azimuthal actions respectively,
and $L_z$ is the $z$-component of the angular momentum vector
(\citetalias{2008gady.book.....B}, sec.\ 3.5.2).

Given a potential $U$ and thus the mapping between $(\bm{r},\bm{v})$ and $\bm{J}$
(denoted as $\bm{J}[\bm{r},\bm{v}]_U$),
the phase-space distribution function can be expressed as 
$f (\bm{r}, \bm{v}) = f(\bm{J}[\bm{r},\bm{v}]_U)=(2\pi)^{-3}p(\bm{J})$.
The corresponding DM density profile $\rho_\mathrm{dm}$ 
can be obtained by integration over velocities (\refeqnalt{eqn:rhofj}).
As the potential itself depends on $\rho_\mathrm{dm}$,
again we have to resort to iterations to solve $\rho_\mathrm{dm}$.
This method was first developed to study the profile of star clusters 
with a growing central massive black hole \citep{1980ApJ...242.1232Y}
and later applied to model the adiabatic contraction of DM haloes
due to the concentrated baryons (\citealt{1986ApJ...301...27B,2005ApJ...634...70S,2020MNRAS.495...12C}).

The iterative solution is very similar to Method I.
The main conceptual difference is that the distribution function $f$ 
here is expressed as an invariant function of actions, $f(\bm{J})$,
while $f$ in Method I is expressed as a function of energy that evolves with time.
The invariant $f(\bm{J})$ is constructed from the initial equilibrium
by the \citet{1916MNRAS..76..572E} inversion (\refeqnalt{eqn:eddington})
with the mapping $f(\bm{J})=f(E[\bm{J}]_{U_0})$
under the total potential, $U_0=U_\mathrm{dm,i}+U_{\mathrm{g,i}}$.
Then the DM density,
\begin{gather}
  \rho_{\mathrm{dm},k} (r) = \int f (\bm{J}[\bm{r}, \bm{v}]_{U_{k-1}}) \dif ^3 \bm{v},
\label{eqn:rhofj}
\end{gather}
and associated potential, $U_k$ (\refeqnalt{eqn:newU}), are computed iteratively until convergence.
We adopt the numerical implementation from the package \texttt{Agama} \citep{2019MNRAS.482.1525V}
for the iterative construction of self-consistent solutions.

\subsection{Method III: Empirical relation between DM and total mass ratios}
\label{sec:method3}

This method has been motivated by a surprising finding by \citet{2020MNRAS.494.4291C} 
concerning the adiabatic contraction of DM due to added baryons.
Comparing hydrodynamic (hydro) cosmological simulations to their dark-matter-only (dmo) counterparts of the same initial conditions,
they define the ratio of the DM mass profile in their hydro runs versus their DM-only runs, 
$\zeta_{\rm dm}(r)=M_{\rm dm, hydro}(r)/M_{\rm dm, dmo}(r)$, and a similar ratio for the profiles of the total mass,
$\zeta_{\rm tot}(r)=M_{\rm tot, hydro}(r)/M_{\rm tot, dmo}(r)$.
They report a tight power-law relation,
$\zeta_\mathrm{dm}(r)=A\zeta_\mathrm{tot}^B(r)$ with $A = 1.02$ and $B = 0.54$ for any radius $r\in [1 \kpc, \rvir]$.
This relation is consistent within 5\% with the prediction by 
the \citet{2004ApJ...616...16G} model, a widely used empirical model of halo adiabatic contraction.

Without a clear physical motivation for this tight relation in hand,
we test the possible validity of this empirical relation using our N-body simulations for a variety of cases of mass change.
As shown in \reffig{fig:mass_ratio}, the following power-law
\begin{gather}
  M_{\mathrm{dm,f}} / M_{\mathrm{dm,i}} = A (M_{\mathrm{tot,f}} / M_{\mathrm{tot,i}})^B
\label{eqn:power-law}
\end{gather}
with $A = 1$ and $B = 0.6$ can describe the simulation results within $10\%$
at radii where the initial gas mass ratio $M_\mathrm{g,i}/M_\mathrm{tot,i} \lesssim 0.4$.

The DM mass enters the two sides of the equation, as $M_\mathrm{tot}=M_\mathrm{g}+M_\mathrm{dm}$.
With this relation, once we know the initial mass profiles, $M_\mathrm{g,i}(r)$ and $M_\mathrm{dm,i}(r)$,
and the final gas profile $M_\mathrm{g,f}(r)$,
the final DM mass distribution $M_\mathrm{dm,f}$ can be extracted by a root-finding algorithm at each radius.

The validity of a similar tight power-law relation both in our simulations and in \citet{2020MNRAS.494.4291C}
is very interesting and worth a physical understanding. 
The small difference in the obtained slope, $B=0.6$ in our simulations versus $B=0.54$,
is worth understanding as well.

\subsection{Method IV: Energy conservation of shells}
\label{sec:method4}

As introduced in \refsec{sec:intro},
\citetalias{2020MNRAS.491.4523F} assumed during the relaxation energy conservation of shells that encompass a fixed DM mass.
The method CuspCore I has been applied to the formation of flat cores in low-mass DM haloes and the origin of UDGs 
from outflow episodes driven by supernova feedback \citepalias{2020MNRAS.491.4523F}.
It has also been integrated into a hybrid scenario where post-compaction infalling satellites heat up the DM cusps by dynamical friction, allowing AGN-driven outflows to generate cores in more massive haloes with $\mvir \ge 10^{12}\msun$ at $z\sim 2$
\citep{2021MNRAS.508..999D}.
In the following, we first refer to the original method
and then present two variants that are possibly more accurate and better justified.

The specific energy of a DM shell at radius $r_\mathrm{i}$ in the initial equilibrium state is
\begin{equation}
  E_\mi (r_\mi) = U_\mi (r_\mi) + K_\mi (r_\mi),
\end{equation}
where $U_\mi (r)=U_\mathrm{dm,\mi} (r)+U_\mathrm{g,\mi} (r)$ is the potential profile
and $K_\mi (r)$ is the specific kinetic energy of the shell.
For a spherical system in equilibrium, 
$K(r)$ can be solved by the Jeans equation.
Specifically, given a density profile $\rho_\mathrm{dm}(r)$ in the potential $U (r)$, we have
\begin{equation}
  K (r) = \frac{3 - 2 \beta}{2} \sigma_r^2(r),
  \label{eqn:K(r)}
\end{equation}
where $\beta$ is the velocity anisotropy and
$\sigma_r$ is the radial velocity dispersion determined by the Jeans equation (\citetalias{2008gady.book.....B}, eq.\ 4.216),
\begin{equation}
  \sigma_r^2(r) = \frac{1}{r^{2 \beta} \rho_\mathrm{dm} (r)} \int_r^{\infty} d r' {r'}^{2 \beta} \rho_\mathrm{dm} (r') \frac{\dif U}{\dif r'}.
\end{equation}
We set $\beta=0$ because the system is taken to be isotropic.

When the gas potential becomes $U_\mathrm{g,f}(r)=U_\mathrm{g,i}(r)+\Delta U_\mi(r)$ due to an instantaneous mass change,
the energy of a shell in the transitional state immediately turns into $E_\mt(r_\mi) = E_\mi (r_\mi) + \Delta U_\mi (r_\mi)$.

After relaxation to the final equilibrium, 
the DM profile becomes $\rho_\mathrm{dm,f}$
(with a contribution to the potential, $U_\mathrm{dm,f}$) which is assumed to follow a DZ profile
whose parameters $(\alpha,c)$ are to be determined.
A shell encompassing a given mass has moved to a final radius $r_\mf$
that satisfies 
$M_{\mathrm{dm}, \mf} (< r_\mf) = M_{\mathrm{dm}, \mi} (< r_\mi)$.
The final energy of this shell is 
\begin{equation}
  E_\mf (r_\mf) = U_\mf (r_\mf) + K_\mf (r_\mf),
\end{equation}
where $U_\mf = U_\mathrm{dm,\mf} + U_\mathrm{g,\mf}$
and the kinetic energy $K_\mf$ is again set by the Jeans equation with the same anisotropy $\beta=0$
(see \citetalias{2020MNRAS.491.4523F} and Appendix \ref{sec:vel_beta} for justification of $\beta$).
The assumed energy conservation during the relaxation corresponds to $E_\mt=E_\mf$, i.e.,
\begin{gather}
  E_\mi (r_\mi) + \Delta U_\mi (r_\mi) = E_\mf (r_\mf).
  \label{eqn:energy_conserv}
\end{gather}
With this assumption,
the parameters that determine the final DM profile $\rho_\mathrm{dm,f}$ can be solved. 
In practice, we find the best-fit parameters by minimizing the mean square of $\abs{ E_\mt (r_\mi) - E_\mf (r_\mf) }$ 
for a hundred shells equally spaced in $\log_{10}(r_\mi/\rvir)$ from $-2$ to 0.

\begin{figure*}
\centering
\includegraphics[width=1\textwidth]{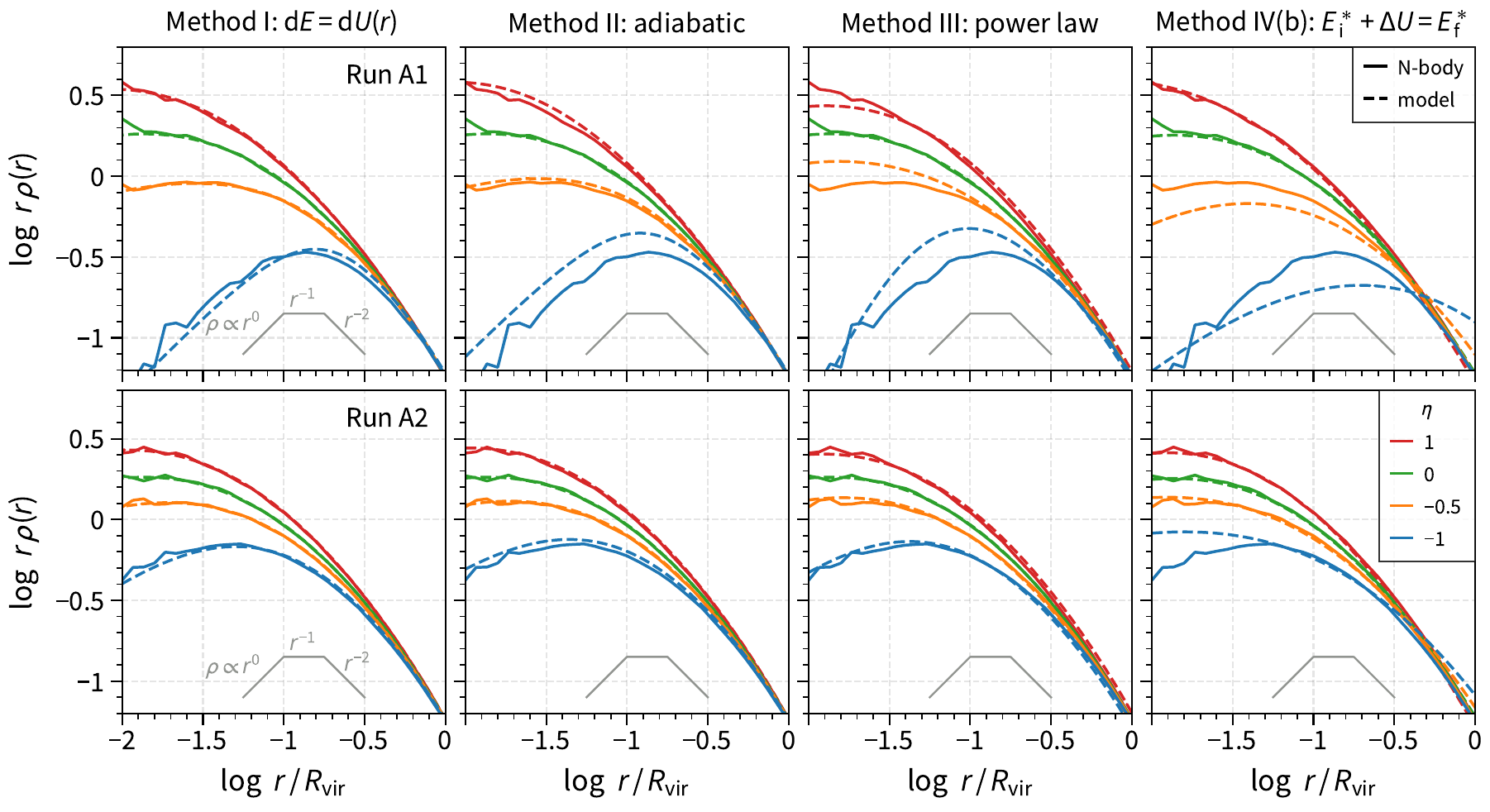}
\vspace{-2em}
\caption{%
Model performance in matching the simulations for the four methods. 
Shown are the model prediction for the relaxed DM profiles against 
the final simulation snapshots.
The profiles are of $r \rho(r)$, where a density core is represented by a rising curve of slope unity (grey lines at the bottom for reference).
The simulations have {the same initial NFW DM profile but}
different gas profiles (rows, {concentrated and diffuse gas profiles for Run A1 and A2 respectively}) and different fractional gas change $\eta$ (colors).
See \reffig{fig:res1} for results of several other test cases.
All the methods recover the qualitative behavior and provide a fair match for moderate gas loss. Method I provides an excellent match to the simulation, even for an extreme mass loss that leads to an extended core.
}
\label{fig:res_cmp}
\end{figure*}

The energy conservation applied in \refeqn{eqn:energy_conserv}
was not formally justified. 
In fact, while the total energy of the \emph{system} is conserved during relaxation, 
the direct sum of the energies of all shells or particles,
$\mathcal{E}_\mathrm{tot}=\int \rho_\mathrm{dm}(r) E(r) 4\pi r^2\dif r$,
is not conserved, owing to the double counting of the potential energy between each DM shell/particle pair 
(see Appendix \ref{sec:Edef}).
From another angle, 
as shown in \refsec{sec:basic_res}, the energies of particles continue to evolve due to the redistribution of DM,
making the average energy of shells evolve as well.

Within the framework of CuspCore I, the above flaw can possibly be remedied by replacing
the energy in \refeqn{eqn:energy_conserv} with an energy-like quantity that is better conserved for shells.
Here we have tested the original method and two variants 
with alternative energy definitions (see Appendix \ref{sec:Edef} for their rationale), as follows.

\begin{enumerate}[(a),leftmargin=*]

\item $E=K+U$, the original method (\citetalias{2020MNRAS.491.4523F}) with $U=U_\mathrm{dm} + U_\mathrm{g}$. 

\item $E^\ast = K + \frac{1}{2} U_\mathrm{dm} + U_\mathrm{g}$, 
for which the total energy of the system is conserved.
The factor of half multiplying $U_\mathrm{dm}$ is because 
the potential energy between each DM particle pair has been counted twice in $\mathcal{E}_\mathrm{tot}$.

\item $E_\mathrm{in}=K-GM_\mathrm{dm}(<\!r)/r+U_\mathrm{g}$,
for which the total energy is conserved as well (see Appendix \ref{sec:Edef}).
Specifically, for a set of \emph{non-crossing} spherical shells, 
a shell experiences no net gravitational force from any outer shells
(shell theorem, \citetalias{2008gady.book.....B}, sec.\ 2.2)
and thus the quantity $K-GM/r$ is conserved (e.g., \citealt{1987ApJ...318...15R}).
In our problem, we have to further include the external gas potential $U_\mathrm{g}$.
\end{enumerate}

The motivation for any of the above variants of energy conservation is vague, and it becomes more so when shell crossing occurs during the relaxation process. 
While the direct sum of $E^\ast$ or $E_\mathrm{in}$ over all shells is indeed conserved,
nothing guarantees such ``energy'' to be conserved for individual shells. 
The performance of these approximations should be verified using the simulations. 

As shown in Appendix \ref{sec:Edef}, 
among the three "energy" variants, $E^\ast$ is best conserved for shells,
although the conservation becomes worse towards the halo center when the gas ejection is strong.
We therefore address variant (b), with $E^\ast$, in the main text below, 
and refer to the other two variants in Appendix \ref{sec:compare2}, \reffig{fig:res1}.

\begin{figure*}
\centering
\includegraphics[width=1\textwidth]{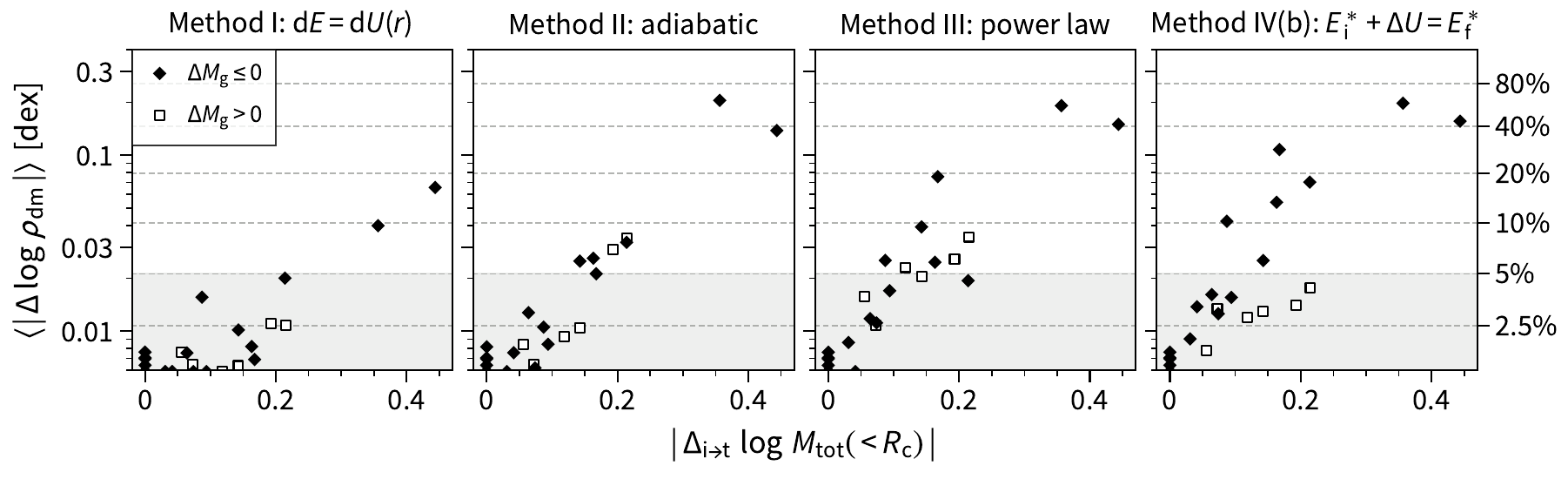}
\vspace{-2em}
\caption{%
Model performance in matching the simulations for the four methods:
the average (root-mean-square) deviation between the model prediction
and N-body simulations for the final DM profile in $[0.015, 0.3]\rvir$.
Each symbol represents a test case (24 in total),
whose fractional mass change within $R_\mathrm{c}=0.067\rvir$ due to the gas removal/addition 
is shown in $x$-axis in terms of 
$\left|\log M_\mathrm{tot,t}(<R_\mathrm{c})-\log M_\mathrm{tot,i}(<R_\mathrm{c})\right|$.
The numbers on the right indicate the relative errors converted 
from $\Delta \log \rho$ for reference.
The gray shade in each panel displays the region with relative errors smaller than 10\%.
When the mass change is small, the DM response is nearly adiabatic 
and all the methods provide accurate predictions.
It becomes more challenging with greater mass changes,
and only Method I reproduces the simulated profiles within $15\%$ for all the test cases.
}
\label{fig:res_score}
\end{figure*}

\section{Testing the models with simulations}
\label{sec:result}

We apply the above four methods to the test cases described in  \refsec{sec:cases}
and compare the model prediction of the relaxed DM profiles
to the final snapshots of the N-body simulations.

We first show the model prediction of Method I against all the 24 simulations in 
\reffig{fig:prof_nbody}.
We then compare the four methods against simulations A1 and A2 with different gas changes in \reffig{fig:res_cmp}
and provide the results for several other simulations in \reffig{fig:res1} (Appendix \ref{sec:compare2}).
All the methods produce the general trend in all cases, 
with better accuracy for gas addition or moderate gas ejection.

For a quantitative comparison,
we present the performance of each method in \reffig{fig:res_score} 
in terms of the average deviation (root-mean-square error) between the model prediction
and simulation for the logarithm DM profiles in {$[0.015, 0.3]\rvir$}.
The performance is shown as a function of the strength of gas changes
represented by the logarithm mass difference within the typical core size of high-$z$ massive galaxies, $R_\mathrm{c}=0.067\rvir$.
We summarize the lesson from the comparisons of model and simulations in Figs. \ref{fig:prof_nbody}, \ref{fig:res_cmp}, \ref{fig:res_score}
as follows.

\textbullet~ \emph{Method I}.
The method based on tracing the energy diffusion exhibits the best accuracy among the four methods tested.
It can reproduce the simulated profiles 
to within $\sim 15\%$ for the two extreme cases (A1 and B1 with $\eta=-1$,
where a gas mass as high as $\sim 60\%$ of the total mass within $R_\mathrm{c}$ is removed)
and to within $5\%$ for the remaining cases.
The latter value mostly reveals the numerical density fluctuations in the simulations.

\textbullet~ \emph{Method II}. 
The method based on adiabatic invariants shows similar precision to Method I
for moderate mass change with $|\Delta \log M_\mathrm{tot}(<R_\mathrm{c})|<0.15$dex,
suggesting that the DM response is nearly adiabatic in these cases.
However, it underestimates the DM expansion for strong gas ejection,
leading to an overestimation of the inner DM density by $\sim 50\%$
in the two extreme cases.

\textbullet~ \emph{Method III}.
The empirical power-law relation makes similar predictions to Method II but with slightly inferior precision.
Their similarity is not surprising recalling that this empirical relation was originally proposed to describe
the (nearly) adiabatic contraction of DM haloes \citep{2020MNRAS.494.4291C}.
It is able to predict the final DM profile 
to within 10\% at radii where the initial gas mass ratio is lower than $\sim$ 0.4 (\reffig{fig:mass_ratio}).
However, similar to Method II, it underestimates the halo expansion for strong gas ejection.
As an empirical relation, it does not always ensure a physical solution.
For example, it predicts a density profile decreasing towards the center for the simulation A1 with $\eta=-1$ (\reffig{fig:res_cmp}).

\textbullet~ \emph{Method IV}.
The original version [variant (a)] based on energy conservation of shells
underestimates the DM response systematically in most cases (\reffig{fig:res1}).
For example,
the predicted central density is four times higher than the simulated density in the simulations A1, A2, and B1 with $\eta=-1$.
The variant (b) with the alternative ``energy'' definition, $E^\ast = K + \frac{1}{2} U_\mathrm{dm} + U_\mathrm{g}$,
shows improved performance, especially for cases with gas addition or small gas removal.
But its predictions are inferior to Method I and II for stronger gas ejection.
In particular, it fails to produce cored profiles.%
\footnote{
   Though recall that when allowing a negative central slope $\alpha$,
   CuspCore I  with the variant (a) did produce cores in low-mass galaxies \citepalias{2020MNRAS.491.4523F} and in high-mass galaxies if the DM has been preheated \citep{2021MNRAS.508..999D}.
}

\section{Discussion}
\label{sec:discuss}

\subsection{Applying to multiple components}
\label{sec:two_comp}

In a follow-up study, our goal will be to study the differential response of stars and DM in a multi-component system,
which can shed light on the formation of dark-matter deficient galaxies with cores.
{Our hypothesis is that the DM is spatially more extended and kinematically hotter than stars, thus being more susceptible to be pushed out.}
As distribution-function based approaches, Method I and II are naturally generalizable to multiple components, 
as each component has its own distribution function and contributes to the total potential.
Method III and IV may also be generalized to two components, where 
we can write the equations for the two components separately 
(\refeqnalt{eqn:power-law} or \ref{eqn:energy_conserv}), 
connected implicitly by the total mass or potential.
The application of these methods to two components is deferred to a future work.

\subsection{Connection to adiabatic processes}

It is known that the DM response to an impulsive change of the potential
is stronger than the response to an adiabatic (slow) change 
\citep[e.g.,][]{2012MNRAS.421.3464P,2016MNRAS.461.2658D,2021ApJ...921..126B}.
Surprisingly, the DM response is nearly adiabatic 
even after a sudden removal/addition of gas mass as high as 40\% (0.15 dex) of the local total mass.
The difference between the two processes becomes prominent only for greater mass changes.

We thus may expect the general success of the adiabatic approximation
in many astrophysical problems, even if the potential does not change so slowly.
This prompts us to consider the scheme of \citet{2012MNRAS.421.3464P},
where energy is transferred into DM through repeated impulsive (but possibly weak) oscillations of potential.
The above results suggest that, to make the energy transfer irreversible and accumulable for core formation,
either each single impulsive gas removal must be strong enough to break the adiabatic approximation
or the profile of recycled gas must not return to its earlier configuration.

On the other side, it also indicates the direct applicability
of our Methods I and IV to adiabatic problems with moderate potential changes.
For greater adiabatic changes, e.g., the halo contraction due to the condensation of a dominant baryon component, 
we may split the total response into multiple successive steps, each of which 
only solves for a small change of potential.

\subsection{Method I}
\label{sec:discuss1}

\subsubsection{Notes on assumptions}
\label{sec:dEdU_assumps}

Two assumptions that underlie Method I should be spelled out and discussed.
Equations (\ref{eqn:eddington} -- \ref{eqn:f=N/g}) are guaranteed to be valid only 
when the system is in an \emph{equilibrium} state and with an \emph{isotropic} velocity distribution ($\beta=0$).

\emph{Equilibrium state assumption}.
A spherical system is in equilibrium if and only if
the radial phase angle, $\theta_r$, is uniformly distributed {(i.e.\ phase-mixed)} for particles in any orbit (\citealt{2016MNRAS.456.1003H}),
where $\theta_r= \frac{\pi}{T_r} \int_{r_\mathrm{per}}^{r} d r/v_r \in [-\pi, \pi]$
and $T_r$ is the radial period.
The distribution of $\theta_r$ is clearly not uniform during the relaxation
at any specific moment before the complete phase mixing,
as indicated by the caustic features in phase space 
(\refsec{sec:profile}; also cf.\ \citealt{2019MNRAS.485.1008B}).
Nonetheless, as a particle moves continuously between its peri- and apo-center,
$\theta_r$ should be uniformly distributed in a time interval long enough,
which may (at least approximately) justify the usage of 
Equations (\ref{eqn:Nvar}) and (\ref{eqn:f=N/g}) in a time-averaged sense.
The above argument seems supported by the high accuracy of our model prediction.

\emph{Isotropic assumption}.
Even when the initial condition is isotropic
(as adopted in this work, see \citetalias{2020MNRAS.491.4523F} for justification),
the relaxation after the potential change may introduce velocity anisotropy.
This is not a major issue for two reasons.
First, the anisotropy developed during the relaxation is usually small.
The simulations end with $\beta < 0.15$ even for our extreme cases
(Appendix \ref{sec:vel_beta}, also cf.\ \citetalias{2020MNRAS.491.4523F}).
Second, the density profile is mostly determined by 
the energy distribution $N(E)$ with only weak dependence on anisotropy
(\citetalias{2008gady.book.....B} sec. 4.3.2d, see also discussion in \citealt{2021A&A...653A.140B}). 

\subsubsection{Possible extensions}
\label{sec:extension}

Below we list several possible future extensions of Method I. 

\emph{Velocity anisotropy}.
It would be straightforward to incorporate angular momentum and thus velocity anisotropy into Method I
when necessary.
Despite the unavoidable additional complexity in numerical implementation,
conceptually the only thing we need is to use $f (E, L)$ instead of $f(E)$ for the distribution function
along with the corresponding $N(E,L)$ and $g(E,L)$ in Equations (\ref{eqn:Nvar}--\ref{eqn:rhof}).

\emph{Additional energy source terms}.
It is possible to inject kinetic energy as a function of radius, $\Delta K(r)$,
by simply replacing $E$ with $E - \Delta K(r)$ on the right-hand side of Equation (\ref{eqn:Nvar}).
Such energy injection can be,
e.g., dynamical friction heating from accreted satellites
\citep{2001ApJ...560..636E,2021MNRAS.508..999D}, tidal heating from the environment 
\citep{1980ApJ...241..946D,1999ApJ...514..109G,2019MNRAS.487.5272J},
{or heating due to spatial fluctuations \citep{2016MNRAS.461.1745E,2022arXiv220908631H}.}
For reference, see \citet{2021MNRAS.508..999D} for an example of combining dynamical friction heating and gas ejection
using both semi-analytic models (SatGen, \citealt{2021MNRAS.502..621J}) and CuspCore I \citepalias{2020MNRAS.491.4523F}.
We may even introduce angular momentum exchange if the relevant model is available
(then we need to use $f (E, L)$ instead of $f(E)$ of course).

\emph{Realistic time stepping}.
Though the iterative procedure of Method I
is tracing the relaxation process and it is able to precisely reproduce the final DM profile,
a single iteration step does not correspond to a specific snapshot in the N-body simulation.
It is because the dynamical time scale, $t_{\mathrm{dyn}}(r)$, is a function of radius. 
For example, when the inner halo has relaxed to a new equilibrium, 
the outer halo may have not yet responded much
because of  its much longer $t_{\mathrm{dyn}}$ (see \reffig{fig:rhot_t}).
To take this effect into account,
we can set a radius-dependent step factor, $\mu(r) \propto \Delta t/t_\mathrm{dyn}(r)$, 
instead of a constant in \refeqn{eqn:newrho}.
This will enable us to predict the distribution function of intermediate stages during the relaxation
and handle external potentials that vary at different rates 
(from adiabatically to instantaneously) or undergo multiple successive changes.
We leave such exploration to future work.

\subsection{Method II}

Method II naturally becomes inaccurate for a strong sudden mass change,
because the DM response cannot be approximated as an adiabatic process any more.
It seems interesting to improve the action-based method beyond an adiabatic process
by tracing the diffusion of actions (i.e., $J_r$ for spherical systems),
in a similar spirit to Method I.
Unfortunately, 
considering the complicated expression of $J_r$ itself,
it is much more difficult to find and implement a proper description for the diffusion of $J_r$ directly
(see \citealt{2013MNRAS.433.2576P,2021MNRAS.508.1404B} for attempts with toy potentials) 
than that of $E$.
On the other hand,
one can actually derive the evolution of $J_r$ from the output of our Method I
though the mapping between $(E,L)$ and $(J_r,L)$.

\subsection{Method III}

This method is the easiest to implement and fastest to compute among the four methods,
which can be particularly attractive for making a module incorporated into semi-analytic models of galaxy evolution
(e.g., \citealt{2021MNRAS.502..621J}).
It is not so accurate for strong gas ejection,
but it might still be useful when considering successive moderate gas ejection/recycling episodes.

Encouraged by the fair success and simple implementation of Method III, 
it might be interesting to examine other similar empirical relations
\citep[e.g.,][in the context of halo contraction]{1986ApJ...301...27B,2004ApJ...616...16G,2010MNRAS.407..435A}, 
though our problem is clearly beyond their original purpose.
We test the widely used \citet{2004ApJ...616...16G} model in Appendix \ref{sec:Gnedin04}.
It shows good precision for moderate gas changes as expected
but fails to produce the central DM density for strong gas ejection due to artificial shell crossing,
which limits its possible application in the problem of core formation.

\subsection{Method IV}
\label{sec:discuss4}

Method IV \citepalias{2020MNRAS.491.4523F} assumes the energy conservation for shells encompassing constant DM mass.
We explicitly examine the energy of such shells with N-body simulations in Appendix \ref{sec:Edef}.
Among the three ``energy'' definitions presented in \refsec{sec:method4}, 
$E$ and $E_\mathrm{in}$ show systematic changes between the transitional and final states,
while $E^\ast$ is better conserved for most shells during the relaxation.
This explains why the model variant (b) with $E^\ast$ has better precision than the other two variants.
However, it still exhibits a clear deviation between the transitional and final states
in the inner halo especially for cases with complete gas removal ($\eta=-1$) which we are interested in most.
Such a local deviation may lead to a global inferior prediction
when fitting with a given functional form of DM profiles.

Nevertheless, the good level of conservation with $E^\ast$ elsewhere 
(for most cases with $\eta>-1$ and for the outer halo with $\eta=-1$) seems encouraging.
The success and failure of this formalism is worth a physical understanding.
A thorough analysis of shell crossing might eventually
uncover the origin of the deviation in the center
and how to correct it, which is beyond the scope of this paper.

Whereas our idealized simulations suggest the CuspCore I model [variant (a)] 
tends to underestimate the halo expansion in general, 
\citetalias{2020MNRAS.491.4523F} finds that it approximates fairly well 
the evolution of the inner DM profile between successive snapshots
in the NIHAO \citep{2015MNRAS.454...83W} cosmological simulations.
This may reflect the fact that the underestimate by Method IV (a) is not large enough to be detected in the comparison. 
It may also be partly due to the fact that \citetalias{2020MNRAS.491.4523F} considered the change in total mass rather than in the gas mass and approximated the ejected mass as a point mass, both enhancing the DM response.

\section{Conclusion}
\label{sec:conclusion}

Following CuspCore I in \citetalias{2020MNRAS.491.4523F},
we propose a novel analytic model, CuspCore II,
for the response of a non-dissipative spherical system (e.g., of DM or stars) to a rapid change of potential,
as a mechanism for the formation of DM-deficient cores in dwarfs and high-$z$ massive galaxies
or the DM contraction due to baryonic inflows. 

The model assumes an instantaneous change of potential due to gas removal/addition, 
followed by a relaxation to a new equilibrium.
The new proposed treatment of the post-change relaxation to a new equilibrium is physically justified, and it provides more accurate predictions even in extreme cases of massive gas ejection where cores are produced.  

By studying the relaxation process in idealized N-body simulations, we find that the relaxation turns out to be a violent relaxation associated with rapid redistribution of DM 
(as a shorthand for general collisionless particles)
during the first orbital period, followed by phase mixing.
Specifically, we find:

\begin{itemize}[wide]
\item 
The DM density at given radius effectively evolves only during the first radial period $T_r$ and it remains roughly constant afterwards, and so do the orbits and corresponding integrals, $E$ and $J_r$, of individual particles.

\item 
For particles of the same initial orbit, the change of energy $\Delta E$ depends on the phase within the orbit,
manifesting a diffusion of $E$.
The cumulative $\Delta E$ is larger for particles that were initially located closer to the center or moving inwards with higher velocity.

\item 
The radial action $J_r$  of individual particles can either increase or decrease with large variance,
as expected for a non-adiabatic process. But the ensemble distribution, $p(J_r)$, changes much less,
making the DM response nearly adiabatic for moderate gas ejection.
\end{itemize}

We have introduced four different methods for treating the relaxation process, and compared the model predictions with a suite of N-body simulations. Our results can be summarized as follows.
\begin{itemize}[wide]

\item 
By tracing the energy diffusion and
updating the phase-space distribution function iteratively during the relaxation,
the CuspCore II model reproduces the simulated DM profiles with $\sim$10\% accuracy or better,
performing the best among the four methods tested.

\item 
We test the possible validity of two adiabatic methods for comparison.
The exact solution using adiabatic invariants (\citealt{1980ApJ...242.1232Y})
shows similar precision to CuspCore II for moderate mass change, suggesting that 
the relaxation is nearly adiabatic.
An empirical power-law relation between mass ratios 
\citep[with parameters slightly adapted]{2020MNRAS.494.4291C} makes somewhat inferior predictions
but may still be useful because of its simplicity.
However, as might be expected, the two adiabatic methods underestimate the DM response to strong gas ejection.

\item 
The ad-hoc assumption adopted in \citetalias{2020MNRAS.491.4523F}, of energy conservation for shells encompassing a fixed DM mass, turns out to underestimate the DM response.
A variant [Method \hyperref[sec:method4]{IV (b)}] using an alternative ``energy'' definition for shells 
improves the model accuracy significantly for moderate gas change
but fails to reproduce the DM density profile for strong gas ejection.
\end{itemize}

The CuspCore II model provides a simple understanding of the formation of DM cores and UDGs by feedback outflow,
and it enables multiple extensions for practical concerns.
The model can apply to successive inflow/outflow episodes associated with a star formation history,
which presents a more realistic description than a single bursty event
(as envisioned by \citealt{2012MNRAS.421.3464P,2016MNRAS.459.2573R,2016MNRAS.461.1745E}; \citetalias{2020MNRAS.491.4523F}).
Moreover, it will enable the study of the differential response of a multi-component system of stars and DM in the formation of DM-deficient galaxies (\refsec{sec:two_comp}).
It can also combine gas outflows and additional heating sources such as dynamical friction heating 
from accreted satellites (\refsec{sec:extension}),
which is crucial for core formation in high-$z$ massive galaxies \citep{2021MNRAS.508..999D}.
It is possible to incorporate CuspCore II and above extensions into semi-analytic models of galaxy formation,
which will allow us to trace the evolution of the DM profile
as a function of the history of star formation and merger events for a cosmic galaxy sample.

In this paper, we focus on the methodology and do not attempt to compare with observations.
Nevertheless, our current analysis may offer some useful insights with observable implications.
A flat DM core of $\sim 0.1\rvir$, comparable to the typical core size of high-$z$ massive galaxies,
may form from an NFW cusp by removing a gas mass as high as about 60\% of the total mass within $0.1\rvir$
(though more realistic discussion should include stellar component and dynamical friction preheating as aforementioned).
In contrast, an initially contracted DM halo is more resistant to the same gas mass change.
We also see a diversity of the response of the central slope depending on the detailed mass profiles.
The change of the inner slope defined at $0.01\rvir$ is mainly determined by the local $\Delta M_\mathrm{g}$.
Therefore, the inner slope does not necessarily change at the same level
as the overall DM deficit in the core region.

More direct and detailed comparisons between model predictions and observations, 
concerning, e.g., the inner slope and density, DM mass deficit, and core size,
are left to future studies,
where the model extensions mentioned earlier, 
including the successive inflow/outflow episodes, the differential response of stars and DM,
and additional heating sources, will be incorporated.
Before such a comparison, 
it might be helpful to test the extended model with zoom-in cosmological hydrodynamic simulations that
properly resolve the feedback-driven outflows
and the dynamical friction preheating by compact satellites,
though complexities of disentangling the different physical processes are expected.

Finally, CuspCore II presents a novel accurate and self-consistent approach for modeling violent relaxation,
which may apply to other similar problems with proper adaptations,
e.g., the evolution of the spatial distribution of {DM particles and} satellite galaxies 
in response to the non-adiabatic growth of host halo \citep{2021MNRAS.503.1233O},
{and the early dynamical evolution of star clusters due to the dispersal of initial gas \citep{1978A&A....70...57T}}.
Taking advantage of the capability to handle unbound particles (Appendix \ref{sec:der_Nvar})
and additional heating sources (\refsec{sec:extension}),
another possible application is to model the relaxation of satellite galaxies
after the tidal truncation \citep[e.g.,][]{2021MNRAS.505...18E,2021arXiv211101148A,2022arXiv220700604S}
and tidal heating \citep{1980ApJ...241..946D,1999ApJ...514..109G}
which are believed to be crucial processes in 
the evolution of satellites and particularly the formation of UDGs in groups
\citep{2018MNRAS.480L.106O,2019MNRAS.487.5272J,2019MNRAS.485..382C}.

\section*{Acknowledgements}

We thank Maarten Baes, Marius Cautun, Benoit Famaey, Feihong He, Jiaxin Han, Fangzhou Jiang, 
and Guillaume Thomas for the helpful discussion,
{and the anonymous referee for the constructive suggestions}.
We thank Eugene Vasiliev for the elaborate documentation
and the enthusiastic help regarding the package \texttt{Agama}.
This work was supported by ISF grants 861/20 (AD) and 3061/21 (NM; ZZL).
This work is done on the super cluster Moriah at HUJI.

This research made use of the following software:
\texttt{Agama} \citep{2019MNRAS.482.1525V}, 
\texttt{Jupyter} \citep{jupyter},
\texttt{KDEpy} \citep{kdepy}, 
\texttt{Matplotlib} \citep{matplotlib},
\texttt{NEMO} \citep{1995ASPC...77..398T},
\texttt{Numpy} \citep{numpy},
\texttt{ProPlot} \citep{proplot},
and \texttt{Scipy} \citep{scipy}.

\section*{Data Availability}

We provide our implementation of the methods at \url{https://github.com/syrte/CuspCore2}.
The N-body simulations performed in this work will be available on reasonable request to the authors.
A fast Python script for loading NEMO snapshots can be found at
\url{https://github.com/syrte/snapio}.




\bibliographystyle{mnras}
\bibliography{cuspcore_methods,software} 

\begin{thebibliography}{}
\makeatletter
\relax
\def\mn@urlcharsother{\let\do\@makeother \do\$\do\&\do\#\do\^\do\_\do\%\do\~}
\def\mn@doi{\begingroup\mn@urlcharsother \@ifnextchar [ {\mn@doi@}
  {\mn@doi@[]}}
\def\mn@doi@[#1]#2{\def\@tempa{#1}\ifx\@tempa\@empty \href
  {http://dx.doi.org/#2} {doi:#2}\else \href {http://dx.doi.org/#2} {#1}\fi
  \endgroup}
\def\mn@eprint#1#2{\mn@eprint@#1:#2::\@nil}
\def\mn@eprint@arXiv#1{\href {http://arxiv.org/abs/#1} {{\tt arXiv:#1}}}
\def\mn@eprint@dblp#1{\href {http://dblp.uni-trier.de/rec/bibtex/#1.xml}
  {dblp:#1}}
\def\mn@eprint@#1:#2:#3:#4\@nil{\def\@tempa {#1}\def\@tempb {#2}\def\@tempc
  {#3}\ifx \@tempc \@empty \let \@tempc \@tempb \let \@tempb \@tempa \fi \ifx
  \@tempb \@empty \def\@tempb {arXiv}\fi \@ifundefined
  {mn@eprint@\@tempb}{\@tempb:\@tempc}{\expandafter \expandafter \csname
  mn@eprint@\@tempb\endcsname \expandafter{\@tempc}}}

\bibitem[\protect\citeauthoryear{{Abadi}, {Navarro}, {Fardal}, {Babul}  \&
  {Steinmetz}}{{Abadi} et~al.}{2010}]{2010MNRAS.407..435A}
{Abadi} M.~G.,  {Navarro} J.~F.,  {Fardal} M.,  {Babul} A.,   {Steinmetz} M.,
  2010, \mn@doi [\mnras] {10.1111/j.1365-2966.2010.16912.x}, \href
  {https://ui.adsabs.harvard.edu/abs/2010MNRAS.407..435A} {407, 435}

\bibitem[\protect\citeauthoryear{{Amorisco}}{{Amorisco}}{2021}]{2021arXiv211101148A}
{Amorisco} N.~C.,  2021, arXiv e-prints, \href
  {https://ui.adsabs.harvard.edu/abs/2021arXiv211101148A} {p. arXiv:2111.01148}

\bibitem[\protect\citeauthoryear{{An} \& {Evans}}{{An} \&
  {Evans}}{2006}]{2006ApJ...642..752A}
{An} J.~H.,  {Evans} N.~W.,  2006, \mn@doi [\apj] {10.1086/501040}, \href
  {https://ui.adsabs.harvard.edu/abs/2006ApJ...642..752A} {642, 752}

\bibitem[\protect\citeauthoryear{{Baes} \& {Camps}}{{Baes} \&
  {Camps}}{2021}]{2021MNRAS.503.2955B}
{Baes} M.,  {Camps} P.,  2021, \mn@doi [\mnras] {10.1093/mnras/stab634}, \href
  {https://ui.adsabs.harvard.edu/abs/2021MNRAS.503.2955B} {503, 2955}

\bibitem[\protect\citeauthoryear{{Baes} \& {Dejonghe}}{{Baes} \&
  {Dejonghe}}{2021}]{2021A&A...653A.140B}
{Baes} M.,  {Dejonghe} H.,  2021, \mn@doi [\aap] {10.1051/0004-6361/202141463},
  \href {https://ui.adsabs.harvard.edu/abs/2021A&A...653A.140B} {653, A140}

\bibitem[\protect\citeauthoryear{{Baes}, {Camps}  \& {Vandenbroucke}}{{Baes}
  et~al.}{2021}]{2021A&A...652A..36B}
{Baes} M.,  {Camps} P.,   {Vandenbroucke} B.,  2021, \mn@doi [\aap]
  {10.1051/0004-6361/202141281}, \href
  {https://ui.adsabs.harvard.edu/abs/2021A&A...652A..36B} {652, A36}

\bibitem[\protect\citeauthoryear{{Binney} \& {Tremaine}}{{Binney} \&
  {Tremaine}}{2008}]{2008gady.book.....B}
{Binney} J.,  {Tremaine} S.,  2008, {Galactic Dynamics: Second Edition}

\bibitem[\protect\citeauthoryear{{Blumenthal}, {Faber}, {Flores}  \&
  {Primack}}{{Blumenthal} et~al.}{1986}]{1986ApJ...301...27B}
{Blumenthal} G.~R.,  {Faber} S.~M.,  {Flores} R.,   {Primack} J.~R.,  1986,
  \mn@doi [\apj] {10.1086/163867}, \href
  {https://ui.adsabs.harvard.edu/abs/1986ApJ...301...27B} {301, 27}

\bibitem[\protect\citeauthoryear{{Bouch{\'e}} et~al.,}{{Bouch{\'e}}
  et~al.}{2022}]{2022A&A...658A..76B}
{Bouch{\'e}} N.~F.,  et~al., 2022, \mn@doi [\aap]
  {10.1051/0004-6361/202141762}, \href
  {https://ui.adsabs.harvard.edu/abs/2022A&A...658A..76B} {658, A76}

\bibitem[\protect\citeauthoryear{{Boylan-Kolchin} \& {Ma}}{{Boylan-Kolchin} \&
  {Ma}}{2004}]{2004MNRAS.349.1117B}
{Boylan-Kolchin} M.,  {Ma} C.-P.,  2004, \mn@doi [\mnras]
  {10.1111/j.1365-2966.2004.07585.x}, \href
  {https://ui.adsabs.harvard.edu/abs/2004MNRAS.349.1117B} {349, 1117}

\bibitem[\protect\citeauthoryear{{Boylan-Kolchin}, {Bullock}  \&
  {Kaplinghat}}{{Boylan-Kolchin} et~al.}{2011}]{2011MNRAS.415L..40B}
{Boylan-Kolchin} M.,  {Bullock} J.~S.,   {Kaplinghat} M.,  2011, \mn@doi
  [\mnras] {10.1111/j.1745-3933.2011.01074.x}, \href
  {https://ui.adsabs.harvard.edu/abs/2011MNRAS.415L..40B} {415, L40}

\bibitem[\protect\citeauthoryear{{Bullock} \& {Boylan-Kolchin}}{{Bullock} \&
  {Boylan-Kolchin}}{2017}]{2017ARA&A..55..343B}
{Bullock} J.~S.,  {Boylan-Kolchin} M.,  2017, \mn@doi [\araa]
  {10.1146/annurev-astro-091916-055313}, \href
  {https://ui.adsabs.harvard.edu/abs/2017ARA&A..55..343B} {55, 343}

\bibitem[\protect\citeauthoryear{{Burger} \& {Zavala}}{{Burger} \&
  {Zavala}}{2019}]{2019MNRAS.485.1008B}
{Burger} J.~D.,  {Zavala} J.,  2019, \mn@doi [\mnras] {10.1093/mnras/stz496},
  \href {https://ui.adsabs.harvard.edu/abs/2019MNRAS.485.1008B} {485, 1008}

\bibitem[\protect\citeauthoryear{{Burger} \& {Zavala}}{{Burger} \&
  {Zavala}}{2021}]{2021ApJ...921..126B}
{Burger} J.~D.,  {Zavala} J.,  2021, \mn@doi [\apj] {10.3847/1538-4357/ac1a0f},
  \href {https://ui.adsabs.harvard.edu/abs/2021ApJ...921..126B} {921, 126}

\bibitem[\protect\citeauthoryear{{Burger}, {Pe{\~n}arrubia}  \&
  {Zavala}}{{Burger} et~al.}{2021}]{2021MNRAS.508.1404B}
{Burger} J.~D.,  {Pe{\~n}arrubia} J.,   {Zavala} J.,  2021, \mn@doi [\mnras]
  {10.1093/mnras/stab2568}, \href
  {https://ui.adsabs.harvard.edu/abs/2021MNRAS.508.1404B} {508, 1404}

\bibitem[\protect\citeauthoryear{{Burkert}}{{Burkert}}{1995}]{1995ApJ...447L..25B}
{Burkert} A.,  1995, \mn@doi [\apjl] {10.1086/309560}, \href
  {https://ui.adsabs.harvard.edu/abs/1995ApJ...447L..25B} {447, L25}

\bibitem[\protect\citeauthoryear{{Callingham}, {Cautun}, {Deason}, {Frenk},
  {Grand}, {Marinacci}  \& {Pakmor}}{{Callingham}
  et~al.}{2020}]{2020MNRAS.495...12C}
{Callingham} T.~M.,  {Cautun} M.,  {Deason} A.~J.,  {Frenk} C.~S.,  {Grand} R.
  J.~J.,  {Marinacci} F.,   {Pakmor} R.,  2020, \mn@doi [\mnras]
  {10.1093/mnras/staa1089}, \href
  {https://ui.adsabs.harvard.edu/abs/2020MNRAS.495...12C} {495, 12}

\bibitem[\protect\citeauthoryear{{Carleton}, {Errani}, {Cooper}, {Kaplinghat},
  {Pe{\~n}arrubia}  \& {Guo}}{{Carleton} et~al.}{2019}]{2019MNRAS.485..382C}
{Carleton} T.,  {Errani} R.,  {Cooper} M.,  {Kaplinghat} M.,  {Pe{\~n}arrubia}
  J.,   {Guo} Y.,  2019, \mn@doi [\mnras] {10.1093/mnras/stz383}, \href
  {https://ui.adsabs.harvard.edu/abs/2019MNRAS.485..382C} {485, 382}

\bibitem[\protect\citeauthoryear{{Cautun} et~al.,}{{Cautun}
  et~al.}{2020}]{2020MNRAS.494.4291C}
{Cautun} M.,  et~al., 2020, \mn@doi [\mnras] {10.1093/mnras/staa1017}, \href
  {https://ui.adsabs.harvard.edu/abs/2020MNRAS.494.4291C} {494, 4291}

\bibitem[\protect\citeauthoryear{{Chan}, {Kere{\v{s}}}, {Wetzel}, {Hopkins},
  {Faucher-Gigu{\`e}re}, {El-Badry}, {Garrison-Kimmel}  \&
  {Boylan-Kolchin}}{{Chan} et~al.}{2018}]{2018MNRAS.478..906C}
{Chan} T.~K.,  {Kere{\v{s}}} D.,  {Wetzel} A.,  {Hopkins} P.~F.,
  {Faucher-Gigu{\`e}re} C.~A.,  {El-Badry} K.,  {Garrison-Kimmel} S.,
  {Boylan-Kolchin} M.,  2018, \mn@doi [\mnras] {10.1093/mnras/sty1153}, \href
  {https://ui.adsabs.harvard.edu/abs/2018MNRAS.478..906C} {478, 906}

\bibitem[\protect\citeauthoryear{Davis}{Davis}{2021}]{proplot}
Davis L. L.~B.,  2021, ProPlot, \mn@doi{10.5281/zenodo.5602155}

\bibitem[\protect\citeauthoryear{{Dehnen}}{{Dehnen}}{2000}]{2000ApJ...536L..39D}
{Dehnen} W.,  2000, \mn@doi [\apjl] {10.1086/312724}, \href
  {https://ui.adsabs.harvard.edu/abs/2000ApJ...536L..39D} {536, L39}

\bibitem[\protect\citeauthoryear{{Dehnen}}{{Dehnen}}{2001}]{2001MNRAS.324..273D}
{Dehnen} W.,  2001, \mn@doi [\mnras] {10.1046/j.1365-8711.2001.04237.x}, \href
  {https://ui.adsabs.harvard.edu/abs/2001MNRAS.324..273D} {324, 273}

\bibitem[\protect\citeauthoryear{{Dehnen}}{{Dehnen}}{2002}]{2002JCoPh.179...27D}
{Dehnen} W.,  2002, \mn@doi [Journal of Computational Physics]
  {10.1006/jcph.2002.7026}, \href
  {https://ui.adsabs.harvard.edu/abs/2002JCoPh.179...27D} {179, 27}

\bibitem[\protect\citeauthoryear{{Dekel} \& {Silk}}{{Dekel} \&
  {Silk}}{1986}]{1986ApJ...303...39D}
{Dekel} A.,  {Silk} J.,  1986, \mn@doi [\apj] {10.1086/164050}, \href
  {https://ui.adsabs.harvard.edu/abs/1986ApJ...303...39D} {303, 39}

\bibitem[\protect\citeauthoryear{{Dekel}, {Lecar}  \& {Shaham}}{{Dekel}
  et~al.}{1980}]{1980ApJ...241..946D}
{Dekel} A.,  {Lecar} M.,   {Shaham} J.,  1980, \mn@doi [\apj] {10.1086/158409},
  \href {https://ui.adsabs.harvard.edu/abs/1980ApJ...241..946D} {241, 946}

\bibitem[\protect\citeauthoryear{{Dekel}, {Ishai}, {Dutton}  \&
  {Maccio}}{{Dekel} et~al.}{2017}]{2017MNRAS.468.1005D}
{Dekel} A.,  {Ishai} G.,  {Dutton} A.~A.,   {Maccio} A.~V.,  2017, \mn@doi
  [\mnras] {10.1093/mnras/stx486}, \href
  {https://ui.adsabs.harvard.edu/abs/2017MNRAS.468.1005D} {468, 1005}

\bibitem[\protect\citeauthoryear{{Dekel}, {Lapiner}  \& {Dubois}}{{Dekel}
  et~al.}{2019}]{2019arXiv190408431D}
{Dekel} A.,  {Lapiner} S.,   {Dubois} Y.,  2019, arXiv e-prints, \href
  {https://ui.adsabs.harvard.edu/abs/2019arXiv190408431D} {p. arXiv:1904.08431}

\bibitem[\protect\citeauthoryear{{Dekel} et~al.,}{{Dekel}
  et~al.}{2021}]{2021MNRAS.508..999D}
{Dekel} A.,  et~al., 2021, \mn@doi [\mnras] {10.1093/mnras/stab2416}, \href
  {https://ui.adsabs.harvard.edu/abs/2021MNRAS.508..999D} {508, 999}

\bibitem[\protect\citeauthoryear{{Di Cintio}, {Brook}, {Macci{\`o}}, {Stinson},
  {Knebe}, {Dutton}  \& {Wadsley}}{{Di Cintio}
  et~al.}{2014}]{2014MNRAS.437..415D}
{Di Cintio} A.,  {Brook} C.~B.,  {Macci{\`o}} A.~V.,  {Stinson} G.~S.,  {Knebe}
  A.,  {Dutton} A.~A.,   {Wadsley} J.,  2014, \mn@doi [\mnras]
  {10.1093/mnras/stt1891}, \href
  {https://ui.adsabs.harvard.edu/abs/2014MNRAS.437..415D} {437, 415}

\bibitem[\protect\citeauthoryear{{Di Cintio}, {Brook}, {Dutton}, {Macci{\`o}},
  {Obreja}  \& {Dekel}}{{Di Cintio} et~al.}{2017}]{2017MNRAS.466L...1D}
{Di Cintio} A.,  {Brook} C.~B.,  {Dutton} A.~A.,  {Macci{\`o}} A.~V.,  {Obreja}
  A.,   {Dekel} A.,  2017, \mn@doi [\mnras] {10.1093/mnrasl/slw210}, \href
  {https://ui.adsabs.harvard.edu/abs/2017MNRAS.466L...1D} {466, L1}

\bibitem[\protect\citeauthoryear{{Dutton}, {Macci{\`o}}, {Frings}, {Wang},
  {Stinson}, {Penzo}  \& {Kang}}{{Dutton} et~al.}{2016a}]{2016MNRAS.457L..74D}
{Dutton} A.~A.,  {Macci{\`o}} A.~V.,  {Frings} J.,  {Wang} L.,  {Stinson}
  G.~S.,  {Penzo} C.,   {Kang} X.,  2016a, \mn@doi [\mnras]
  {10.1093/mnrasl/slv193}, \href
  {https://ui.adsabs.harvard.edu/abs/2016MNRAS.457L..74D} {457, L74}

\bibitem[\protect\citeauthoryear{{Dutton} et~al.,}{{Dutton}
  et~al.}{2016b}]{2016MNRAS.461.2658D}
{Dutton} A.~A.,  et~al., 2016b, \mn@doi [\mnras] {10.1093/mnras/stw1537}, \href
  {https://ui.adsabs.harvard.edu/abs/2016MNRAS.461.2658D} {461, 2658}

\bibitem[\protect\citeauthoryear{{Eddington}}{{Eddington}}{1916}]{1916MNRAS..76..572E}
{Eddington} A.~S.,  1916, \mn@doi [\mnras] {10.1093/mnras/76.7.572}, \href
  {https://ui.adsabs.harvard.edu/abs/1916MNRAS..76..572E} {76, 572}

\bibitem[\protect\citeauthoryear{{El-Zant}, {Shlosman}  \& {Hoffman}}{{El-Zant}
  et~al.}{2001}]{2001ApJ...560..636E}
{El-Zant} A.,  {Shlosman} I.,   {Hoffman} Y.,  2001, \mn@doi [\apj]
  {10.1086/322516}, \href
  {https://ui.adsabs.harvard.edu/abs/2001ApJ...560..636E} {560, 636}

\bibitem[\protect\citeauthoryear{{El-Zant}, {Freundlich}  \&
  {Combes}}{{El-Zant} et~al.}{2016}]{2016MNRAS.461.1745E}
{El-Zant} A.~A.,  {Freundlich} J.,   {Combes} F.,  2016, \mn@doi [\mnras]
  {10.1093/mnras/stw1398}, \href
  {https://ui.adsabs.harvard.edu/abs/2016MNRAS.461.1745E} {461, 1745}

\bibitem[\protect\citeauthoryear{{Errani} \& {Navarro}}{{Errani} \&
  {Navarro}}{2021}]{2021MNRAS.505...18E}
{Errani} R.,  {Navarro} J.~F.,  2021, \mn@doi [\mnras]
  {10.1093/mnras/stab1215}, \href
  {https://ui.adsabs.harvard.edu/abs/2021MNRAS.505...18E} {505, 18}

\bibitem[\protect\citeauthoryear{{Errani}, {Pe{\~n}arrubia}, {Laporte}  \&
  {G{\'o}mez}}{{Errani} et~al.}{2017}]{2017MNRAS.465L..59E}
{Errani} R.,  {Pe{\~n}arrubia} J.,  {Laporte} C. F.~P.,   {G{\'o}mez} F.~A.,
  2017, \mn@doi [\mnras] {10.1093/mnrasl/slw211}, \href
  {https://ui.adsabs.harvard.edu/abs/2017MNRAS.465L..59E} {465, L59}

\bibitem[\protect\citeauthoryear{{Flores} \& {Primack}}{{Flores} \&
  {Primack}}{1994}]{1994ApJ...427L...1F}
{Flores} R.~A.,  {Primack} J.~R.,  1994, \mn@doi [\apjl] {10.1086/187350},
  \href {https://ui.adsabs.harvard.edu/abs/1994ApJ...427L...1F} {427, L1}

\bibitem[\protect\citeauthoryear{{Freundlich}, {Dekel}, {Jiang}, {Ishai},
  {Cornuault}, {Lapiner}, {Dutton}  \& {Macci{\`o}}}{{Freundlich}
  et~al.}{2020a}]{2020MNRAS.491.4523F}
{Freundlich} J.,  {Dekel} A.,  {Jiang} F.,  {Ishai} G.,  {Cornuault} N.,
  {Lapiner} S.,  {Dutton} A.~A.,   {Macci{\`o}} A.~V.,  2020a, \mn@doi [\mnras]
  {10.1093/mnras/stz3306}, \href
  {https://ui.adsabs.harvard.edu/abs/2020MNRAS.491.4523F} {491, 4523}

\bibitem[\protect\citeauthoryear{{Freundlich} et~al.,}{{Freundlich}
  et~al.}{2020b}]{2020MNRAS.499.2912F}
{Freundlich} J.,  et~al., 2020b, \mn@doi [\mnras] {10.1093/mnras/staa2790},
  \href {https://ui.adsabs.harvard.edu/abs/2020MNRAS.499.2912F} {499, 2912}

\bibitem[\protect\citeauthoryear{{Garrison-Kimmel} et~al.,}{{Garrison-Kimmel}
  et~al.}{2017}]{2017MNRAS.471.1709G}
{Garrison-Kimmel} S.,  et~al., 2017, \mn@doi [\mnras] {10.1093/mnras/stx1710},
  \href {https://ui.adsabs.harvard.edu/abs/2017MNRAS.471.1709G} {471, 1709}

\bibitem[\protect\citeauthoryear{{Genzel} et~al.,}{{Genzel}
  et~al.}{2020}]{2020ApJ...902...98G}
{Genzel} R.,  et~al., 2020, \mn@doi [\apj] {10.3847/1538-4357/abb0ea}, \href
  {https://ui.adsabs.harvard.edu/abs/2020ApJ...902...98G} {902, 98}

\bibitem[\protect\citeauthoryear{{Gnedin} \& {Zhao}}{{Gnedin} \&
  {Zhao}}{2002}]{2002MNRAS.333..299G}
{Gnedin} O.~Y.,  {Zhao} H.,  2002, \mn@doi [\mnras]
  {10.1046/j.1365-8711.2002.05361.x}, \href
  {https://ui.adsabs.harvard.edu/abs/2002MNRAS.333..299G} {333, 299}

\bibitem[\protect\citeauthoryear{{Gnedin}, {Hernquist}  \& {Ostriker}}{{Gnedin}
  et~al.}{1999}]{1999ApJ...514..109G}
{Gnedin} O.~Y.,  {Hernquist} L.,   {Ostriker} J.~P.,  1999, \mn@doi [\apj]
  {10.1086/306910}, \href
  {https://ui.adsabs.harvard.edu/abs/1999ApJ...514..109G} {514, 109}

\bibitem[\protect\citeauthoryear{{Gnedin}, {Kravtsov}, {Klypin}  \&
  {Nagai}}{{Gnedin} et~al.}{2004}]{2004ApJ...616...16G}
{Gnedin} O.~Y.,  {Kravtsov} A.~V.,  {Klypin} A.~A.,   {Nagai} D.,  2004,
  \mn@doi [\apj] {10.1086/424914}, \href
  {https://ui.adsabs.harvard.edu/abs/2004ApJ...616...16G} {616, 16}

\bibitem[\protect\citeauthoryear{{Governato} et~al.,}{{Governato}
  et~al.}{2010}]{2010Natur.463..203G}
{Governato} F.,  et~al., 2010, \mn@doi [\nat] {10.1038/nature08640}, \href
  {https://ui.adsabs.harvard.edu/abs/2010Natur.463..203G} {463, 203}

\bibitem[\protect\citeauthoryear{Granger \& Pérez}{Granger \&
  Pérez}{2021}]{jupyter}
Granger B.~E.,  Pérez F.,  2021, \mn@doi [Computing in Science Engineering]
  {10.1109/MCSE.2021.3059263}, 23, 7

\bibitem[\protect\citeauthoryear{{Guo} et~al.,}{{Guo}
  et~al.}{2020}]{2020NatAs...4..246G}
{Guo} Q.,  et~al., 2020, \mn@doi [Nature Astronomy]
  {10.1038/s41550-019-0930-9}, \href
  {https://ui.adsabs.harvard.edu/abs/2020NatAs...4..246G} {4, 246}

\bibitem[\protect\citeauthoryear{{Han}, {Wang}, {Cole}  \& {Frenk}}{{Han}
  et~al.}{2016}]{2016MNRAS.456.1003H}
{Han} J.,  {Wang} W.,  {Cole} S.,   {Frenk} C.~S.,  2016, \mn@doi [\mnras]
  {10.1093/mnras/stv2707}, \href
  {https://ui.adsabs.harvard.edu/abs/2016MNRAS.456.1003H} {456, 1003}

\bibitem[\protect\citeauthoryear{Harris et~al.,}{Harris et~al.}{2020}]{numpy}
Harris C.~R.,  et~al., 2020, \mn@doi [Nature] {10.1038/s41586-020-2649-2}, 585,
  357

\bibitem[\protect\citeauthoryear{{Hashim}, {El-Zant}, {Freundlich}, {Read}  \&
  {Combes}}{{Hashim} et~al.}{2022}]{2022arXiv220908631H}
{Hashim} M.,  {El-Zant} A.,  {Freundlich} J.,  {Read} J.,   {Combes} F.,  2022,
  arXiv e-prints, \href {https://ui.adsabs.harvard.edu/abs/2022arXiv220908631H}
  {p. arXiv:2209.08631}

\bibitem[\protect\citeauthoryear{{Hayashi}, {Chiba}  \& {Ishiyama}}{{Hayashi}
  et~al.}{2020}]{2020ApJ...904...45H}
{Hayashi} K.,  {Chiba} M.,   {Ishiyama} T.,  2020, \mn@doi [\apj]
  {10.3847/1538-4357/abbe0a}, \href
  {https://ui.adsabs.harvard.edu/abs/2020ApJ...904...45H} {904, 45}

\bibitem[\protect\citeauthoryear{Hunter}{Hunter}{2007}]{matplotlib}
Hunter J.~D.,  2007, \mn@doi [Computing in Science \& Engineering]
  {10.1109/MCSE.2007.55}, 9, 90

\bibitem[\protect\citeauthoryear{{Jackson} et~al.,}{{Jackson}
  et~al.}{2021}]{2021MNRAS.502.4262J}
{Jackson} R.~A.,  et~al., 2021, \mn@doi [\mnras] {10.1093/mnras/stab077}, \href
  {https://ui.adsabs.harvard.edu/abs/2021MNRAS.502.4262J} {502, 4262}

\bibitem[\protect\citeauthoryear{{Jiang}, {Dekel}, {Freundlich}, {Romanowsky},
  {Dutton}, {Macci{\`o}}  \& {Di Cintio}}{{Jiang}
  et~al.}{2019}]{2019MNRAS.487.5272J}
{Jiang} F.,  {Dekel} A.,  {Freundlich} J.,  {Romanowsky} A.~J.,  {Dutton}
  A.~A.,  {Macci{\`o}} A.~V.,   {Di Cintio} A.,  2019, \mn@doi [\mnras]
  {10.1093/mnras/stz1499}, \href
  {https://ui.adsabs.harvard.edu/abs/2019MNRAS.487.5272J} {487, 5272}

\bibitem[\protect\citeauthoryear{{Jiang}, {Dekel}, {Freundlich}, {van den
  Bosch}, {Green}, {Hopkins}, {Benson}  \& {Du}}{{Jiang}
  et~al.}{2021}]{2021MNRAS.502..621J}
{Jiang} F.,  {Dekel} A.,  {Freundlich} J.,  {van den Bosch} F.~C.,  {Green}
  S.~B.,  {Hopkins} P.~F.,  {Benson} A.,   {Du} X.,  2021, \mn@doi [\mnras]
  {10.1093/mnras/staa4034}, \href
  {https://ui.adsabs.harvard.edu/abs/2021MNRAS.502..621J} {502, 621}

\bibitem[\protect\citeauthoryear{{Lapiner}, {Dekel}  \& {Dubois}}{{Lapiner}
  et~al.}{2021}]{2021MNRAS.505..172L}
{Lapiner} S.,  {Dekel} A.,   {Dubois} Y.,  2021, \mn@doi [\mnras]
  {10.1093/mnras/stab1205}, \href
  {https://ui.adsabs.harvard.edu/abs/2021MNRAS.505..172L} {505, 172}

\bibitem[\protect\citeauthoryear{{Lazar} et~al.,}{{Lazar}
  et~al.}{2020}]{2020MNRAS.497.2393L}
{Lazar} A.,  et~al., 2020, \mn@doi [\mnras] {10.1093/mnras/staa2101}, \href
  {https://ui.adsabs.harvard.edu/abs/2020MNRAS.497.2393L} {497, 2393}

\bibitem[\protect\citeauthoryear{{Liao} et~al.,}{{Liao}
  et~al.}{2019}]{2019MNRAS.490.5182L}
{Liao} S.,  et~al., 2019, \mn@doi [\mnras] {10.1093/mnras/stz2969}, \href
  {https://ui.adsabs.harvard.edu/abs/2019MNRAS.490.5182L} {490, 5182}

\bibitem[\protect\citeauthoryear{{Lim} et~al.,}{{Lim}
  et~al.}{2020}]{2020ApJ...899...69L}
{Lim} S.,  et~al., 2020, \mn@doi [\apj] {10.3847/1538-4357/aba433}, \href
  {https://ui.adsabs.harvard.edu/abs/2020ApJ...899...69L} {899, 69}

\bibitem[\protect\citeauthoryear{{Lovell} et~al.,}{{Lovell}
  et~al.}{2018}]{2018MNRAS.481.1950L}
{Lovell} M.~R.,  et~al., 2018, \mn@doi [\mnras] {10.1093/mnras/sty2339}, \href
  {https://ui.adsabs.harvard.edu/abs/2018MNRAS.481.1950L} {481, 1950}

\bibitem[\protect\citeauthoryear{{Lynden-Bell}}{{Lynden-Bell}}{1967}]{1967MNRAS.136..101L}
{Lynden-Bell} D.,  1967, \mn@doi [\mnras] {10.1093/mnras/136.1.101}, \href
  {https://ui.adsabs.harvard.edu/abs/1967MNRAS.136..101L} {136, 101}

\bibitem[\protect\citeauthoryear{{Mancera Pi{\~n}a}, {Aguerri}, {Peletier},
  {Venhola}, {Trager}  \& {Choque Challapa}}{{Mancera Pi{\~n}a}
  et~al.}{2019}]{2019MNRAS.485.1036M}
{Mancera Pi{\~n}a} P.~E.,  {Aguerri} J.~A.~L.,  {Peletier} R.~F.,  {Venhola}
  A.,  {Trager} S.,   {Choque Challapa} N.,  2019, \mn@doi [\mnras]
  {10.1093/mnras/stz238}, \href
  {https://ui.adsabs.harvard.edu/abs/2019MNRAS.485.1036M} {485, 1036}

\bibitem[\protect\citeauthoryear{{Mancera Pi{\~n}a}, {Fraternali}, {Oosterloo},
  {Adams}, {Oman}  \& {Leisman}}{{Mancera Pi{\~n}a}
  et~al.}{2022}]{2022MNRAS.512.3230M}
{Mancera Pi{\~n}a} P.~E.,  {Fraternali} F.,  {Oosterloo} T.,  {Adams} E. A.~K.,
   {Oman} K.~A.,   {Leisman} L.,  2022, \mn@doi [\mnras]
  {10.1093/mnras/stab3491}, \href
  {https://ui.adsabs.harvard.edu/abs/2022MNRAS.512.3230M} {512, 3230}

\bibitem[\protect\citeauthoryear{{Mart{\'\i}nez-Delgado}
  et~al.,}{{Mart{\'\i}nez-Delgado} et~al.}{2016}]{2016AJ....151...96M}
{Mart{\'\i}nez-Delgado} D.,  et~al., 2016, \mn@doi [\aj]
  {10.3847/0004-6256/151/4/96}, \href
  {https://ui.adsabs.harvard.edu/abs/2016AJ....151...96M} {151, 96}

\bibitem[\protect\citeauthoryear{{Moore}}{{Moore}}{1994}]{1994Natur.370..629M}
{Moore} B.,  1994, \mn@doi [\nat] {10.1038/370629a0}, \href
  {https://ui.adsabs.harvard.edu/abs/1994Natur.370..629M} {370, 629}

\bibitem[\protect\citeauthoryear{{Navarro}, {Eke}  \& {Frenk}}{{Navarro}
  et~al.}{1996a}]{1996MNRAS.283L..72N}
{Navarro} J.~F.,  {Eke} V.~R.,   {Frenk} C.~S.,  1996a, \mn@doi [\mnras]
  {10.1093/mnras/283.3.L72}, \href
  {https://ui.adsabs.harvard.edu/abs/1996MNRAS.283L..72N} {283, L72}

\bibitem[\protect\citeauthoryear{{Navarro}, {Frenk}  \& {White}}{{Navarro}
  et~al.}{1996b}]{1996ApJ...462..563N}
{Navarro} J.~F.,  {Frenk} C.~S.,   {White} S. D.~M.,  1996b, \mn@doi [\apj]
  {10.1086/177173}, \href
  {https://ui.adsabs.harvard.edu/abs/1996ApJ...462..563N} {462, 563}

\bibitem[\protect\citeauthoryear{{Nestor Shachar} et~al.,}{{Nestor Shachar}
  et~al.}{2022}]{2022arXiv220912199N}
{Nestor Shachar} A.,  et~al., 2022, arXiv e-prints, \href
  {https://ui.adsabs.harvard.edu/abs/2022arXiv220912199N} {p. arXiv:2209.12199}

\bibitem[\protect\citeauthoryear{Odland}{Odland}{2018}]{kdepy}
Odland T.,  2018, KDEpy: Kernel Density Estimation in Python,
  \mn@doi{10.5281/zenodo.2392268}

\bibitem[\protect\citeauthoryear{{Ogiya}}{{Ogiya}}{2018}]{2018MNRAS.480L.106O}
{Ogiya} G.,  2018, \mn@doi [\mnras] {10.1093/mnrasl/sly138}, \href
  {https://ui.adsabs.harvard.edu/abs/2018MNRAS.480L.106O} {480, L106}

\bibitem[\protect\citeauthoryear{{Ogiya} \& {Nagai}}{{Ogiya} \&
  {Nagai}}{2022}]{2022MNRAS.514..555O}
{Ogiya} G.,  {Nagai} D.,  2022, \mn@doi [\mnras] {10.1093/mnras/stac1311},
  \href {https://ui.adsabs.harvard.edu/abs/2022MNRAS.514..555O} {514, 555}

\bibitem[\protect\citeauthoryear{{Ogiya}, {Taylor}  \& {Hudson}}{{Ogiya}
  et~al.}{2021}]{2021MNRAS.503.1233O}
{Ogiya} G.,  {Taylor} J.~E.,   {Hudson} M.~J.,  2021, \mn@doi [\mnras]
  {10.1093/mnras/stab361}, \href
  {https://ui.adsabs.harvard.edu/abs/2021MNRAS.503.1233O} {503, 1233}

\bibitem[\protect\citeauthoryear{{Ogiya}, {van den Bosch}  \&
  {Burkert}}{{Ogiya} et~al.}{2022}]{2022MNRAS.510.2724O}
{Ogiya} G.,  {van den Bosch} F.~C.,   {Burkert} A.,  2022, \mn@doi [\mnras]
  {10.1093/mnras/stab3658}, \href
  {https://ui.adsabs.harvard.edu/abs/2022MNRAS.510.2724O} {510, 2724}

\bibitem[\protect\citeauthoryear{{Oh}, {de Blok}, {Brinks}, {Walter}  \&
  {Kennicutt}}{{Oh} et~al.}{2011a}]{2011AJ....141..193O}
{Oh} S.-H.,  {de Blok} W.~J.~G.,  {Brinks} E.,  {Walter} F.,   {Kennicutt}
  Robert~C. J.,  2011a, \mn@doi [\aj] {10.1088/0004-6256/141/6/193}, \href
  {https://ui.adsabs.harvard.edu/abs/2011AJ....141..193O} {141, 193}

\bibitem[\protect\citeauthoryear{{Oh}, {Brook}, {Governato}, {Brinks}, {Mayer},
  {de Blok}, {Brooks}  \& {Walter}}{{Oh} et~al.}{2011b}]{2011AJ....142...24O}
{Oh} S.-H.,  {Brook} C.,  {Governato} F.,  {Brinks} E.,  {Mayer} L.,  {de Blok}
  W.~J.~G.,  {Brooks} A.,   {Walter} F.,  2011b, \mn@doi [\aj]
  {10.1088/0004-6256/142/1/24}, \href
  {https://ui.adsabs.harvard.edu/abs/2011AJ....142...24O} {142, 24}

\bibitem[\protect\citeauthoryear{{Oh} et~al.,}{{Oh}
  et~al.}{2015}]{2015AJ....149..180O}
{Oh} S.-H.,  et~al., 2015, \mn@doi [\aj] {10.1088/0004-6256/149/6/180}, \href
  {https://ui.adsabs.harvard.edu/abs/2015AJ....149..180O} {149, 180}

\bibitem[\protect\citeauthoryear{{Pe{\~n}arrubia}}{{Pe{\~n}arrubia}}{2013}]{2013MNRAS.433.2576P}
{Pe{\~n}arrubia} J.,  2013, \mn@doi [\mnras] {10.1093/mnras/stt935}, \href
  {https://ui.adsabs.harvard.edu/abs/2013MNRAS.433.2576P} {433, 2576}

\bibitem[\protect\citeauthoryear{{Penoyre} \& {Haiman}}{{Penoyre} \&
  {Haiman}}{2018}]{2018MNRAS.473..498P}
{Penoyre} Z.,  {Haiman} Z.,  2018, \mn@doi [\mnras] {10.1093/mnras/stx2469},
  \href {https://ui.adsabs.harvard.edu/abs/2018MNRAS.473..498P} {473, 498}

\bibitem[\protect\citeauthoryear{{Pontzen} \& {Governato}}{{Pontzen} \&
  {Governato}}{2012}]{2012MNRAS.421.3464P}
{Pontzen} A.,  {Governato} F.,  2012, \mn@doi [\mnras]
  {10.1111/j.1365-2966.2012.20571.x}, \href
  {https://ui.adsabs.harvard.edu/abs/2012MNRAS.421.3464P} {421, 3464}

\bibitem[\protect\citeauthoryear{{Power}, {Navarro}, {Jenkins}, {Frenk},
  {White}, {Springel}, {Stadel}  \& {Quinn}}{{Power}
  et~al.}{2003}]{2003MNRAS.338...14P}
{Power} C.,  {Navarro} J.~F.,  {Jenkins} A.,  {Frenk} C.~S.,  {White} S.~D.~M.,
   {Springel} V.,  {Stadel} J.,   {Quinn} T.,  2003, \mn@doi [\mnras]
  {10.1046/j.1365-8711.2003.05925.x}, \href
  {https://ui.adsabs.harvard.edu/abs/2003MNRAS.338...14P} {338, 14}

\bibitem[\protect\citeauthoryear{{Price} et~al.,}{{Price}
  et~al.}{2021}]{2021ApJ...922..143P}
{Price} S.~H.,  et~al., 2021, \mn@doi [\apj] {10.3847/1538-4357/ac22ad}, \href
  {https://ui.adsabs.harvard.edu/abs/2021ApJ...922..143P} {922, 143}

\bibitem[\protect\citeauthoryear{{Read}, {Agertz}  \& {Collins}}{{Read}
  et~al.}{2016a}]{2016MNRAS.459.2573R}
{Read} J.~I.,  {Agertz} O.,   {Collins} M.~L.~M.,  2016a, \mn@doi [\mnras]
  {10.1093/mnras/stw713}, \href
  {https://ui.adsabs.harvard.edu/abs/2016MNRAS.459.2573R} {459, 2573}

\bibitem[\protect\citeauthoryear{{Read}, {Iorio}, {Agertz}  \&
  {Fraternali}}{{Read} et~al.}{2016b}]{2016MNRAS.462.3628R}
{Read} J.~I.,  {Iorio} G.,  {Agertz} O.,   {Fraternali} F.,  2016b, \mn@doi
  [\mnras] {10.1093/mnras/stw1876}, \href
  {https://ui.adsabs.harvard.edu/abs/2016MNRAS.462.3628R} {462, 3628}

\bibitem[\protect\citeauthoryear{{Rom{\'a}n} \& {Trujillo}}{{Rom{\'a}n} \&
  {Trujillo}}{2017}]{2017MNRAS.468..703R}
{Rom{\'a}n} J.,  {Trujillo} I.,  2017, \mn@doi [\mnras] {10.1093/mnras/stx438},
  \href {https://ui.adsabs.harvard.edu/abs/2017MNRAS.468..703R} {468, 703}

\bibitem[\protect\citeauthoryear{{Ryden} \& {Gunn}}{{Ryden} \&
  {Gunn}}{1987}]{1987ApJ...318...15R}
{Ryden} B.~S.,  {Gunn} J.~E.,  1987, \mn@doi [\apj] {10.1086/165349}, \href
  {https://ui.adsabs.harvard.edu/abs/1987ApJ...318...15R} {318, 15}

\bibitem[\protect\citeauthoryear{{Sales}, {Wetzel}  \& {Fattahi}}{{Sales}
  et~al.}{2022}]{2022NatAs...6..897S}
{Sales} L.~V.,  {Wetzel} A.,   {Fattahi} A.,  2022, \mn@doi [Nature Astronomy]
  {10.1038/s41550-022-01689-w}, \href
  {https://ui.adsabs.harvard.edu/abs/2022NatAs...6..897S} {6, 897}

\bibitem[\protect\citeauthoryear{{Sellwood} \& {McGaugh}}{{Sellwood} \&
  {McGaugh}}{2005}]{2005ApJ...634...70S}
{Sellwood} J.~A.,  {McGaugh} S.~S.,  2005, \mn@doi [\apj] {10.1086/491731},
  \href {https://ui.adsabs.harvard.edu/abs/2005ApJ...634...70S} {634, 70}

\bibitem[\protect\citeauthoryear{{Sharma}, {Salucci}  \& {van de Ven}}{{Sharma}
  et~al.}{2022}]{2022A&A...659A..40S}
{Sharma} G.,  {Salucci} P.,   {van de Ven} G.,  2022, \mn@doi [\aap]
  {10.1051/0004-6361/202141822}, \href
  {https://ui.adsabs.harvard.edu/abs/2022A&A...659A..40S} {659, A40}

\bibitem[\protect\citeauthoryear{{Somerville} \& {Dav{\'e}}}{{Somerville} \&
  {Dav{\'e}}}{2015}]{2015ARA&A..53...51S}
{Somerville} R.~S.,  {Dav{\'e}} R.,  2015, \mn@doi [\araa]
  {10.1146/annurev-astro-082812-140951}, \href
  {https://ui.adsabs.harvard.edu/abs/2015ARA&A..53...51S} {53, 51}

\bibitem[\protect\citeauthoryear{{St{\"u}cker}, {Ogiya}, {Angulo},
  {Aguirre-Santaella}  \& {S{\'a}nchez-Conde}}{{St{\"u}cker}
  et~al.}{2022}]{2022arXiv220700604S}
{St{\"u}cker} J.,  {Ogiya} G.,  {Angulo} R.~E.,  {Aguirre-Santaella} A.,
  {S{\'a}nchez-Conde} M.~A.,  2022, arXiv e-prints, \href
  {https://ui.adsabs.harvard.edu/abs/2022arXiv220700604S} {p. arXiv:2207.00604}

\bibitem[\protect\citeauthoryear{{Teuben}}{{Teuben}}{1995}]{1995ASPC...77..398T}
{Teuben} P.,  1995, in {Shaw} R.~A.,  {Payne} H.~E.,   {Hayes} J.~J.~E.,  eds,
  Astronomical Society of the Pacific Conference Series Vol. 77, Astronomical
  Data Analysis Software and Systems IV. p.~398

\bibitem[\protect\citeauthoryear{{Tollet} et~al.,}{{Tollet}
  et~al.}{2016}]{2016MNRAS.456.3542T}
{Tollet} E.,  et~al., 2016, \mn@doi [\mnras] {10.1093/mnras/stv2856}, \href
  {https://ui.adsabs.harvard.edu/abs/2016MNRAS.456.3542T} {456, 3542}

\bibitem[\protect\citeauthoryear{{Tutukov}}{{Tutukov}}{1978}]{1978A&A....70...57T}
{Tutukov} A.~V.,  1978, \aap, \href
  {https://ui.adsabs.harvard.edu/abs/1978A&A....70...57T} {70, 57}

\bibitem[\protect\citeauthoryear{{{\"U}bler} et~al.,}{{{\"U}bler}
  et~al.}{2021}]{2021MNRAS.500.4597U}
{{\"U}bler} H.,  et~al., 2021, \mn@doi [\mnras] {10.1093/mnras/staa3464}, \href
  {https://ui.adsabs.harvard.edu/abs/2021MNRAS.500.4597U} {500, 4597}

\bibitem[\protect\citeauthoryear{{Vasiliev}}{{Vasiliev}}{2018}]{2018arXiv180208255V}
{Vasiliev} E.,  2018, arXiv e-prints, \href
  {https://ui.adsabs.harvard.edu/abs/2018arXiv180208255V} {p. arXiv:1802.08255}

\bibitem[\protect\citeauthoryear{{Vasiliev}}{{Vasiliev}}{2019}]{2019MNRAS.482.1525V}
{Vasiliev} E.,  2019, \mn@doi [\mnras] {10.1093/mnras/sty2672}, \href
  {https://ui.adsabs.harvard.edu/abs/2019MNRAS.482.1525V} {482, 1525}

\bibitem[\protect\citeauthoryear{{Virtanen} et~al.,}{{Virtanen}
  et~al.}{2020}]{scipy}
{Virtanen} P.,  et~al., 2020, \mn@doi [Nature Methods]
  {https://doi.org/10.1038/s41592-019-0686-2}, \href {https://rdcu.be/b08Wh}
  {17, 261}

\bibitem[\protect\citeauthoryear{{Wang}, {Dutton}, {Stinson}, {Macci{\`o}},
  {Penzo}, {Kang}, {Keller}  \& {Wadsley}}{{Wang}
  et~al.}{2015}]{2015MNRAS.454...83W}
{Wang} L.,  {Dutton} A.~A.,  {Stinson} G.~S.,  {Macci{\`o}} A.~V.,  {Penzo} C.,
   {Kang} X.,  {Keller} B.~W.,   {Wadsley} J.,  2015, \mn@doi [\mnras]
  {10.1093/mnras/stv1937}, \href
  {https://ui.adsabs.harvard.edu/abs/2015MNRAS.454...83W} {454, 83}

\bibitem[\protect\citeauthoryear{{Wang} et~al.,}{{Wang}
  et~al.}{2022}]{2022arXiv220612121W}
{Wang} W.,  et~al., 2022, arXiv e-prints, \href
  {https://ui.adsabs.harvard.edu/abs/2022arXiv220612121W} {p. arXiv:2206.12121}

\bibitem[\protect\citeauthoryear{{Wright}, {Tremmel}, {Brooks}, {Munshi},
  {Nagai}, {Sharma}  \& {Quinn}}{{Wright} et~al.}{2021}]{2021MNRAS.502.5370W}
{Wright} A.~C.,  {Tremmel} M.,  {Brooks} A.~M.,  {Munshi} F.,  {Nagai} D.,
  {Sharma} R.~S.,   {Quinn} T.~R.,  2021, \mn@doi [\mnras]
  {10.1093/mnras/stab081}, \href
  {https://ui.adsabs.harvard.edu/abs/2021MNRAS.502.5370W} {502, 5370}

\bibitem[\protect\citeauthoryear{{Young}}{{Young}}{1980}]{1980ApJ...242.1232Y}
{Young} P.,  1980, \mn@doi [\apj] {10.1086/158553}, \href
  {https://ui.adsabs.harvard.edu/abs/1980ApJ...242.1232Y} {242, 1232}

\bibitem[\protect\citeauthoryear{{Zhao}}{{Zhao}}{1996}]{1996MNRAS.278..488Z}
{Zhao} H.,  1996, \mn@doi [\mnras] {10.1093/mnras/278.2.488}, \href
  {https://ui.adsabs.harvard.edu/abs/1996MNRAS.278..488Z} {278, 488}

\bibitem[\protect\citeauthoryear{{Zhao}}{{Zhao}}{2002}]{2002MNRAS.336..159Z}
{Zhao} H.,  2002, \mn@doi [\mnras] {10.1046/j.1365-8711.2002.05722.x}, \href
  {https://ui.adsabs.harvard.edu/abs/2002MNRAS.336..159Z} {336, 159}

\bibitem[\protect\citeauthoryear{{Zhu}, {Marinacci}, {Maji}, {Li}, {Springel}
  \& {Hernquist}}{{Zhu} et~al.}{2016}]{2016MNRAS.458.1559Z}
{Zhu} Q.,  {Marinacci} F.,  {Maji} M.,  {Li} Y.,  {Springel} V.,   {Hernquist}
  L.,  2016, \mn@doi [\mnras] {10.1093/mnras/stw374}, \href
  {https://ui.adsabs.harvard.edu/abs/2016MNRAS.458.1559Z} {458, 1559}

\bibitem[\protect\citeauthoryear{{de Blok}, {McGaugh}, {Bosma}  \& {Rubin}}{{de
  Blok} et~al.}{2001}]{2001ApJ...552L..23D}
{de Blok} W.~J.~G.,  {McGaugh} S.~S.,  {Bosma} A.,   {Rubin} V.~C.,  2001,
  \mn@doi [\apjl] {10.1086/320262}, \href
  {https://ui.adsabs.harvard.edu/abs/2001ApJ...552L..23D} {552, L23}

\bibitem[\protect\citeauthoryear{{de Blok}, {Walter}, {Brinks}, {Trachternach},
  {Oh}  \& {Kennicutt}}{{de Blok} et~al.}{2008}]{2008AJ....136.2648D}
{de Blok} W.~J.~G.,  {Walter} F.,  {Brinks} E.,  {Trachternach} C.,  {Oh}
  S.~H.,   {Kennicutt} R.~C. J.,  2008, \mn@doi [\aj]
  {10.1088/0004-6256/136/6/2648}, \href
  {https://ui.adsabs.harvard.edu/abs/2008AJ....136.2648D} {136, 2648}

\bibitem[\protect\citeauthoryear{{van Dokkum}, {Abraham}, {Merritt}, {Zhang},
  {Geha}  \& {Conroy}}{{van Dokkum} et~al.}{2015}]{2015ApJ...798L..45V}
{van Dokkum} P.~G.,  {Abraham} R.,  {Merritt} A.,  {Zhang} J.,  {Geha} M.,
  {Conroy} C.,  2015, \mn@doi [\apjl] {10.1088/2041-8205/798/2/L45}, \href
  {https://ui.adsabs.harvard.edu/abs/2015ApJ...798L..45V} {798, L45}

\bibitem[\protect\citeauthoryear{{van Dokkum}, {Danieli}, {Abraham}, {Conroy}
  \& {Romanowsky}}{{van Dokkum} et~al.}{2019}]{2019ApJ...874L...5V}
{van Dokkum} P.,  {Danieli} S.,  {Abraham} R.,  {Conroy} C.,   {Romanowsky}
  A.~J.,  2019, \mn@doi [\apjl] {10.3847/2041-8213/ab0d92}, \href
  {https://ui.adsabs.harvard.edu/abs/2019ApJ...874L...5V} {874, L5}

\bibitem[\protect\citeauthoryear{{van den Bosch}, {Ogiya}, {Hahn}  \&
  {Burkert}}{{van den Bosch} et~al.}{2018}]{2018MNRAS.474.3043V}
{van den Bosch} F.~C.,  {Ogiya} G.,  {Hahn} O.,   {Burkert} A.,  2018, \mn@doi
  [\mnras] {10.1093/mnras/stx2956}, \href
  {https://ui.adsabs.harvard.edu/abs/2018MNRAS.474.3043V} {474, 3043}

\makeatother
\end{thebibliography}





\appendix

\renewcommand{\thefigure}{A\arabic{figure}}
\setcounter{figure}{0}
\begin{figure*}
\centering
\includegraphics[width=0.95\textwidth]{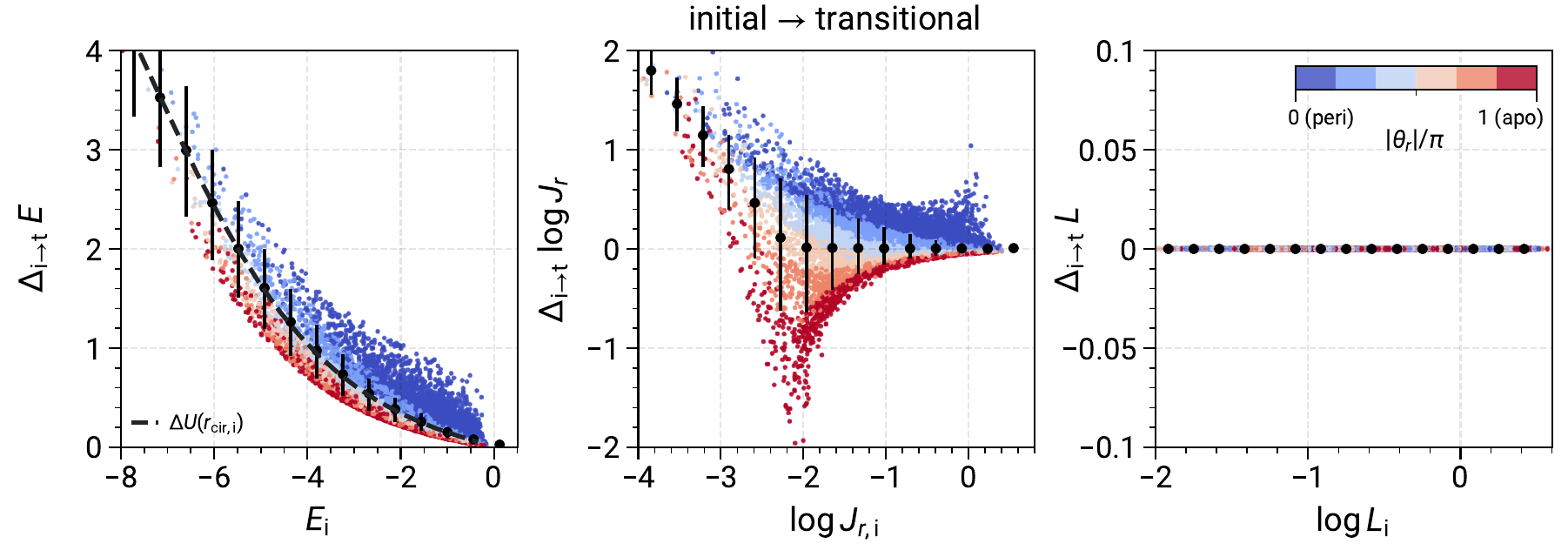}\\ \vspace{2mm}
\includegraphics[width=0.95\textwidth]{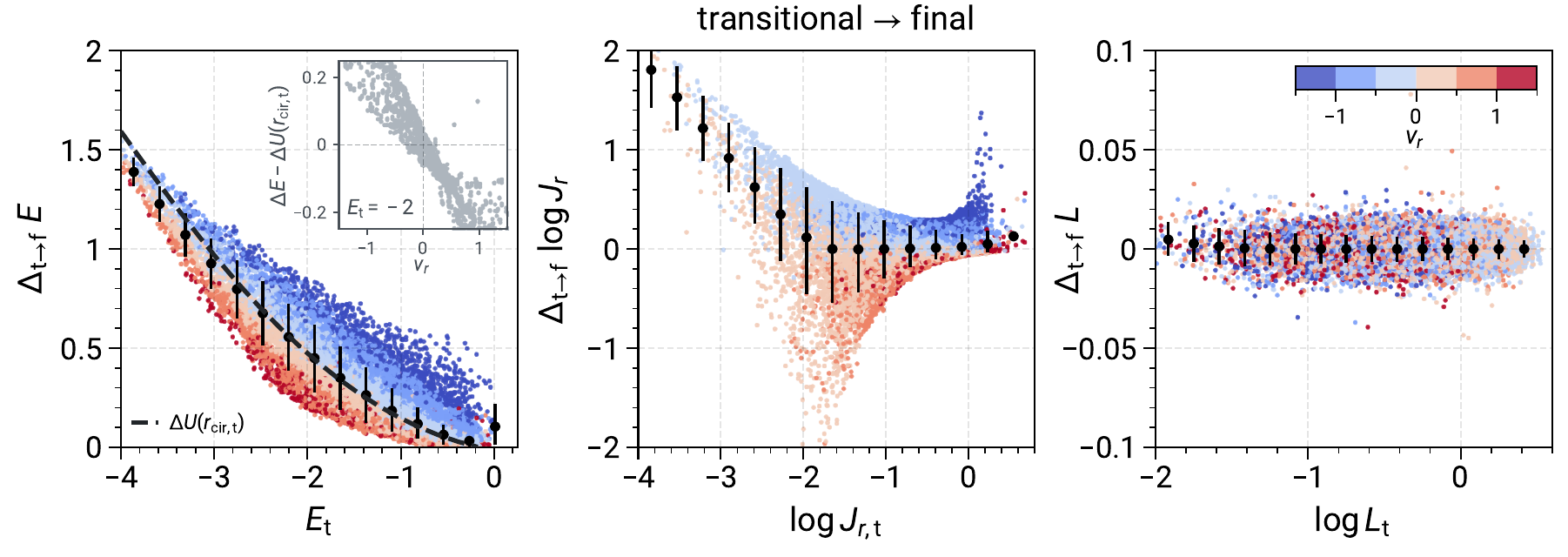}
\vspace{-1em}
\caption{%
Change of the orbit integrals 
due to the initial gas ejection (upper panels) and subsequent DM relaxation (lower panels)
for particles in the Run A1 with $\eta=-1$ (same as \reffig{fig:orbint_pdf}).
A random subset of 1\% particles in the simulation is shown.
The particles in the upper panels are colored by radial angles, $|\theta_r/\pi|\in [0, 1]$ in the initial potential,
while those in the bottom panels are colored by the initial radial velocities $v_r$.
In each panel, the black dots with errorbar show the mean and the 18 and 84-th percentiles 
for binned data points.
In the first column, the potential change at the circular radius 
for given energy, $\Delta U(r_\mathrm{cir})$ is shown as dark dashed lines for reference.
The inset plot in the lower-left panel shows the relation between $v_r$ and the energy change residual,
$\Delta E - \Delta U(r_\mathrm{cir})$, for particles with $E_\mt=-2$.
\vspace{-1em}
}
\label{fig:orbint_diff}
\end{figure*}

\section{Change of orbital integrals}
\label{sec:integral_change}

In \reffig{fig:orbint_diff}, we show the change of orbital integrals 
due to the initial gas ejection and subsequent DM relaxation
for particles in the Run A1 with complete gas removal ($\eta=-1$).
For individual particles,
$L$ of is well conserved because of the spherical symmetry,
while $E$ and $J_r$ suffer from diffusion in both stages.
For given $E$ or $J_r$, the diffusion is mainly determined by the radial phase $\theta_r$ in the first stage
and by the initial radial velocity $v_r$ during the relaxation.
A particle initially located closer to the center or moving inwards with larger $|v_r|$
has a greater positive $\Delta E$ and $\Delta J$.

As shown in the upper-left panel of \reffig{fig:orbint_diff}, 
for a particle collection of given $E$, the average energy gain due to a sudden potential change $\Delta U$
can be well approximated by
\begin{equation}
\avg{\Delta E} = \avg{\Delta U(r)} \simeq \Delta U(r_\mathrm{cir}),
\end{equation}
where $r_\mathrm{cir}(E)$ is the circular orbit radius that
satisfies $E = \frac{1}{2}v^2_\mathrm{c}(r_\mathrm{cir})+ U (r_\mathrm{cir})$ with $v^2_\mathrm{c}(r)=r\frac{\dif U(r)}{2\dif r}$.
Note that an orbit with energy $E$ will always pass through $r_{\mathrm{cir}}(E)$ regardless of the orbit circularity.
The rigorous $\avg{\Delta E}$ can be derived by taking the average $\avg{U(r)}$ along the orbit in principle
\citep[e.g.,][eq.\ 12]{2012MNRAS.421.3464P},
but $r_\mathrm{cir}$ is much easier to compute.
This approximation can also apply to short time intervals $\Delta t$.
It becomes less accurate when directly applied to the whole relaxation,
but nevertheless still providing a quick good estimate (lower-left panel).

As expected for a non-adiabatic process, the radial action $J_r$ is not conserved for individual particles,
suffering from a strong diffusion (see also \citealt{2019MNRAS.485.1008B}).
Except for the nearly circular orbits  ($J_r/\rvir\vvir \lesssim 10^{-2}$),
the average $\Delta J_r$ is, however, very small, 
which largely keeps the ensemble distribution $p(J_r)$ unchanged as shown in \reffig{fig:orbint_pdf}.

\renewcommand{\thefigure}{B\arabic{figure}}
\setcounter{figure}{0}
\begin{figure*}
\centering
\includegraphics[width=1\textwidth]{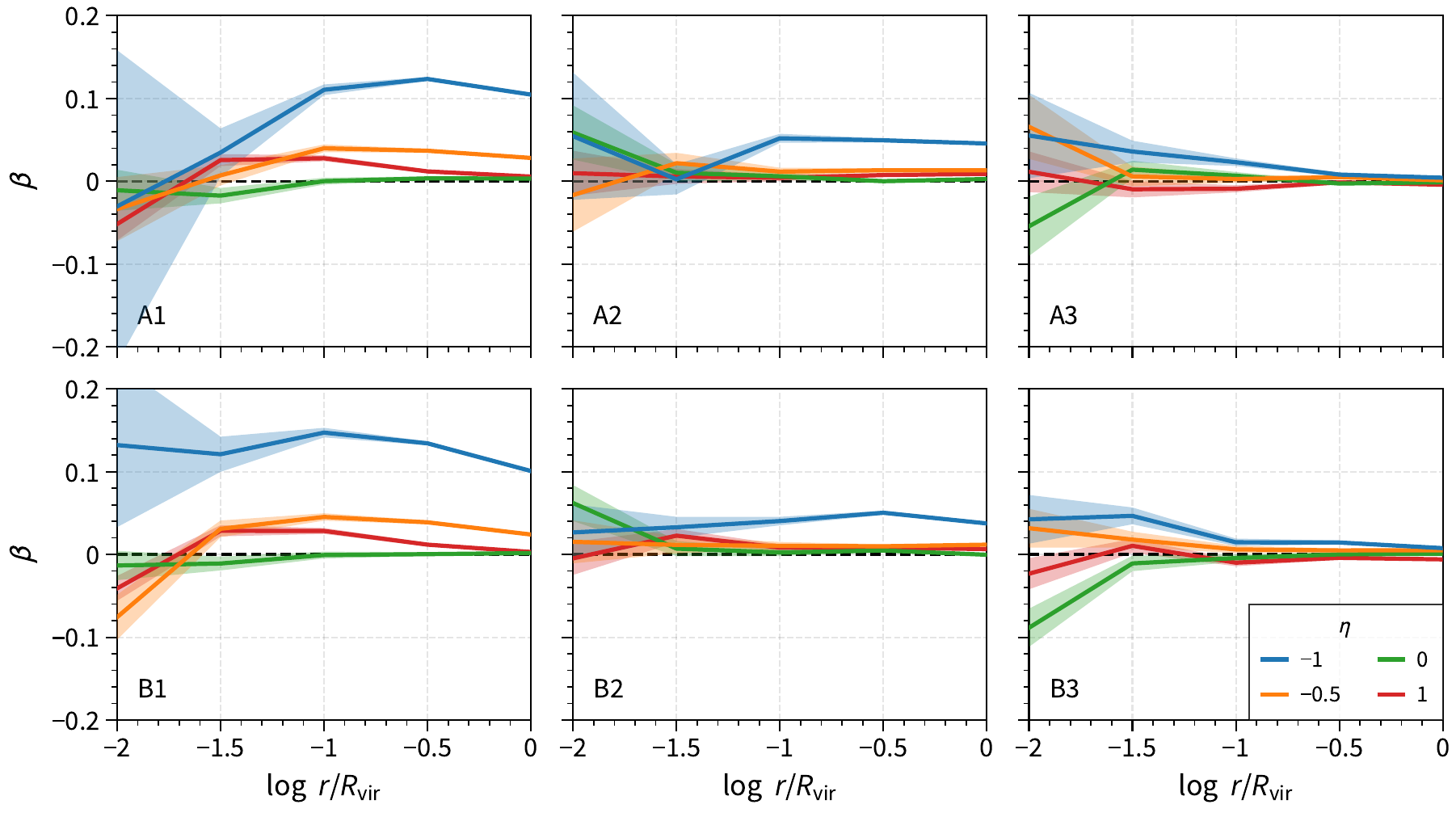}
\vspace{-2em}
\caption{%
The velocity anisotropy profile of the DM in the final snapshots of simulations.
All the simulations have initial condition of $\beta=0$.
The systems with strong gas ejection developed slight radial anisotropy of $\beta\sim 0.05$ to 0.15,
while the remaining simulations remain largely isotropic.
}
\label{fig:beta}
\end{figure*}

\renewcommand{\thefigure}{F\arabic{figure}}
\setcounter{figure}{0}
\begin{figure*}
\centering
\includegraphics[width=1\textwidth]{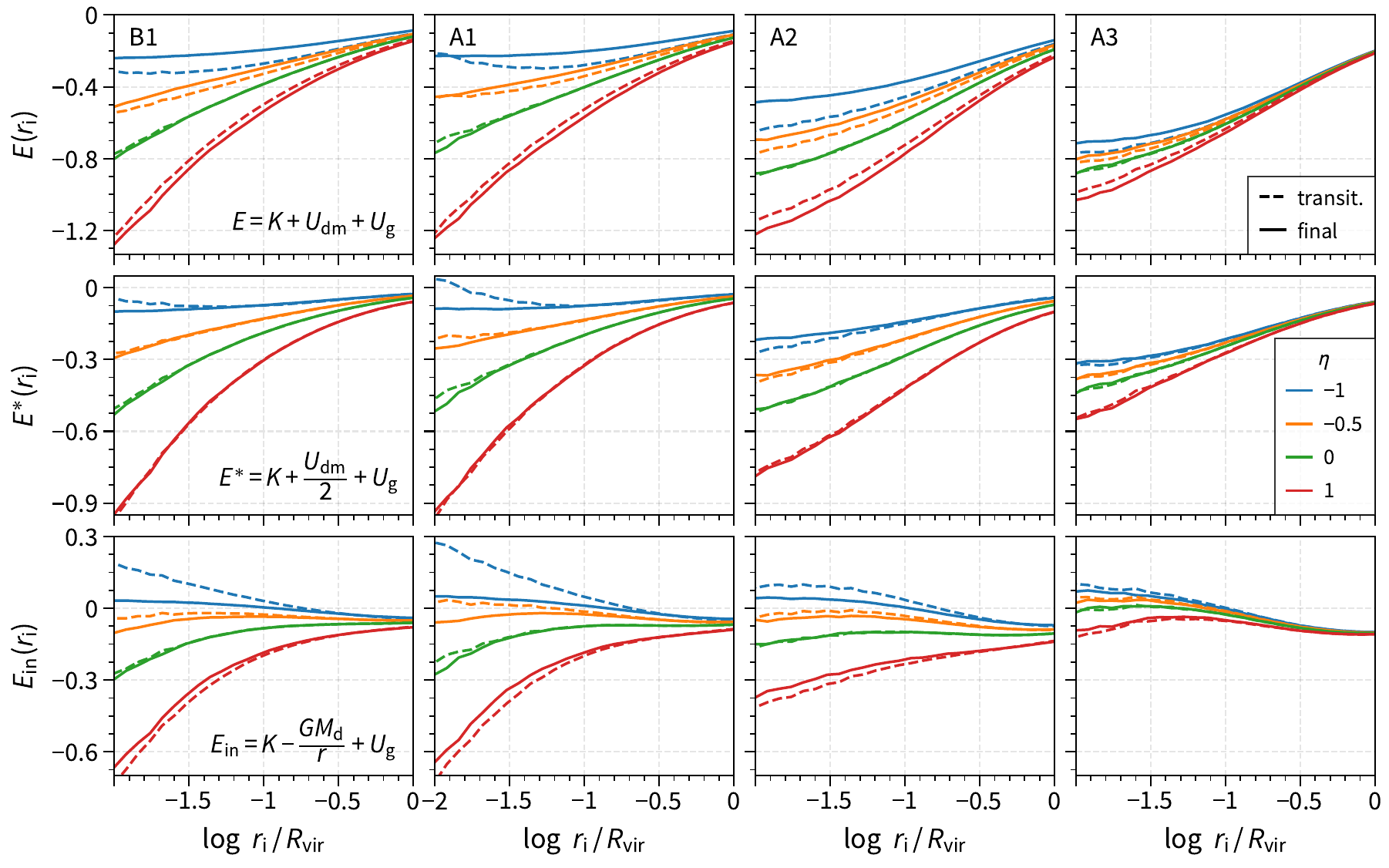}
\vspace{-2em}
\caption{%
The ``energy'' of shells as a function of initial radius $r_i$ in the 
transitional (dashed lines) and final (solid lines) states for simulations B1 and A1--A3 (columns).
The results of three ``energy'' definitions are shown from top to bottom rows
(see \refsec{sec:method4}).
In each column, the energy is scaled by $U_\mathrm{i,0.01}$, the initial potential energy at $0.01\rvir$.
Among the three definitions, $E^\ast$ is best conserved during the relaxation.
}
\label{fig:E(r)}
\end{figure*}

\section{Velocity Anisotropy of simulations}
\label{sec:vel_beta}

\reffig{fig:beta} shows the velocity anisotropy (\citetalias{2008gady.book.....B}, eq.\ 4.61),
$\beta = 1 - \frac{1}{2}\avg{v_t^2}/\avg{v_r^2}$,
as a function of radius in the final snapshots of our 24 simulations,
where ${v_t}$ and ${v_r}$ are the tangential and radial velocities respectively.
The initial conditions of the simulations are taken to be isotropic ($\beta=0$).
The systems with strong gas ejection developed slight radial anisotropy of $\beta\sim 0.05$ to 0.15,
while the remaining simulations remain roughly isotropic.

\section{Energy distribution after the potential change}
\label{sec:der_Nvar}

We start with the joint distribution of the energy and radius, $P (E, r)$.
For an spherical and isotropic system in equilibrium,
a pair of $(E+\dif E, r+\dif r)$ corresponds to a volume 
of $4 \pi r^2 \dif r \times 4 \pi v^2 \dif v$ in phase space.
Recalling 
$f(E) \equiv \dif^6 M / \dif^3 \bm{r} \dif^3 \bm{v}$,
we have
\begin{gather}
  P (E, r) \equiv \frac{\dif^2 M}{\dif E \dif r} = 16 \sqrt{2} \pi^2 f  (E) r^2 \sqrt{E - U (r)},
\end{gather}
where $\frac{\partial E}{\partial v} = v$ and $v = \sqrt{2 (E - U)}$ are used.

We consider the potential change in a small time interval
during which particles do not travel far from their original position.
As the potential changes from $U (r)$ to $U' (r) = U (r) + \Delta U (r)$,
a particle at $r$ with energy $E$ has a new energy $E' = E + \Delta U (r)$. Then
the new energy distribution is
\begin{align}
  N (E') & \equiv \frac{\dif M}{\dif E'} 
           = \int \frac{\dif^2 M}{\dif E \dif r} \left|\frac{\partial (E,r)}{\partial (E',r)}\right| \dif r
           = \int P (E' - \Delta U (r), r) \dif r  \nonumber\\
         & = 16 \sqrt{2} \pi^2 \int_0^{r_{E'}} f (E' - \Delta U (r)) r^2 \sqrt{E' - U'(r)} \dif r,
\end{align}
where $r_{E'}$ is the radius satisfies $U' (r_{E'}) = E'$ and $E' - \Delta U - U = E' - U'$ is used.
If $\Delta U = 0$, the above equation is exactly
the usual $N (E) = f (E) g (E)$, 
where $g(E)$ is the volume of phase space per unit energy (\citetalias{2008gady.book.....B}, eq. 4.56).

A remark: we only consider $N(E')$ for $E'<0$ in above equation,
thus the unbound particles with $E'\ge 0$ are discarded instantaneously.
The automatic treatment of unbound particles might be another merit of CuspCore II,
though the appropriateness and accuracy of such instantaneous removal are to be verified.

\section{Implementation of Method I}
\label{sec:implement}

The implementation of Method I has consulted the public codes of dynamical models, 
\texttt{Agama} \citep{2019MNRAS.482.1525V} and \texttt{SpheCow} \citep{2021A&A...652A..36B}.

In each iteration, we compute $N(E)$, $f(E)$, $g(E)$, $\rho(r)$, and $U(r)$ 
on an equal-spaced grid of $\ln r_E$ or $\ln r$, where $U(r_E) = E$.
Following \citet{2018arXiv180208255V},
a cubic spline interpolation is then used within the grid and a linear extrapolation is used outside 
(thus assuming a power-law function in the very center and outskirt). 
For the inter- and extrapolation, $N(E)$ is expressed as $\ln N(\ln r_E)$ [similarly for $f(E), g(E)$, and $\rho(r)$],
while $U (r)$ is implemented as a bijection between $\ln [1/U(0) - 1/U(r)]$ and $\ln r$.
We compute the integrals in Equations (\ref{eqn:Nvar} -- \ref{eqn:newU}) through 
the Gauss-Legendre quadrature on $\ln r_E$ or $\ln r$ with change of variables.
The integration limits of 0 or infinity are replaced by sufficient small or large values beyond the interpolation grid.
The iteration procedure of the solution stops when the difference in the DM mass profile 
between two adjacent steps satisfies $|M_{k+1}(r)/M_{k}(r)-1|<10^{-5}$ for all radii on the interpolating grid.

The Eddington inversion (\refeqnalt{eqn:eddington}) is more complicated, because it involves
a second derivative, $\dif^2 \rho / \dif U^2$. 
With above interpolation techniques, we rewrite \refeqn{eqn:eddington} in a form
suitable for a Gauss-Jacobi quadrature of the type ($-0.5, 0$),
\begin{gather}
  f (r_E) = \frac{1}{\sqrt{8} \pi^2} \int_{\ln r_E}^{\ln r_{\max}}
  \sqrt{\frac{\ln r - \ln r_E}{U (r) - U (r_E)}}
  \times \frac{\mathcal{D} (\ln r)\ \dif \ln r}{\sqrt{\ln r - \ln r_E}}
\end{gather}
with
{%
\allowdisplaybreaks
\begin{align}
  \mathcal{D} & =  \frac{\dif}{\dif \ln r} \left( \frac{\dif \rho}{\dif U} \right) \notag \\
                      & =  \frac{r \rho}{G M_{\mathrm{tot}}}
    \left[ \frac{\dif^2\ln \rho}{(\dif \ln r)^2} + \frac{\dif\ln \rho}{\dif \ln r} 
    \left(1 + \frac{\dif\ln \rho}{\dif \ln r} - \frac{\dif \ln M_{\mathrm{tot}}}{\dif \ln r}  
    \right)\right],
\end{align}
where $M_{\mathrm{tot}}=\frac{r^2}{G}\frac{\dif U}{\dif r}$ is the total mass profile (including gas)
and $r_{\max}$ is an arbitrary radius where $\rho$ drops sufficiently close to 0.
}

\renewcommand{\thefigure}{G\arabic{figure}}
\setcounter{figure}{0}
\begin{figure*}
\centering
\includegraphics[width=1\textwidth]{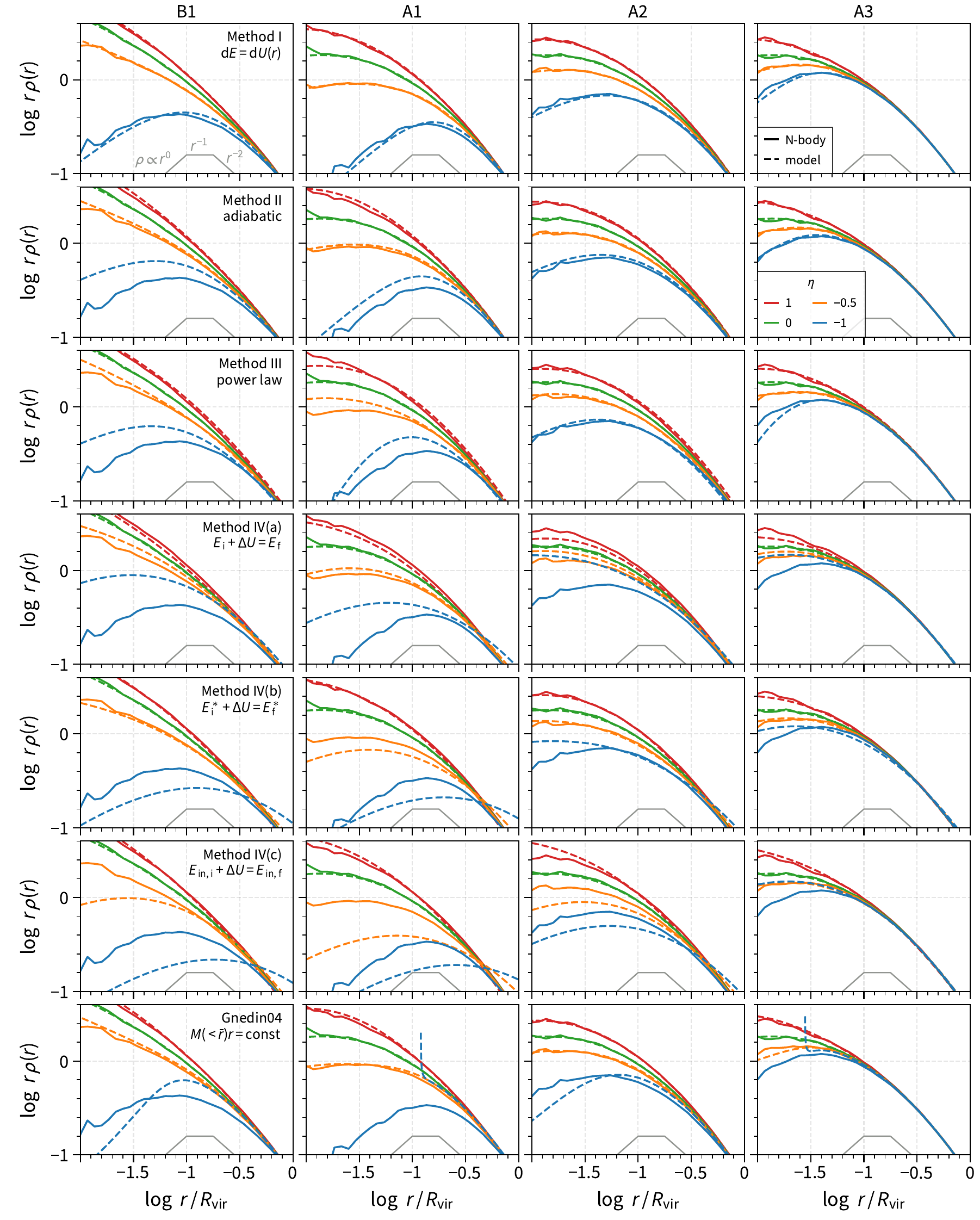}
\vspace{-2em}
\caption{%
Model prediction (dashed lines) for the relaxed DM profiles 
in comparison with simulations (solid lines),
including different initial conditions (columns) and
and fractional gas mass changes $\eta$ (colors, see the legend).
The figure is similar to \reffig{fig:res_cmp}, but showing more methods against more simulations.
The models include:
Method I (energy diffusion, \refsec{sec:method1}),
Method II (adiabatic invariants, \refsec{sec:method2}),
Method III (empirical power-law relation, \refsec{sec:method3}),
three variants of Method IV (a--c) (energy conservation of shells, \refsec{sec:method4}),
and the \citet{2004ApJ...616...16G} model (Appendix \ref{sec:Gnedin04}).
We only show the simulations B1 and A1--A3, because all the methods, except for Method IV (a) and (c), 
work fairly well for Run B2 and B3 where the gas change is weak.
}
\label{fig:res1}
\end{figure*}

\section{Variant solution of Method I using average energy change}
\label{sec:method1_var}

Here we present an alternative solution of Method I.
As shown in  Appendix \ref{sec:integral_change},
the average energy change of particles with given $E$ can be well approximated by
$\avg{\Delta E} \simeq \Delta U(r_\mathrm{cir})$ for short time intervals,
where $r_\mathrm{cir}$ is the circular orbit radius corresponding to $E$.
Therefore, when the potential changes from $U_{k-1}(r)$ to $U_k(r)=U_{k-1}(r)+\Delta U(r)$,
we have
\begin{equation}
  \langle E' \rangle \simeq E + \Delta U (r_{\mathrm{cir}}),
\label{eqn:dE=dU(rcir)}
\end{equation}
with $r_{\mathrm{cir}}=r_\mathrm{cir}(E \mid U_{k-1})$.
Expressing $\Delta U(r_\mathrm{cir})$ as a function of $E$,
we can derive the new energy distribution via change of variables,
\begin{gather}
  N_{k} (E') \simeq N_k(\langle E' \rangle) \simeq \left[ 1 + \frac{\dif \Delta U (E)}{\dif E}  \right]^{- 1} N_{k-1} (E). 
\end{gather}
This can serve as an alternative to Equation (\ref{eqn:Nvar}) for the iterative procedure.
The two approaches give nearly identical results.

Nonetheless, Equation (\ref{eqn:Nvar}) might be favorable
for the physical motivation of tracing the detailed energy diffusion
and for the convenience of handling the velocity anisotropy and additional energy sources
(\refsec{sec:extension}).

\section{Energy definition for shells}
\label{sec:Edef}

The original version of CuspCore (\refsec{sec:method4}) assumes 
the energy of shells that contain a fixed mass is conserved during relaxation.
A necessary condition is thus to conserve the direct sum of all shells' energy,
$\int \rho_\mathrm{dm}(r) E(r) 4\pi r^2\dif r$.
Unfortunately, the conventional $E = K + U$ does not meet this requirement as shown below.

The total energy of an isolated system (which is different from the direct sum of all particles' energy) 
is conserved during the relaxation after an initial gas removal/addition.
The total energy of the DM component under its self-gravity potential ($U_\mathrm{dm}$) 
and a static external gas potential ($U_\mathrm{g}$) is
\begin{align}
  \mathcal{E}_{\mathrm{tot}} & =\mathcal{U}_{\mathrm{tot}} +\mathcal{K}_{\mathrm{tot}} \notag\\
  & = \frac{1}{2}  \int \rho_{\mathrm{dm}} U_{\mathrm{dm}} \dif^3 {\bm{r}}
      + \int \rho_{\mathrm{dm}} U_{\mathrm{g}} \dif^3 {\bm{r}}
      + \int \rho_{\mathrm{dm}} K_\mathrm{dm} \dif^3 {\bm{r}} \notag\\
  & = \int \rho_{\mathrm{dm}}  \left( K_\mathrm{dm} + \frac{1}{2} U_{\mathrm{dm}} + U_{\mathrm{g}} \right) 4 \pi r^2 \dif r
\label{eqn:Etot}
\end{align}
where $\rho_\mathrm{dm}$ and $K_\mathrm{dm}$ are the density and specific kinetic energy profiles respectively.
The factor $1 / 2$ before ${U}_\mathrm{dm}$ is because each pair of DM mass elements has been counted twice
(\citetalias{2008gady.book.....B} eq.\ 2.23).

One can see from Equation \ref{eqn:Etot} that the direct sum of
the conventional energy $E=K + U_\mathrm{dm} + U_\mathrm{g}$ of all DM shells does not conserve in general
(unless $\frac{1}{2} \int \rho_\mathrm{dm} U_\mathrm{dm} \dif^3\bm{r}$ is constant).
Instead, the total energy is conserved if defining the specific ``energy'' $E^\ast$ as
\begin{gather}
  E^\ast = K + \frac{1}{2} U_\mathrm{dm} + U_\mathrm{g}.
\end{gather}
However, $E^\ast$ is not the only form that conserves the total energy.
Using the alternative expression of self-gravity potential energy (\citetalias{2008gady.book.....B} eq.\ 2.24),
one can show that the sum of 
$E_\mathrm{in}=K-GM_\mathrm{dm}(<r)/r+U_\mathrm{g}$ of all DM shells
is also conserved.%
\footnote{%
  One may even find other more complicated forms that conserve the same total energy, e.g., $K + \frac{1}{2} \left( U_\mathrm{dm} + \frac{\rho_\mathrm{g}}{\rho_\mathrm{dm}} U_\mathrm{dm} + U_\mathrm{g} \right)$, or arbitrary linear combination of them.
}
Nothing as we know ensures that such $E^\ast$ or $E_\mathrm{in}$ will be conserved for individual shells.
Therefore, the appropriateness of the energy definition has to be verified with simulations.

We examine the conservation of the three energy definitions, $E^\ast,E$, and $E_\mathrm{in}$, against N-body simulations.
In \reffig{fig:E(r)},
we plot the energy of shells (labeled by initial radii) in the transitional and final states separately.
$E$ exhibits systematic differences between the two states for $\eta \neq 0$.
As expected, it has increased due to the expansion of the DM halo (and thus lowered self-gravity potential)
in cases with $\eta<0$, and vice versa for $\eta>0$.
Among the three definitions, $E^\ast$ is the one best conserved, consistent with the above analysis of total energy.
This is also confirmed by the performance of the model prediction as shown in \reffig{fig:res1}
[Method IV (a--c)].
However, it still exhibits a clear deviation
in the core region for cases with complete gas removal ($\eta=-1$), which we are interested in most.

\section{Comparison between model predictions and N-body simulations}
\label{sec:compare2}

\reffig{fig:res1} shows the model prediction for the relaxed DM profiles
in comparison with simulations. The models include:
Method I (\refsec{sec:method1}),
Method II (\refsec{sec:method2}),
Method III (\refsec{sec:method3}),
three variants of Method IV (a)--(c) (\refsec{sec:method4}),
and the \citet{2004ApJ...616...16G} model (Appendix \ref{sec:Gnedin04}).
We only show the simulations B1 and A1--A3, because all the methods, except for Method IV (a) and (c), 
work fairly well for Run B2 and B3 where the gas change is weak.

\section{Testing Gnedin et al.\ 2004}
\label{sec:Gnedin04}

\citet{2004ApJ...616...16G} is a widely used empirical model of halo adiabatic contraction.
It predicts the final position of shells that encompass a fixed DM mass using the relation,
$M_\mathrm{tot,i}(<\bar r_\mi) r_\mi=M_\mathrm{tot,f}(<\bar r_\mf) r_\mf$,
where $\bar r=0.85\rvir (r/\rvir)^{0.8}$ approximates to the orbit-averaged radius for particles within $r$.

We test the \citet{2004ApJ...616...16G} model against with our N-body simulations in \reffig{fig:res1}.
The model prediction matches the simulations pretty well in most cases,
except for those with complete gas removal ($\eta=-1$, blue dashed curves).
In particular, we get unphysical solutions with zero density in the inner halo 
in Run A1 and A3 with $\eta=-1$, where the initial central potentials were totally dominated by concentrated gas
(see \reffig{fig:profile} for gas fraction).
As explained in \reffig{fig:Gnedin04_ri_rf},
the model fails because the predicted final position of an inner shell becomes greater than that an outer shell.
Nevertheless, we emphasize that these test cases are far beyond the original purpose of the \citet{2004ApJ...616...16G} model.

\renewcommand{\thefigure}{H\arabic{figure}}
\setcounter{figure}{0}
\begin{figure}
\centering
\includegraphics[width=0.95\columnwidth]{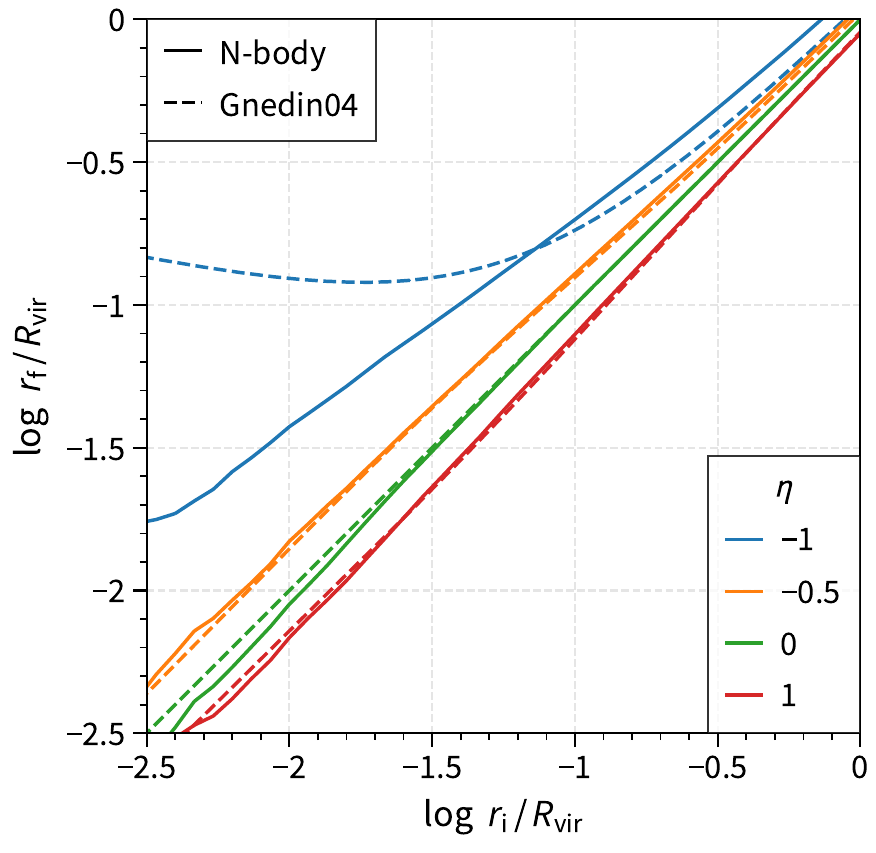}
\vspace{-1em}
\caption{%
Prediction of the \citet{2004ApJ...616...16G} model (dashed lines) 
against A1 simulations (solid lines) for the relation between 
the initial and final location of shells, $r_\mi$ and $r_\mf$.
The model fails to retain any mass (thus leaving a hole) within $\sim 0.1\rvir$
for the case with complete gas ejection ($\eta=-1$).
}
\label{fig:Gnedin04_ri_rf}
\end{figure}



\bsp	
\label{lastpage}
\end{document}